\documentclass[a4paper,12pt,declaration=on]{icldt}
\usepackage{graphicx}
\usepackage{epsfig}
\usepackage{amsmath}
\usepackage{epigraph}
\usepackage{verbatim}
\usepackage{subfig}

\newcommand\beq{\begin{equation}}
\newcommand\eeq{\end{equation}}
\newcommand\bea{\begin{eqnarray}}
\newcommand\eea{\end{eqnarray}}
\newcommand\bi{\begin{itemize}}
\newcommand\ei{\end{itemize}}

\def\D{\Delta}

\def\D{{\mathcal{D}}}

\def\Z{{\bf z}}
\def\t{{\tau}}

\def\q{{\bf q}}

\def\d{{\partial}}
\def\s{{\sigma}}
\def\a{\alpha}
\def\b{\beta}
\def\e{\epsilon}
\def\eps{\epsilon}
\def\z{\xi}

\def\l{\lambda}
\def\Tr{{\rm Tr}}

\newcommand\bra[1]{\left<#1\right|}
\newcommand\ket[1]{\left|#1\right>}
\newcommand\brak[2]{\left<#1|#2\right>}
\newcommand\dif[2]{\frac{\partial #1}{\partial #2}}
\newcommand\diff[2]{\frac{\partial^{2} #1}{\partial #2^{2}}}

\def\half{\frac {1} {2}}
\def\h{\e} 

\usepackage{geometry}
\usepackage{epstopdf}
\author{James M Yearsley}
\title{Aspects of Time in Quantum Theory}
\department{Theoretical Physics}
\newenvironment{preface}{%
\chapter*{Preface}\thispagestyle{plain}}

\dedication{For my parents, to whom I owe everything\\ {\em and}\\For Laura, whose love and support made this possible.
\vspace{50pt}

{\small \em{

but more than all(as all your more than eyes\\tell me)there is a time for timelessness

E.E. Cummings}}}

\begin{document}

\maketitle

\abstract 
\addcontentsline{toc}{chapter}{Abstract}

We consider a number of aspects of the problem of defining time observables in quantum theory. Time observables are interesting quantities in quantum theory because they often cannot be associated with self-adjoint operators. Their definition therefore touches on foundational issues in quantum theory. 

Various operational approached to defining time observables have been proposed in the past. Two of the most common are those based on pulsed measurements in the form of strings of projection operators and continuous measurements in the form of complex potentials. One of the major achievements of this thesis is to prove that these two operational approaches are equivalent. 

However operational approaches are somewhat unsatisfying by themselves. To provide a definition of time observables which is not linked to a particular measurement scheme we employ the decoherent, or consistent, histories approach to quantum theory. We focus on the arrival time, one particular example of a time observable, and we use the relationship between pulsed and continuous measurements to relate the decoherent histories approach to one based on complex potentials. This lets us compute the arrival time probability distribution in decoherent histories and we show that it agrees with semiclassical expectations in the right limit. We do this both for a free particle and for a particle coupled to an environment.

Finally, we consider how the results discussed in this thesis relate to those derived by coupling a particle to a model clock. We show that for a general class of clock models the probabilities thus measured can be simply related to the ideal ones computed via decoherent histories.

\makededication

\preface 
\addcontentsline{toc}{chapter}{Preface}
The work presented in this thesis was carried out in the Theoretical Physics Group, Imperial College, between October 2007 and September 2011. The supervisor was Prof. Jonathan J. Halliwell.

The results presented in Chapters 3 and 7 were published in Journal of Physics A \cite{Ye1} and Physical Review A \cite{Ye2} respectively. The work presented in Chapter 4 was done in collaboration with Jonathan Halliwell and published in Journal of Physics A \cite{HaYe3}. Chapters 5 and 6 are based on work done in collaboration with Jonathan Halliwell and published in Physical Review A \cite{HaYe1} and Physics Letters A \cite{HaYe2}. The work presented in Chapter 8 was done in collaboration with Delius Downs, Jonathan Halliwell and Anna Hashagen and was published in Physical Review A \cite{Yeclocks}.

I am grateful to Carl Bender, Adolfo del Campo, Fay Dowker, Bei-Lok Hu, Seth Lloyd, Gonzalo Muga, Lawrence Schulman and Antony Valentini for their comments and suggestions on the material contained in this thesis.

Some of the work in this thesis was carried out during a short stay at the Apuan Alps Centre for Physics at the Towler Institute, Italy during the conference ``21st Century directions in de Broglie-Bohm theory and beyond.'' I would like to thank Mike Towler and the other conference organisers for their hospitality.

In addition I would like to thank Kate Clements, Ben Hoare, Steven Johnston, Sam Kitchen, Johannes Knoller, Noppadol `Omega' Mekareeya, Tom Pugh and David Weir for many useful, informative and occasionally even relevant discussions.

None of the work set out in this thesis would have been possible without the limitless support and encouragement of my supervisor, Jonathan Halliwell. I am grateful for all his help over the four years of my PhD here at Imperial and I consider myself incredibly fortunate to have had him as a teacher and collaborator.

\tableofcontents
\listoffigures


\chapter{Introduction}
\epigraph{The time is out of joint; O curs\'ed sprite,\\That ever I was born to set it right!}{Shakespeare, Hamlet}

\section{A Brief History of Time in Quantum Theory}\label{sec:1.1}

The treatment of time observables is an important loose end in quantum theory \cite{time, time2}. Whilst there is no indication of any disagreement between the predictions of quantum theory and the outcomes of experiment, time observables constitute a  class of quantities that are observables in classical mechanics, but for which no satisfactory treatment exists in standard quantum theory.

As well as being an important topic in its own right, a consistent treatment of time observables might also shed light on problems in quantum gravity and quantum cosmology. One of the most tricky problems faced by any attempt to quantise gravity is that of the problem of time \cite{Ishtime}. In essence the symmetries of general relativity prevent the identification of any variable to play the role of time in quantum gravity. Whilst there is no generally agreed way of tackling this problem one school of thought holds that any approach to quantising gravity must proceed by treating time and space on an equal footing. A theory treating time and space in a truly symmetric way would have as observables probabilities for entering given regions of spacetime rather than, for example, statements about the position of a particle at a fixed moment of time.
Since these are not the kind of probabilities one normally deals with in quantum theory it is instructive to see if the same exercise can be performed in non-relativistic quantum theory. That is, can one assign probabilities to questions of the form, ``What is the probability of finding a particle in a given region of space, $\Delta$, in a given interval of time, $[t_{0},t_{1}]$?''

\begin{figure}[htbp]
\centering
\subfloat[Probabilities for Space-like Surfaces]{\includegraphics[width=0.5\textwidth]{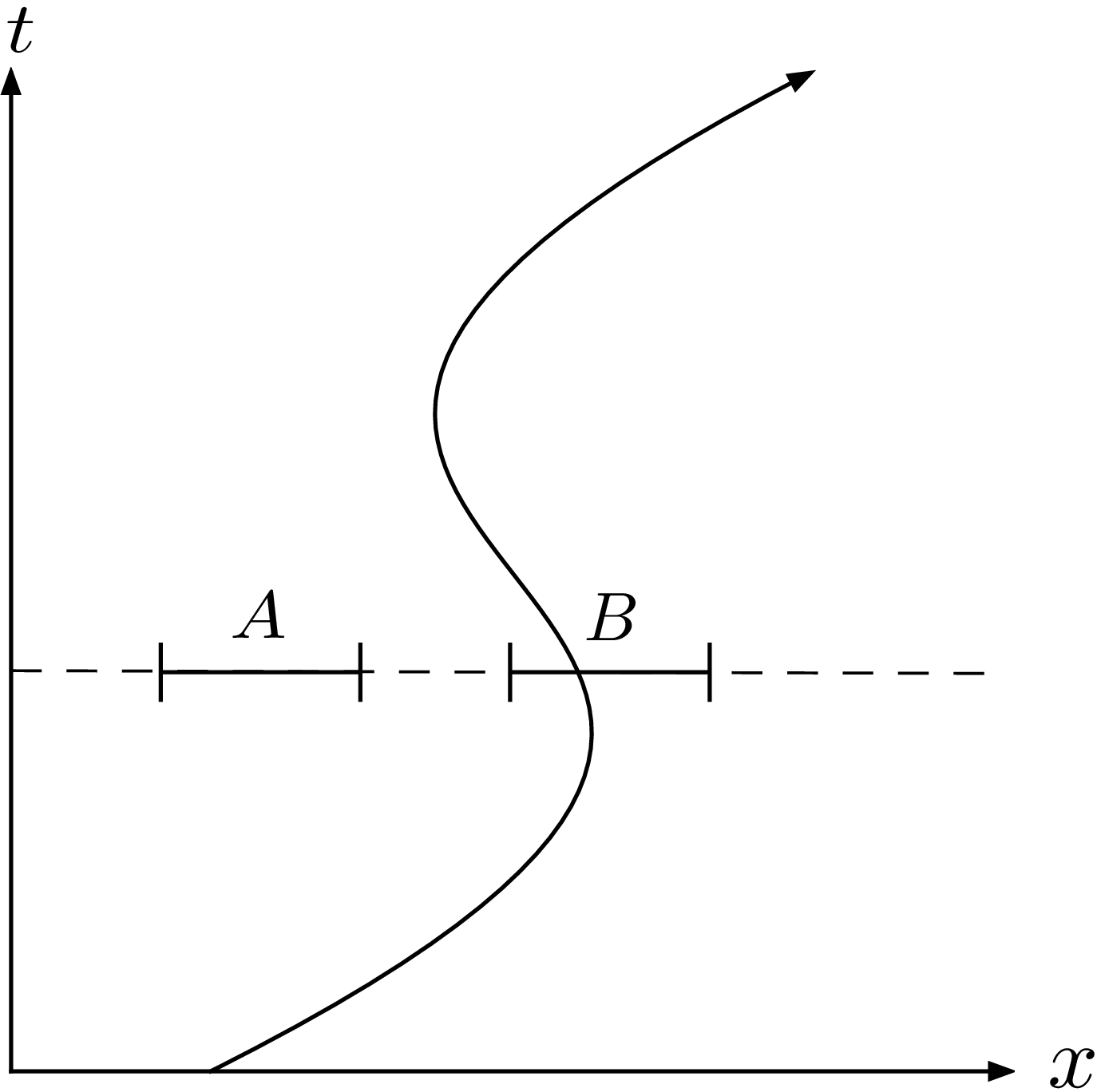} }
\subfloat[Probabilities for Time-like Surfaces]{\includegraphics[width=0.5\textwidth]{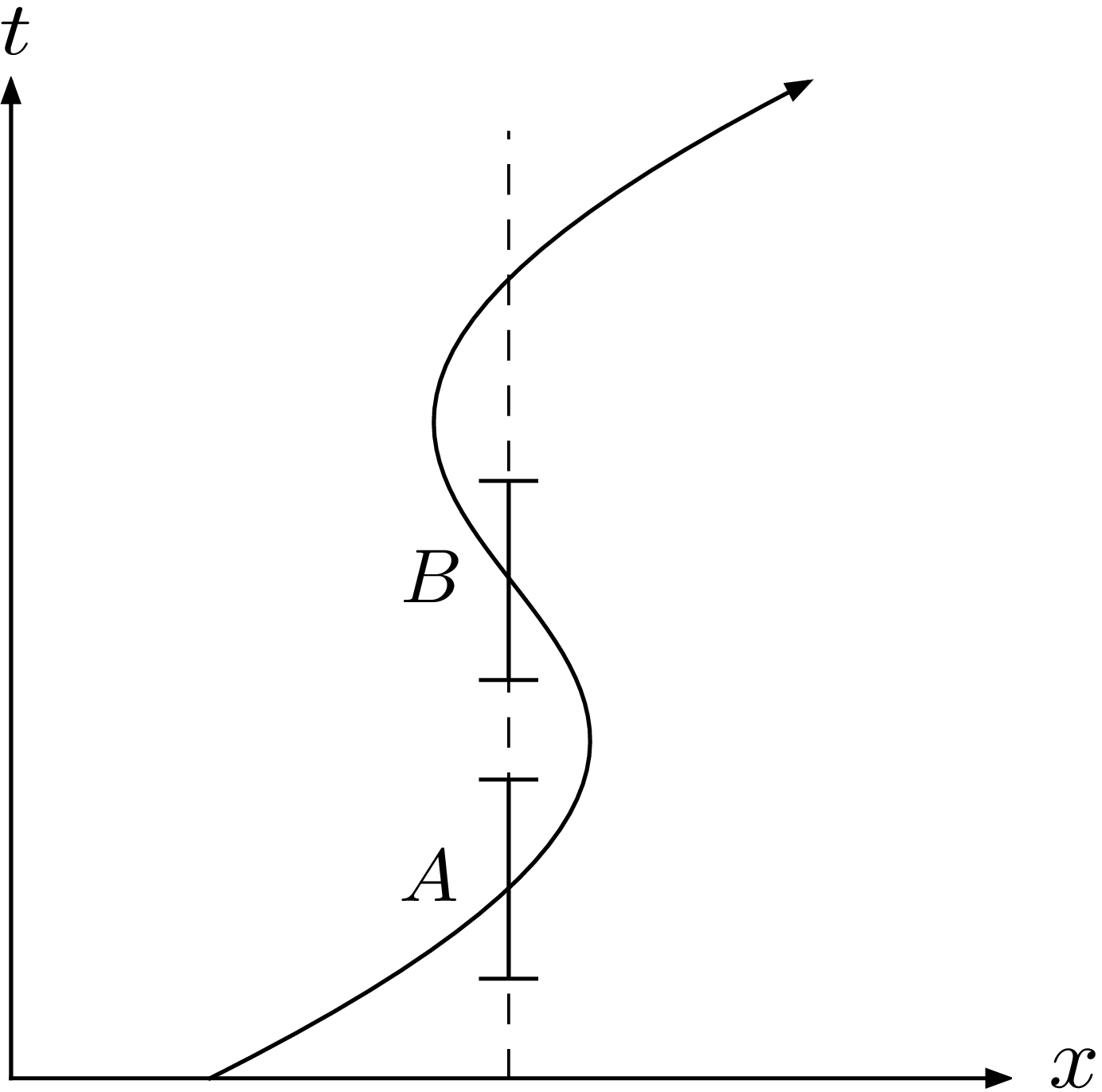}}
\caption[Probabilities for Space-like and Time-like Surfaces]{{\em {\em(a)} Since every trajectory crosses a surface of constant time only once, passing through disjoint intervals $A$ or $B$ at a given time represent exclusive alternatives. {\em(b)} In contrast trajectories may cross a surface of constant position many times. Therefore passing through disjoint intervals $A$ or $B$ are {\em not} exclusive alternatives. }}
\label{fig:space}
\end{figure}

Asking questions of this sort quickly leads us to another problem in non-relativistic quantum theory, which is this: ``What exactly is the status of the variable $t$ that appears in Schr\"odinger's equation?'' It should be clear immediately that it is not related to any operator on the Hilbert space of the system. Indeed it is not really a quantum quantity at all and is probably best thought of as referring to some external classical clock. This hybrid approach to dynamics, having a quantum wavefunction depend on an external classical clock variable, could be seen as troubling. An operationalist might reply that the co-ordinate $x$ also refers to an external classical world, but this line of reasoning is over simplistic for three reasons. The first is that both position and its conjugate variable, momentum, appear as operators in quantum mechanics, whereas energy is represented by an operator in quantum mechanics, but time is not. There is therefore a lack of symmetry in the treatment. The second reason is that quantum mechanics may be written in a representation-free manner in terms of abstract vectors in Hilbert space. Here there need be no mention of position or momentum, and yet the Schr\"odinger equation for evolution in this abstract space still includes the time coordinate. The final reason is the most profound, and again touches on the lack of symmetry in the treatment of space and time coordinates. In non-relativistic quantum mechanics, the object,
\beq
P(x)dx=|\psi(x,t)|^{2}dx
\eeq
gives the probability of finding the particle in the interval $[x,x+dx]$ at the time $t$. What is the corresponding probability for finding the particle at the position $x$ in the interval $[t,t+dt]$? It is not given by
\beq
P(t)dt\neq|\psi(x,t)|^{2}dt,
\eeq
 since finding the particle at a position $x$ at the times $t_{1}$ or $t_{2}$ are not exclusive alternatives, see Fig.(\ref{fig:space}). Yet in the standard interpretations of quantum mechanics the wavefunction is supposed to contain all the information about a given system \cite{IshQM}. If the square norm of a general wavefunction is not a probability distribution on a set of times does this mean there can be no satisfactory treatment of time observables in standard quantum theory?

To get a better idea of where the difficulty lies it is instructive to look more closely at the classical case. Here the solution of a particular problem may be expressed in terms of an equation of motion for the particle, of the form,
\beq
x=f(t,x_{0},p_{0})\label{1.3}
\eeq
where $x_{0}$ represents the initial condition. This is the solution to the question: ``Where is the particle at time $t$?'' Under general conditions one may invert this equation of motion to give,
\beq
t=F(x,x_{0},p_{0})\label{1.4}
\eeq
which is the solution to the question: ``When is the particle at position $x$?'' Crucial to this analysis is the concept of a trajectory and that the particle will with certainty be found somewhere on this trajectory, at any time. This concept is of course lacking in quantum theory. 

In classical mechanics observables are given by functions on phase space, and so Eq.(\ref{1.4}) is equivalent to some phase space function,
\beq
t=\t(x,p).\label{1.5}
\eeq 
Crucially, unlike classical observables like the energy or angular momentum, construction of Eq.(\ref{1.5}) requires knowledge of the equations of motion, Eq.(\ref{1.3}). Strictly speaking then, even in classical mechanics questions such as ``When does the particle reach the origin?'' have a different character to questions like ``What is the energy of the particle?''. The observable coresponding to the former is a function on phase space which is constructed using the equations of motion. In a sense, therefore, the surprise is not that quantum theory struggles to deal with time observables, but that classical mechanics deals with them so easily. This is essentially a consequence of the realist nature of classical physics. 

Approaches to the problem of defining time observables in quantum theory fall roughly into three camps. Although it is not the intention of this thesis to provide a complete overview of the literature on time observables (see Ref.\cite{MugLev} for a detailed review in the case of arrival time), the following characterisations will be useful in understanding how the approaches taken in this work relate to previous studies.

\subsection*{Trajectory Before Quantisation}
These are approaches which in some way borrow the classical structure of a trajectory and use it in the quantum analysis. 
One way to do this is to begin with the classical description, in terms of Eq.(\ref{1.3}) and use this to construct the classical phase space observable Eq.(\ref{1.5}), as outlined above. One then tries to quantise this quantity to obtain the operator,
\beq
\hat t=\t(\hat x, \hat p)\label{1.6}
\eeq
This category of approaches includes the Aharanov-Bohm time operator \cite{ABtime}. Apart from the obvious issues surrounding whether Eq.(\ref{1.6}) is well defined for a given classical observable, there is something troubling about this analysis. Since the quantum system we are trying to analyse will not be explicable in terms a particle following a single trajectory, it is not clear whether using the classical trajectory to obtain the operator Eq.(\ref{1.6}) is a valid procedure. Even if the resulting operator is well defined the relation between this observable and the results of measurement remains unclear.  Since this is essentially the question of how to understand the quantisation of a given classical system we are unlikely to make much headway with this issue here. It does, however, motivate trying to find an alternative way to proceed.

Other approaches that fall into this category include the analysis of Kijowski and its variants, see eg Refs.\cite{Kij,HegMug}. Here one begins by deciding what properties one would expect the quantum probability distribution to possess, such as time translation covariance or positivity, based on the properties of the classical observable. One then uses this ``wish list'' of properties to determine the correct operator, with the eventual hope that a sufficiently detailed list of properties might uniquely determine this operator. It is not clear that this approach should work in all cases however: it is easy to think of quantum observables, such as angular momentum or the energy levels of a harmonic oscillator, that have a very different character from their classical counterparts. 

The general attitude of this thesis is that whilst any of the analyses above might arrive at the correct definition of a time observable in a particular case, they cannot be relied upon to do so in all cases.

\subsection*{Trajectory After Quantisation}
These approaches include those based on the de Broglie-Bohm (dBB) interpretation of quantum theory \cite{Holland}, analyses based on the WKB approximation \cite{jjhqcnotes} and also the decoherent histories approach presented in this thesis. Here one first quantizes the system and then looks for structures to play the role of the classical trajectories used to define the classical observables. For example for gaussian states the center of the wavepacket will follow simple curve in configuration space. Provided one does not perform measurements that disturb this state this limited notion of trajectory can be used to find approximate arrival and dwell times for the particle, limited by the width of the wavepacket. Of course these states form only a limited class of possible states, and it is a general feature of this category of approaches that they may fail for highly non-classical states.

In the DH approach, described in more detail below, the aim is to assign probabilities to coarse grained histories of a quantum system. Provided these probabilities can be assigned consistently one can then use these histories in much the same way as the classical trajectories to discus time observables. 

In the dBB interpretation, the wavefunction is supplemented by a real particle following a trajectory computed from the wavefunction. Since a quantum particle always comes equipped with a unique trajectory computation of time observables is, at least conceptually, as trivial as in the classical case. The price one pays, however, is that the behavior of these trajectories generally depends on whether or not one choses to carry out a measurement on the system. Thus although in dBB probabilities for time observables can be obtained without the need for an observer, the probabilities obtained in this way do depend on whether or not there is an observer.  We will not have much to say in this thesis about dBB, for more information see Ref.\cite{Holland}.

\subsection*{Trajectory Free Approaches}
These approaches may also be called ``operational.'' Here one takes a completely different tack and instead looks for other ways to define time observables classically, and then tries to repeat this analysis quantum mechanically. 

These approaches include studies based on model clocks, see Ref.\cite{clock} and Chapter 8 of this thesis. Here the idea is to couple the system we wish to measure to some additional degrees of freedom which function as a clock. In this context a clock is simply a quantum system coupled to our system of interest in such a way that the final state of the clock is correlated with the time observable we wish to measure. More details about this scheme can be found in Chapter 8.

In classical mechanics many time observables can be defined by imposing absorbing boundary conditions on the system. See Chapter 5 for a full discussion in the arrival time case. The analogue of these absorbing boundary conditions in quantum theory in the inclusion of a complex absorbing potential, see Ref.\cite{complex} and also Chapter 5 of this thesis. 

Typically the probability distributions obtained from these approaches are complicated functions of the properties of both the particle and measuring apparatus. There may or may not exist limits in which dependence on the measurement parameters drops out, depending on the set up.
\newline

As well as the technical differences between these approaches the different approaches display very different attitudes towards the observables we are trying to define. In particular the first and third approaches assume that an observable such as the arrival time of a particle is always a well defined quantity and that for any system one may obtain the probability distribution of that observable. The second class of approaches, in contrast, are based on recovering the classical notion of trajectory after quantisation, and thus they are generally only well defined for semiclassical systems. Proponents of these approaches would claim that quantities such as the arrival time distribution {\it are only defined} for suitably classical systems\footnote{In much the same way as it is claimed that time itself is only {\it emergent}, from an ultimately {\it timeless} theory of quantum gravity \cite{IshBut}.}. 

An obvious question is whether there is anything linking these three different approaches to time observables. This is an important question, firstly since although any particular approach might claim to have the solution to these problems the validity of these approaches themselves is questionable. If we could demonstrate that for a particular time observable the same results are obtained via a number of approaches this would give us confidence that this result is correct. Secondly, the different approaches discussed above have their roots in different interpretations of quantum theory. Although we will try and avoid philosophical questions about the foundations of quantum theory in this thesis if we were to show that, for example, expressions for time observables in de Broglie-Bohm theory were identical to those obtained via decoherent histories this might shed interesting light on the relationship between different interpretations of quantum theory \footnote{Not to mention the fact that although any interpretation of quantum theory is constrained by the requirement that it must agree with the predictions of standard Copenhagen quantum theory, since Copenhagen quantum theory struggles to incorporate many time observables there is a possibility that different interpretations of quantum theory may be free to make different predictions for time observables. There may therefore be experimental tests that could be done to distinguish between them. Exciting as this line of thought is, we shall not dwell on it in this thesis.}.

\section{The Quantum Zeno Effect}\label{sec:1.2}

There are a number of interesting quantum effects relevant to the definition of time observables, the most important of which is probably the quantum Zeno effect. The effect was first noticed by Allcock in his seminar series of papers on the arrival time problem \cite{All}, but the first serious study was undertaken by Misra and Sudershan in Ref.\cite{Zeno}. 

The quantum Zeno effect is often expressed as a ``freezing'' of evolution due to repeated measurement.  
Much has been learned about the phenomenon since the early investigations however and it is now understood that there is more to the effect than simply preventing evolution. Repeated projections onto a subspace allow for unitary evolution within that subspace, the effect of the repeated measurement being simply equivalent to imposing ``hard-wall'' boundary conditions, preventing the particle from leaving the subspace \cite{Sch2}.  It is also known that there also exists an ``anti-Zeno'' effect, that is, a speeding up of evolution due to repeated measurement \cite{Aze}, although we shall have nothing to say about the anti-Zeno effect in this thesis.

The quantum Zeno effect plays an important role in study of time observables because, at least for the ``trajectory after quantisation'' and ``trajectory free'' approaches outlined in the previous section, the definition of these observables involves statements about the state of the system at multiple times. This means one has to consider sequential measurements on the system of interest, if these are performed too frequently the quantum Zeno effect will result. However the situation is complicated by the fact that most discussions of the quantum Zeno effect take place in a setting appropriate for quantum optics or quantum information, ie a system with a small number of discrete energy levels, not in the context of particles in 1D. In the rest of this section we therefore outline the standard discussion of the quantum Zeno effect and the way in which this discussion might be modified to make it relevant to time observables. Our discussion will be brief, since the details of the Zeno limit in the arrival time problem etc are worked out more thoroughly in Chapter 4.

We begin by introducing the Zeno effect in the way it appears in the original discussion of Misra and Sudarshan, Ref.\cite{Zeno}. The idea is to utilise some form of product formula to show that the repeated process of evolution followed by measurement is equivalent to restricted evolution in the subspace defined by the measurement. We consider a system with initial state $\ket{\psi}$ which evolves under a Hamiltonian $H$ for a total time $T$, interspersed with $N$ projective measurements $P$. The state at time $T$ is given by,
\bea
\ket{\psi_{T}}&=&V_{N}(T)\ket{\psi}=e^{-iHT/N}Pe^{-iHT/N}...e^{-iHT/N}P\ket{\psi}\nonumber\\
&=&[\exp(-iHT/N)P]^{N}\ket{\psi}\label{1.7a}
\eea
One then considers the limit $N\to\infty$, and one can show, under certain conditions on $H$ and $P$,
\beq
V_{\infty}(T)=\lim_{N\to\infty}V_{N}(T)=\exp(-iH_{Z}T)P\label{1.7},
\eeq
where the Zeno hamiltonian $H_{Z}=PHP$. This shows that evolution under repeated measurement leads to unitary evolution in the subspace defined by the measurements. 

If we consider the evolution in Eq.(\ref{1.7a}), with $P=\theta(\hat x)$ and $H$ the free hamiltonian for a particle in 1D, we have a simple model of the type of measurement set up we might use to determine the time at which  the particle arrived at $x=0$. Without concerning ourselves too much with the details, since these are covered in Section \ref{sec:1.4} and the rest of this thesis, it is clear that the Zeno effect sets a lower limit on the resolution with which we can define an arrival time using this set up. 

Although this analysis is mathematically precise this precision is obtained at the expense of providing any great physical insight. A key question concerns the time scale on which the Zeno effect begins to occur. That is, what is the minimum spacing between measurements, $T/N$, that I can take without causing significant reflection from the measuring aparatus. This is important since we can hardly expect to achieve $N\to\infty$ in the real word. 
One way to approach this is to look for the time scale on which the state may be said to leave the subspace defined by $P$. The idea is that projections more frequent than this will ``interrupt'' the state in the process of exiting the region, so that this time scale provides an upper bound on the separation between projections necessary to give rise to the Zeno effect.  We follow the analysis of Peres \cite{Peres} and consider an initial state, with evolution for a time $t$ followed by a measurement which projects back onto the original state. The survival probability, $S(t)$, is given by,
\bea
S(t)&=&||\bra{\psi}\exp(-iHt)\ket{\psi}||^{2}\nonumber\\
&=&1-t^{2}(\Delta H)^{2}+O(t^{4})=1-(t/\t_{z})^{2}+O(t^{4})\label{1.8.1}
\eea
where we have defined the ``Zeno time''\footnote{Note that there are many different notions of the time it takes a system to make a transition. See \cite{Sch} for a more detailed discussion.},
\beq
\t_{z}^{-2}=(\Delta H)^{2}=\bra{\psi}H^{2}\ket{\psi}-\bra{\psi}H\ket{\psi}^{2}.\label{1.8}
\eeq
We see from this expression that the short time behaviour is quadratic in $t$, thus we confirm that in the limit of infinitely frequent measurements the state does not evolve. Secondly, we see that there is a characteristic time scale, $\t_{Z}=(\Delta H)^{-1}$, on which the evolution takes place.

Although we have arrived here at what is normally called the ``Zeno'' time, it is not clear that this really is the time scale on which the projections give rise to reflection and thus the Zeno effect.  For a minimum uncertainty gaussian wavepacket with spatial width $\s$ peaked around some momentum $p_{0}$ Eq.(\ref{1.8}) gives $\t_{z}=m\s/|p_{0}|$. This is indeed the time scale on which the state crosses the origin, however this is an essentially classical timescale and it seems unlikely that this is relevant for reflection, which is a quantum effect. It is also worth pointing out that if we really are limited in the accuracy of our description by the time scale  $1/\Delta H$ then it is impossible to formulate a description of crossing time probabilities in the normal sense. This is because our minimum temporal resolution is the same order as the time taken for our wavepacket to cross the origin, so the probabilities for crossing will essentially be 1 for one time interval and 0 for the rest.

Projecting back onto the original state captures the notion of the time on which the state changes, and this is certainly the same as leaving {\em some} subspace of the full Hilbert space, but we are interested in a particular subspace, that of states with support only in $x>0$ and this analysis does not capture that. There is another problem, which is that Eq.(\ref{1.8.1}) is a Taylor expansion of the initial state about $t=0$. It is known that for general wavefunctions the behaviour  may not be analytic at $t=0$. Put another way, the Zeno time is expressed in terms of moments of the wavefunction, and these may not exist for general wavefunctions. This is particularly apparent when one tries to extend this analysis to more general measurements.

 One of the major achievements in this thesis is to show, in Chapter 4, that the minimum temporal resolution is in fact set by $1/E$, where $E$ is the energy of the system, and not by Eq.(\ref{1.8}).

\section{The Backflow Effect}\label{sec:1.3}
Another quantum effect relevant to the understanding of time observables is the backflow effect \cite{BrMe,back}. Consider an initial state $\ket{\psi}$ consisting entirely of negative momenta evolved freely for a time $t$ and compute the standard Schr\"odinger probability current at the origin, defined in such a way that classically it should be positive. This can be written in operator form as,
\beq
J(t)=\frac{(-1)}{2m}\bra{\psi}e^{iHt}\{\hat p \delta(\hat x)+\delta(\hat x)\hat p\}e^{-iHt}\ket{\psi}\label{1.back.cur}.
\eeq
(We will discuss the current in much more detail in later sections.) Now the operator $(-)(\hat p \delta(\hat x)+\delta(\hat x)\hat p)$ is not positive, even when restricted to act on states with negative momentum. This means although Eq.(\ref{1.back.cur}) should be positive classically, it need not be quantum mechanically. What is remarkable is that very little is known about this effect. One thing that is known is that the amount of probability that can flow backwards is bounded, in the sense that,
\beq
\int_{t_{1}}^{t_{2}}dtJ(t)\geq\l,
\eeq
where $\l$ is some dimensionless constant computed numerically to be $\l\approx-0.0384517$ \cite{back}. The existence of such a bound is somewhat unexpected, Bracken and Melloy in Ref.\cite{BrMe} conjecture that it is a ``new dimensionless quantum number.'' Certainly there is no dependence on $\hbar$ in this constant and as such it has no obvious classical limit. 

One interesting set of questions concern the state for which this bound is obtained, does it have an analytical expression? Does it have an obvious physical interpretation? 
Another question concerns the typicality, or otherwise, of backflow. Do relatively standard wavefunctions such as gaussian wavepackets or superpositions of these give backflow, or do we require more pathalogical states? If we can indeed produce backflow with more familiar states, is it possible to get close to the maximum amount of backflow? 

Some work towards addressing these questions is in progress \cite{YeBack}, but much remains unclear. A better understanding of the backflow effect would shed much light on the problem of defining time observables in quantum theory. This is because, together with the Zeno effect, it is these quantum effects that provide the fundamental limitation on the accuracy with which time observables can be defined.

\section{The Arrival Time Problem in Quantum Mechanics}\label{sec:1.4}
\subsection{General Theory}

\begin{figure}[h] 
   \centering
   \includegraphics[width=10cm]{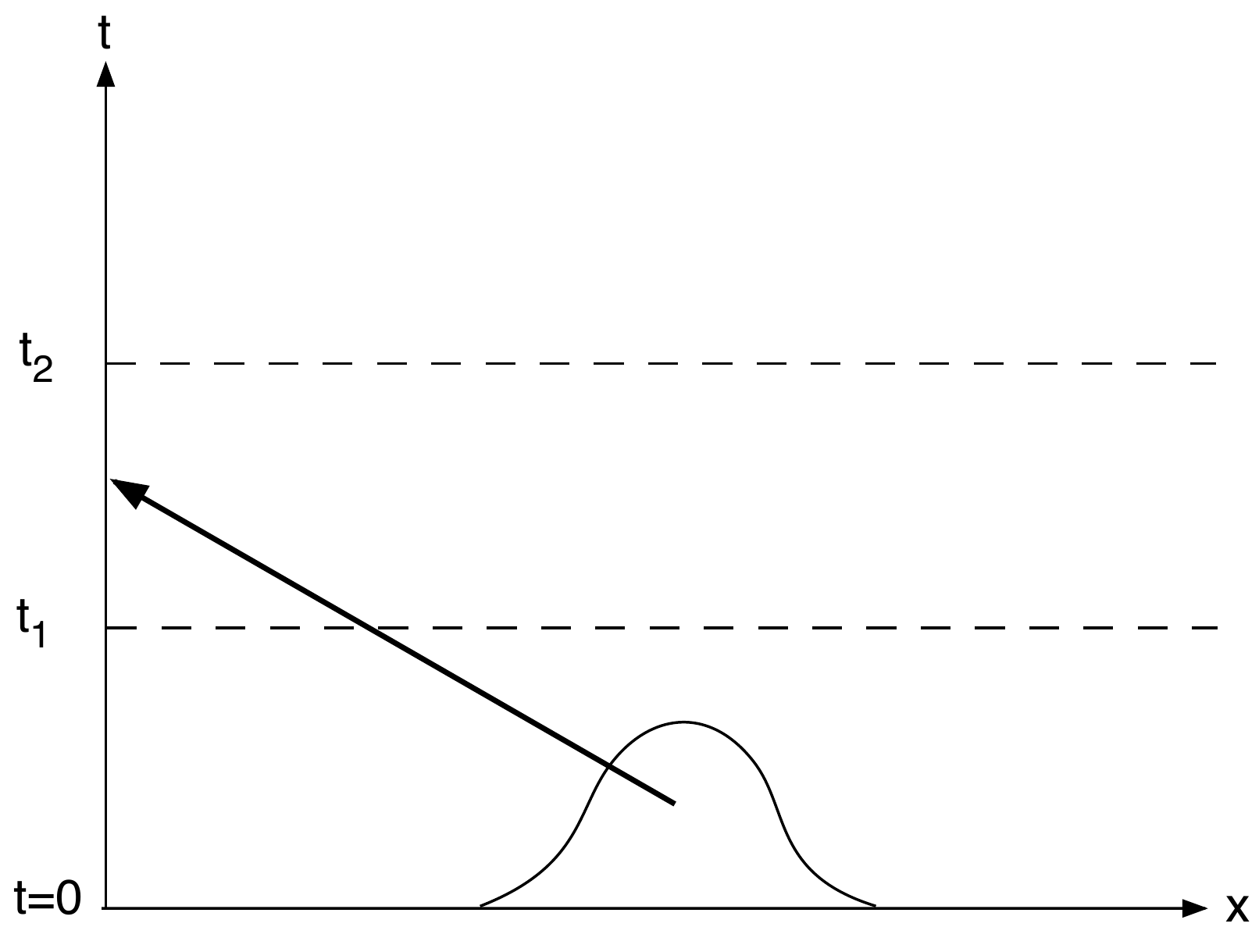} 
   \caption[The Arrival Time Problem]{{\em The Arrival Time Problem: What is the probability that an incoming wavepacket crosses the origin during the time interval $[t_{1},t_{2}]$?}}
   \label{fig:f1}
\end{figure}

Much of this thesis will be concerned, directly or otherwise, with the question of analysing the arrival time problem in quantum theory, pictured in Fig.(\ref{fig:f1}). It therefore seems useful to give a brief account of the problem here and also discuss some of the features any solution must posses. 

The classical ``arrival time problem'' is the following, ``Given a free particle which, at $t=0$ has position $q_{0}$ and momentum $p_{0}$, what is the time at which the particle crosses the origin?'' The solution, of course, is that the particle crosses at a time,
\beq
\t_{c}=-\frac{m q_{0}}{p_{0}}.\label{1.12}
\eeq
In order to compare the definition of time observables in classical and quantum physics it is useful to work in terms of classical phase space distributions. 
For a free classical particle in 1D the phase space probability distribution $w_{t}(q,p)$ obeys the following,
\beq
\frac{\partial w}{\partial t}=-\frac{p}{m}\frac{\partial w}{\partial q}\label{1.weq}.
\eeq
The solutions of this equation are of the form,
\beq
w_{\t}(q,p)=w_{0}(q-p\t/m,p),
\eeq
so that evolution is just a linear transformation in phase space.

We can write the probability of crossing at a time $\t$ in terms of this distribution as,
\beq
\Pi_{c}(\t)=\int dpdq \delta(\t+mq/p)w_{0}(q,p)=\int dpdq \frac{|p|\delta(q)}{m} w_{\t}(q,p)\label{1.13a}
\eeq
For states consisting of entirely negative momenta this may also be written as,
\beq
\Pi_{c}(\t)=-\frac{\partial}{\partial \t}\int dpdq \theta(q) w_{\t}(q,p)\label{1.14a},
\eeq
which can be seen by using Eq.(\ref{1.weq}) in Eq.(\ref{1.14a}) and then integrating by parts.

Eqs.(\ref{1.13a}), (\ref{1.14a}) express the arrival time distribution in two distinct ways. Eq.(\ref{1.13a}) tells us the arrival time distribution is the expectation value of a certain phase space distribution, whilst Eq.(\ref{1.14a}) tells us the arrival time distribution can be defined in terms of the ``survival probability.'' More precisely the arrival time probability is just the rate at which probability leaves the region $x>0$, ie the flux across $x=0$. We will see below that these two expressions suggest different ways of obtaining the quantum arrival time distribution.

The quantum arrival time problem is essentially the following question, ``For a given wavefunction what is the probability distribution on the set of times at which the system may be said to arrive at $x=0$?'' For most situations of interest we can put the following extra constraints on the wavefunction; at some initial time $t=0$ the wavefunction consists only of negative momenta and the particle is concentrated in $x>0$.

The challenge is to understand the correct quantum analogue, $\Pi(\t)$, of Eqs.(\ref{1.13a}), (\ref{1.14a}), if indeed it exists. We need to do this subject to the following constraints,
\begin{itemize}
\item $\Pi(\t)$ must be a valid probability distribution, ie $\int_{-\infty}^{\infty} d\t \Pi(\t)=1$ and $\Pi(\t)\geq0$.
\item We must recover Eqs.(\ref{1.13a}), (\ref{1.14a}) in some classical limit.
\end{itemize}

There are other constraints we could impose, indeed in Ref.\cite{Kij} is was shown that demanding that $\Pi(\t)$ satisfy some set of conditions abstracted from the classical case is, almost, enough to fix it identically. This is an example of a ``trajectory before quantisation'' approach, see Section \ref{sec:1.1}. 

In some ways the most natural way to quantise $\Pi_{c}(\t)$ is via Eq.(\ref{1.13a}). This suggests the (formal) quantum expression,
\beq
\Pi(\t)=\bra{\psi_{0}}\delta(\t-\hat \t)\ket{\psi_{0}}\label{1.15a}
\eeq
where the arrival time operator,
\beq
\hat \t=-m \widehat{\left(\frac{q}{p}\right)}
\eeq
is some quantisation of Eq.(\ref{1.12}). We hit a problem, however, which is that it has proven surprisingly difficult to find a way of defining $\hat\t$ as a self-adjoint operator. This means Eq.(\ref{1.15a}) is merely formal and cannot be used to define $\Pi(\t)$. Work has been done to try and get round this problem \cite{MugLev,ABtime}, but we will not consider this further in this thesis. At any rate, the criticisms inherent to any approach of this type, as laid out in Section \ref{sec:1.1} apply here.

\subsection{Standard Forms of the Arrival Time Distribution}

We review here some of the standard, mainly semiclassical,
formulae for arrival time. It is useful to collect these expressions in one place as we will refer to them repeatedly throughout this thesis. None of these semiclassical expressions are entirely satisfactory, but any more fundamental approach to defining arrival times must reproduce these expressions for some suitable class of states and time intervals. 

We consider a free particle
described by an initial wavepacket with entirely negative momenta
concentrated in $x>0$.  A widely discussed candidate for the
distribution $p(t_{1},t_{2})$  is the natural quantum version of Eq.(\ref{1.14a}), the integrated probability current density \cite{MugLev,cur,HaYe1},
\beq
p(t_{1},t_{2})=\int_{t_{1}}^{t_{2}}dtJ(t)\label{1.16a}.
\eeq

Defining $P(t)=\theta(\hat x_{t})$, and introducing the Wigner function \cite{Wig}, defined for a state $\rho(x,y)$ by
\beq
W_{t}(p,q)=\frac{1}{2\pi}\int d\z e^{-ip\z}\rho(q+\frac{\z}{2},q-\frac{\z}{2}),\label{1.wigdef}
\eeq
$J(t)$ can be written in any of the following forms:
\bea
J(t)&=&-\frac{\partial}{\partial t}\bra{\psi_{0}}e^{iHt}\theta(\hat x)e^{-iHt}\ket{\psi_{0}}\nonumber\\
 &=& \frac{(-1)}{2m}\bra{\psi_{t}}
\left(  \hat{p} \delta(\hat x)+ \delta(\hat x)\hat{p}\ \right) \ket{\psi_{t}}\nonumber\\
&=&\frac{i}{2m}\left(\psi^{*}(0,t)\frac{\partial
\psi(0,t)}{\partial x}-\frac{\partial \psi^{*}(0,t)}{\partial
x}\psi(0,t)\right)\nonumber\\
&=&\int dpdq \frac{ |p|\delta(q)}{m}W_{t}(p,q)\label{1.8.cur}
\eea
This probability is normalised to 1 when $t_{1}=0$ and $t_{2}\to\infty$, and has the correct semiclassical limit  \cite{HaYe1}. This expression arises from considering the survival probability, that is, the probability that the particle is still in $x>0$ at some time $t$,
\beq
p(x>0,t)=\bra{\psi_{t}}\theta(\hat x)\ket{\psi_{t}}.
\eeq
The rate at which probability leaves $x>0$ is then given by Eq.(\ref{1.8.cur}) and this is a candidate for the arrival time probability distribution.
There is a problem with Eq.(\ref{1.8.cur}) however, which is that it need not be positive, even for states consisting entirely of negative momenta. This is the backflow effect, discussed in Section \ref{sec:1.4}. This means we cannot regard Eq.(\ref{1.16a}) as the correct arrival time distribution.  

The backflow effect arises because of interference between portions of the state that have crossed the origin, and that part which has yet to cross. In a limited sense then, this problem is similar to that of defining an arrival time distribution for a classical particle in a potential, where the particle may cross the origin and then re-cross at some later time.  In the classical problem one imposes absorbing boundary conditions to remove that part of the state that has crossed the origin. The analogue in the quantum case is the use of complex absorbing potentials. This is an example of a ``trajectory free'' approach, as outlined in Section \ref{sec:1.1}.  Under suitable conditions this approach does indeed yield the the arrival time distribution as Eq.(\ref{1.16a}). See Chapter 5 for a detailed discussion of this.

Note however that wavefunctions consisting of a single gaussian wavepacket never display backflow. This means that Eq.(\ref{1.16a}) may indeed be valid for these states. These wavefunctions are also WKB states and the WKB interpretation, an example of a ``trajectory after quantisation'' approach as laid out in Section \ref{sec:1.1}, does indeed give Eq.(\ref{1.16a}) as the correct arrival time distribution, at least for these states.

We mention here briefly that the arrival time analysis of Kijowski \cite{Kij} gives,
\beq
\Pi(t)=\frac{1}{m}\bra{\psi_{t}}|\hat p|^{1/2}\delta(\hat x)|\hat p|^{1/2}\ket{\psi_{t}}\label{1.Kij}
\eeq
which can be obtained by an operator re-ordering of $J(t)$, but which has the virtue of being positive. We will not have anything to say about this distribution in this thesis, since it seems to be unrelated either to the histories analysis, Chapter 5, or to the behavior of ideal clocks, Chapter 8. Other authors have, however, shown that the distribution Eq.(\ref{1.Kij}) does arise naturally in analyses of the time of arrival based on POVMs, see Ref.\cite{Kijref} for details.

For arrival time probabilities defined by measurements, considered later in this
thesis, one might expect a very different result in the regime of
strong measurements, since most of the
incoming wavepacket will be reflected at $x=0$.  This is the
essentially the Zeno effect \cite{Zeno}. It was found in a
complex potential model that the arrival time distribution
in this regime is the kinetic energy density
\beq
\Pi(t)=
C\bra{\psi_{t}}\hat p\delta(\hat x)\hat
p\ket{\psi_{t}}.\label{1.8.Zeno}
\eeq
where $C$ is a constant which
depends strongly on the underlying measurement model \cite{Ech,Hal2}. (See Ref.\cite{Muga2} for a discussion of kinetic energy density.) 
However because
the majority of the incoming wavepacket is reflected, it is
natural to normalise this distribution by dividing by the
probability that the particle is ever detected, that is,
\bea
\Pi_N (t ) &=& \frac { \Pi ( t) } {  \int_0^\infty d s \Pi (s )  }
\nonumber \\
&=& \frac { 1 } { m | \langle p \rangle| }
\langle \psi_t |\hat p \delta ( \hat x ) \hat p
| \psi_t \rangle
\label{1.8.normalized}
\eea
where $ \langle p \rangle $ is the average momentum of the initial
state. This normalised
probability distribution does not depend on the details of the
detector. This suggests that the form
Eq.(\ref{1.8.normalized}) may be the generic result in this regime,
although a general argument for this is yet to be found.

It is found in practice that measurement models for
arrival times lead to distributions
depending on both the initial state of the particle
and the details of the clock or measuring device, typically of the form
\beq
\Pi_{C}(t)= \int_{-\infty}^{\infty} ds \ R (t,s) \Pi (s) \label{1.8.best},
\eeq
where $\Pi(t)$ is one of the ideal
distributions discussed above and the response function  $R (t,s) $
is some function of the clock
variables. (In some cases this expression will be a convolution).
However, it is of interest to coarse grain
by considering probabilities $p(t_1, t_2)$ for arrival times lying in some interval $[t_{1},t_{2}]$. The resolution
function $R$ will have some resolution time scale associated with
it, and  if the interval $t_{2}-t_{1}$ is much larger than this
time scale, we expect the dependence on $R$ to drop out, so that
\beq
p(t_{1},t_{2})=\int_{t_{1}}^{t_{2}}dt\ \Pi_{C}(t)\approx \int_{t_{1}}^{t_{2}}dt\ \Pi(t).
\label{1.8.cg}
\eeq
This is the sense in which many different models are in agreement
with semi-classical formulae at coarse-grained scales.

In addition to a lack of positivity, there is a more fundamental problem with Eq.(\ref{1.16a}), which is that probabilities in quantum theory should be expressible in the form \cite{NC}
\beq
p(\a)=\Tr(P_{\a}\rho).\label{1.formofprob}
\eeq
Here $P_{\a}$ is a projection operator, or more generally a POVM, associated with the outcome $\a$. Eq.(\ref{1.16a}) cannot be expressed in this form and we therefore conclude that it is not a fundamental expression in quantum theory.

The conclusion then is that whilst there is no axiomatic approach that yields Eq.(\ref{1.16a}) as the correct arrival time distribution, there is extremely good evidence for it from semiclassical and other approaches. One aim of this thesis is to show how Eq.(\ref{1.16a}) may be derived from an underlying axiomatic approach to quantum theory, thus providing the justification for these semiclassical and operational approaches.

\section{The Decoherent Histories Approach to Quantum Theory}\label{sec:1.5}

The following is a brief review of the aspects of the decoherent histories approach to quantum theory relevant for this thesis. More extensive discussions can be found Refs.\cite{GeH1, GeH2, Gri, Omn, Hal2, DoH}. 

Alternatives at fixed moments of time in quantum theory are
represented by a set of projection operators $\{ P_a \}$,
satisfying the conditions
\bea
\sum_a P_a &=& 1,\label{1.14}\\
P_a P_b &=& \delta_{ab} P_a,\label{1.15}
\eea 
where we take $a$ to run over some finite range. In the decoherent histories approach to quantum
theory histories are represented by class operators $C_{\a}$ which are time-ordered
strings of projections,
\beq
C_{\a} = P_{a_n} (t_n) \cdots P_{a_1}(t_1),
\label{1.16}
\eeq
or sums of such strings\footnote{The proper framework in which expressions such as Eq.(\ref{1.16}) are to be understood is the temporal logic framework of Isham and coworkers \cite{Ish}. Note also that class operators may be defined involving an {\em infinite} number of times, provided appropriate care is taken over the definition of the infinite products involved \cite{IshLin}.} \cite{Ish}. Here the projections are in the Heisenberg
picture and $ \a $ denotes the string $ (a_1, \cdots a_n)$. 
All class operators satisfy the condition
\beq
\sum_{\a} C_{\a} = 1.
\label{1.17}
\eeq
Note that in the same way that Eqs.(\ref{1.14}), (\ref{1.15}) do not define a unique decomposition of the state of a system at a moment of time, Eqs.(\ref{1.16}), (\ref{1.17}) do not define a unique decomposition of the possible set of histories of the system. It is important to decide upon the set $\{C_{\a}\}$ to be used at the beginning of the analysis, and to avoid working with class operators from more than one set\footnote{Strictly speaking, one cannot generate logical contradictions by considering class operators from more than one set but doing so can nevertheless lead to confusion. This is similar to the choice of a family of projectors at a single time, Eq.(\ref{1.15}). Consider a single spin-$\frac{1}{2}$ particle. Suppose one finds that,
\beq
\begin{array}{cc}
\mbox{prob}(s_{z}=\uparrow)=\frac{1}{2},&\mbox{prob}(s_{z}=\downarrow)=\frac{1}{2}\\
\mbox{prob}(s_{x}=\uparrow)=\frac{1}{2},&\mbox{prob}(s_{x}=\downarrow)=\frac{1}{2}
\end{array}
\eeq
where $\mbox{prob}(s_{z}=\uparrow)$ is the probability that the spin will be measured to be ``up'' in the $z$-direction. One cannot conclude from this that $\mbox{prob}(s_{z}=\uparrow \mbox{and } s_{x}=\uparrow)=\frac{1}{4}$ for example. 

In the same way, consider two different decoherent sets of histories represented by class operators $\{C_{\a}, C_{\b}...\}$ and $\{C_{\a'}, C_{\b'}...\}$. Now suppose that using the first set we can conclude that $p(\a)=1$ and using the second that $p(\a')=1$. There is no difficulty here. The two sets of histories provide {\em complimentary}, not {\em contradictory} descriptions of the system. Note also that Eq.(\ref{1.18}) implies that the probability of a given history is independent of the set of histories in which it appears, and so we could never derive the genuinely contradictory result that $p(\a)$ is 1 in one set of histories and 0 in another.}
 \cite{Grifbook}.

Probabilities are assigned to histories via the formula
\beq
p(\a) = {\rm Tr} \left( C_{\a} \rho C_{\a}^{\dag} \right).
\label{1.18}
\eeq
These probabilities are clearly real and positive,
however probabilities assigned in this way do not necessarily obey the
probability sum rules, because of quantum interference.
We therefore introduce the decoherence functional
\beq
D(\a, \b) = {\rm Tr} \left( C_{\a} \rho C_{\b}^{\dag} \right),
\eeq
which may be thought of as a measure of interference between pairs of histories.
We require that sets of histories satisfy the condition of
decoherence, which is
\beq
D(\a, \b) = 0, \ \ \ \a \ne \b\label{1.19d}.
\eeq
Decoherence implies the weaker
condition of consistency, which is that 
\beq
 {\rm Re } D(\a,\a') = 0, \quad \a \ne \b,
 \eeq
and this is equivalent to the requirement that the above probabilities satisfy the probability sum rules. In most situations of physical interest both the real and imaginary parts of $  D(\a,\b)$ vanish for $ \a \ne \b$, and so we shall require Eq.(\ref{1.19d}) throughout this thesis.  This condition is related to the existence of records \cite{GeH2,Hal5}. Decoherence is only approximate in general which raises the question of how to measure approximate decoherence. The decoherence functional satisfies the inequality \cite{DoH}
\beq
| D( \a, \b ) |^2 \le p (\a) p ( \b).
\eeq
This suggests that a sensible measure of approximate decoherence is
\beq
| D( \a, \b ) |^2  \ \ll \  p (\a) p ( \b).
\label{DA}
\eeq

We note briefly that when there is decoherence Eq.(\ref{1.17}) and Eq.(\ref{1.19d}) together imply that the probabilities $p(\a)$ are given by the simpler expressions
\beq
q(\a) = {\rm Tr} \left( C_{\a} \rho \right).
\label{1.20}
\eeq
Decoherence ensures that $q(\a)$ is real and positive, even though it
is not in general. 


A few comments are in order about the place of decoherent histories in the foundations of quantum theory. Although some early workers tried to present decoherent histories as an ``interpretation'' of quantum theory, see in particular \cite{Gri, Omn}, the more modern view is rather that the histories approach is a valuable tool in understanding how to apply quantum theory to closed systems, but not an ``interpretation'' in the usual sense. The histories approach does not try to explain the contextual, non-local nature of quantum theory in the way that, for example, the de Broglie-Bohm approach does, nor does it try to introduce new physics to solve the problems inherent in the quantum description of measurement, in the way that collapse theories do. Rather the histories approach is one in which there is no contridiction between the essential quantum nature of reality and the existence of a classical realm of our experience. The histories approach sees classical physics, together with classical logic, as emergent from the true quantum description of our world. One can of course carry out any particular measurement at any time and thereby obtain information about the properties of a system at one or more times, however the histories approach gives the conditions under which one may reason classically with these properties. In particular, if a set of histories satisfies the condition of decoherence, then one may treat the system {\it as if} \footnote{Whist again we stress that it is beyond the scope of this thesis to enter into a detailed discussion of the philosophical underpinnings of quantum mechanics in general and the decoherent histories approach in particular, the choice of words is important here. Claiming that it is consistent to act {\it as if} a system possesses one of a number of properties is not the same as claiming that it really does possess them.} it possesses the properties associated with these histories, regardless of whether one carries out a measurement to check this or not.

For more details of the way in which decoherent histories may be thought of as an ``interpretation'' of quantum theory see \cite{Grifbook, CQO}.

\section{The Decoherent Histories Approach to the Arrival Time Problem: Introduction and History}\label{sec:1.6}

Now that we have introduced the general framework of decoherent histories, we turn to the question of applying it to the specific case of time observables. The aim is to derive the correct class operators for questions such as ``Did the particle cross the origin in the interval $[t_{1},t_{2}]$?'' Since the solution to this problem is the subject of Chapter 5 of this thesis, we will not present the correct class operators here, but rather we will discuss in general terms how they may be defined. This is partly to motivate the use of the histories approach, since it will quickly become apparent that time observables are easily framed in this language. We also wish, however, to discuss the history of the decoherent histories approach to the arrival time problem and in particular some of the misleading results that have been obtained in the past by considering class operators defined via path integrals. We will see that the reason for these previous misunderstandings was a lack of appreciation of the role of the Zeno effect in the definition of class operators for time observables and it is a major goal of this thesis to correct these issues.

Let us begin with the general issue of defining class operators for entering some region of spacetime $\Delta$. Some early papers on the decoherent histories approach gave the following analysis of the problem \cite{Har,MiH,HaZa,Ya1,YaT}. Consider the propagator between two general spacetime points, considered as a path integral. Partition the set of paths summed over according to whether they ever enter the region $\Delta$ or not,
\bea
g(x_{1},t_{1}|x_{0},t_{0})&=&\int Dx e^{i S}=\int_{c} Dx e^{iS}+\int_{nc}Dx e^{iS}\nonumber\\
&=&\int_{c}Dx e^{iS}+g_{r}(x_{1},t_{1}|x_{0},t_{0})
\eea
where $\int_{c}$ denotes the sum over all paths that enter $\Delta$, and $\int_{nc}$ denotes the sum over all paths that do not enter $\Delta$. Here $g_{r}$ is the restricted propagator, defined as a sum over all paths which never enter the region $\Delta$. The path integral over paths that do enter $\Delta$ would appear to be the class operator we are looking for. We may therefore define the class operators for entering and not entering $\Delta$ as,
\bea
C_{nc}(t_{1},t_{0})&=&g_{r}(t_{1},t_{0})\\
C_{c}(t_{1},t_{0})&=&e^{-iH(t_{1}-t_{0})}-g_{r}(t_{1},t_{0}).
\eea
These definitions were used by earlier worker on decoherent histories, in particular in the series of papers by Yamada and Takagi \cite{YaT}. 

However there is a problem with these definitions, which is that they suffer from the Zeno effect. To see why this is the case consider the specific example of the arrival time problem. Let $ P = \theta ( \hat x ) $ denote the projection onto the positive $x$-axis.  The restricted propagator in this case may be written as,
\beq
g_r (\tau, 0 ) =  \exp \left( - i P H P \tau \right)P\label{1.35}
\eeq
so is unitary in the Hilbert space of states with support only in $x>0$ \cite{Wall,Sch2}. Compare Eq.(\ref{1.35}) with Eq.(\ref{1.7}).

Differently put, an incoming wave packet evolving according to the restricted propagator undergoes total reflection, so never crosses $x=0$. This means the probability of not crossing the origin is 1, whatever the initial state.  In the work of Yamada and Takagi \cite{YaT} and also in later work by Wallden \cite{Wall} this manifests itself as the inability to assign probabilities to anything other than states symmetric about the origin, for which the arrival probabilities vanish. Clearly these results are unphysical.

A moment's reflection reveals that this result is not that suprising. We are demanding that the class operator for not crossing is a sum over paths that spend exactly zero time in $x<0$. Our intuition would suggest that this restriction is too sharp and that we should instead sum over all paths spending a time less than some time $\delta t$ in $x<0$. 

These heuristic notions are very difficult to make precise in the path integral formulation of decoherent histories however. Instead, it is more helpful to attempt to define $C_{nc}$ in the manner of Eq.(\ref{1.16}). A natural starting point is
\beq
C^{\epsilon}_{nc} =P e^{ - i H \epsilon } P \cdots e^{ - i H \epsilon } P,
\label{1.19}
\eeq
where there are $N$ projectors and $\t=N\e$. We recover the restricted propagator if we take the limit $\e\to0$, compare Eq.(\ref{1.19}) with Eq.(\ref{1.7a}). In light of this, one idea is to simply decline to take the limit $\epsilon \rightarrow 0 $
and  define the class operator for not crossing to be Eq.(\ref{1.19}) for some $\e$.
 Clearly if $\epsilon$ is large enough
the system will be monitored sufficiently infrequently to let the wave packet
cross $x=0$ without too much reflection. Eq.(\ref{1.19}) is an important expression for the rest of this thesis, however as it stands this proposal is somewhat vague. It is a major goal of this thesis to turn this heuristic proposal into a concrete way of defining class operators. The key question concerns the timescale $\e$. These issues will form the subject of Chapter 4.

A second option is to take the limit of $\e\to0$ in Eq.(\ref{1.19}), but ``soften'' the projections
to POVMs, that is, to replace $P = \theta (\hat x)$ with a function which is
approximately $1$ for large positive $x$, approximately $0$ for large negative $x$,
and with a smooth transition in between. Although we have introduced the decoherent histories approach to quantum theory in Section \ref{sec:1.5} above in terms of projection operators, the discussion is readily generalised to include POVMs \footnote{The reason we have not discussed POVMs is the following. Consider two distinct histories, $\{\a,\b\}$, defined by strings of projection operators. Now turn the projection operators into POVMs by smearing them with a gaussian of finite width and call these new histories $\{\tilde\a,\tilde\b\}$.  These smeared histories are no longer distinct and so even if $D(\a,\b)=0$ we will have $D(\tilde\a,\tilde\b)\neq0$. This makes it very hard to decide between histories that interfere, and histories that simply overlap. }. It would be interesting to pursue this idea, but we shall not do so in this thesis\footnote{In fact in some ways we do exactly this in later sections, since the complex potential introduced in Chapters 4 and 5 is very much like a POVM. Our motivation for introducing it there is rather different, however.}.

\section{Summary and Overview of the Rest of This Thesis}

The major goal of this thesis is to give a decoherent histories analysis of the arrival time problem, that will point to the way to a description of general time observables in quantum theory. We wish firstly to know what the correct quantum analogues of Eqs.(\ref{1.13a}), (\ref{1.14a}) are, and how they may be obtained from first principles. We also wish to understand how the semiclassical result, Eq.(\ref{1.16a}), emerges in some limit from this quantum result.

The key question concerns the definition of the appropriate class operators for these observables. In Section \ref{sec:1.5} we saw that the Zeno effect limits the usefulness of the path integral definitions of the class operators, so our challenge is to work with class operators defined in the manner of Eq.(\ref{1.19}). These class operators represent evolution in the presence of pulsed measurements. One of the major achievements of this thesis will come in Chapter 4, where we will show that such evolution is, under appropriate conditions, equivalent to evolution under continuous measurement in the form of a complex step potential. Before we can do this we will need some results about propagation in the presence of these step potentials and Chapters 2 and 3 will be devoted to explaining the necessary path integral methods. 

Complex potentials also arise in other standard approaches to the arrival time problem in classical and quantum mechanics. These potentials are generally of the form $V(x)=-iV_{0}\chi(x)$, where $\chi(x)$ is the characteristic function of some region of configuration space. If we are to work with these complex potentials we must be able to deal with the propagator between two arbitary points in configuration space in the presence of these simple complex potentials. In Chapter 2 we introduce the Path Decomposition Expansion (PDX), a useful technique for evaluating path integrals with piecewise defined potentials. In Chapter \ref{props} we then use this technique to evaluate the propagator for the step and Dirac delta function potentials.
As well as introducing the full expression for the PDX, which is exact, we also introduce a useful semi-classical approximation, valid when the height of the potential is small compared with the energy of the incoming state.

In Chapter 4 we ask the following question: Are the schemes of pulsed and continuous measurement, represented by complex potentials and strings of projection operators respectively, in any sense equivalent? We will find that the two approaches are indeed related, in the sense that the propagators for evolution under the two measurement schemes are equivalent under certain conditions. 

In Chapter 5 of this thesis we use the results of Chapters 2, 3 and 4 to attack the arrival time problem. Once we have proven that evolutions under pulsed and continuous measurements are equivalent we use this to replace the class operator, Eq.(\ref{1.19}) with evolution in the presence of a complex potential. In Chapter 5 this will allow us to undertake a thorough examination of the arrival time problem and to obtain the arrival time distribution $\Pi(t)$. 

However, despite the apparent complexity of the approach, the final results will be remarkably simple and will suggest that there may be a less rigorous but more intuitive way to arrive at them. In Chapter 6 we will therefore present a simplified derivation of the class operators for the arrival time problem, using only a semiclassical approximation. We will also show how these class operators may be extended in a straightforward way to situations where there are states incident on the origin with both positive and negative momenta.

With the main goal of this thesis achieved, in Chapters 7 and 8 we turn to some related questions that lie somewhat outside the main scope of this work. In Chapter 7 we look at arrival times for open quantum systems, trying to understand the way in which the current, Eq.(\ref{1.16a}), emerges as the classical arrival time distribution, both in general terms, and by extending the decoherent histories analysis to this situation. In Chapter 8 we turn to the  issue of arrival and dwell times defined via ideal clocks, finding agreement with the decoherent histories analysis in the appropriate limit, but also achieving an understanding of the emergence of the current as the correct arrival time distribution for a very general class of model clocks.

We summarize the results of this thesis in Chapter 9, and outline some possible areas for further study.

\chapter{The Path Decomposition Expansion (PDX)}\label{pdx}
\epigraph{There is surely no greater wisdom than to mark well the beginnings and endings of things.}{Francis Bacon, Essays 1625, Of Delays}

\section{Introduction}
In this chapter we describe some useful path integral techniques. We will make frequent use of the results in this chapter throughout the rest of this thesis. Since this chapter is essentially technical background, the majority of the material presented here is not original research. In Section 2.2 we introduce the Path Decomposition Expansion, then in Section 2.3 we use this to compute the scattering states for a complex step potential. In Section 2.4 we then introduce a useful semiclassical approximation, valid when the height of the potential is small compared with the energy of the incoming state.

\section{The Path Decomposition Expansion (PDX)}

In this section we introduce the Path Decomposition Expansion (PDX), a useful technique for evaluating path integrals with piecewise defined potentials. 
Although for the most part the application of these results will be to the simple situation where we have a particle incoming on an imaginary potential of step function form, in this chapter we will deal with real potentials and we will begin by assuming the more general form $V(x)=V_{0}\theta(-x)f(x)$. 

We wish to evaluate the propagator
\beq
g(x_1, \tau | x_0 ,0 ) = \langle x_1 | \exp  \left( -  i H_0 \tau
-   i V_0 \theta (- \hat x) f(\hat x) \tau \right)
| x_0 \rangle,
\label{12}
\eeq
for arbitrary $x_1$ and $ x_0 > 0$. This may be calculated using a sum over paths,
\beq
g(x_1, \tau | x_0,0 ) = \int {\cal D} x \exp \left( i S \right),
\label{13}
\eeq
where
\beq
S = \int_0^{\tau} dt \left( \half m \dot x^2 - V_0 \theta (-x) f(x) \right),
\label{14}
\eeq
and the sum is over all paths $ x(t)$ from $x(0) = x_0$ to $x(\tau) = x_1$.

\begin{figure}[h]
\begin{center}
  \includegraphics[width=4in]{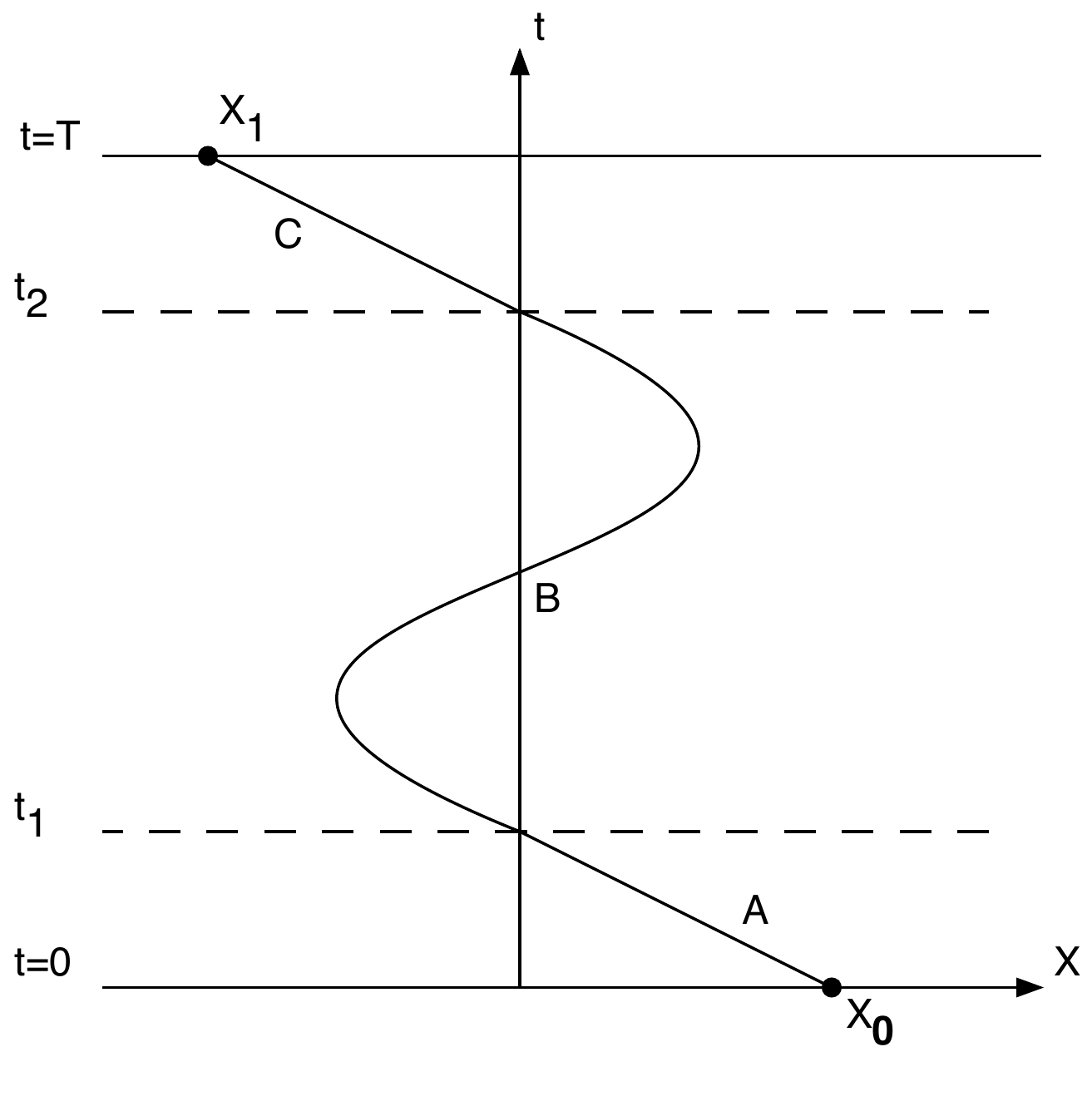} 
\caption[The Path Decomposition Expansion (PDX)]{{\em A typical path from $x_{0}$ to $x_{1}$}}\label{pdxfig}
\end{center}
\end{figure}

To deal with the step function form of the potential we need to split off the sections
of the paths lying entirely in $x>0$ or $x<0$. The way to do this is to use
the path decomposition expansion
or PDX \cite{pdx, pdx2,HaOr,Hal3}.
Consider first paths from $x_0 > 0 $ to $ x_1 < 0 $. A typical such path may cross $x=0$ many times, but the set of paths may be partitioned by their {\it first} or {\it last} crossing times. We therefore split every path into three parts: (A) a restricted part that starts at $x_{0}$ and does not cross $x=0$, but that ends on $x=0$ at time $t_{1}$, (B) an unrestricted part from $x=0$ to $x=0$ that may cross $x=0$ many times and, (C) a further restricted part from $x=0$ to $x_{1}$ that does not re-cross $x=0$, Fig.(\ref{pdxfig}).
As a consequence
of this, it is possible to derive the formula,
\beq
g(x_1, \tau | x_0,0 ) = \frac {i } {2m} \int_{0}^{\tau} dt_1
\ g (x_1, \tau | 0, t_1) \frac {\partial g_r } { \partial x} (x,t_1| x_0,0) \big|_{x=0}.
\label{PDX1}
\eeq
Here, $g_r (x,t|x_0,0)$ is the restricted propagator given by a
sum over paths of the form (\ref{13}) but with all paths restricted to
$x(t) >0$. It vanishes when either end point is the origin but its derivative
at $x=0$ is non-zero (and in fact the derivative of $g_r$ corresponds to
a sum over all paths in $x>0$ which end on $x=0$ \cite{HaOr}).

It is also useful
to record a PDX formula involving the last crossing time $t_2$,
for $x_0>0$ and $x_1 < 0 $,
\beq
g(x_1, \tau | x_0,0 ) =  - \frac {i  } {2m} \int_{0}^{\tau} dt_2
\ \frac {\partial g_r} {\partial x} (x_1, \tau | x, t_2) \big|_{x=0} \ g (0,t_2|x_0,0).
\label{16}
\eeq
These two formulae may be combined to give a first and last crossing version of the PDX,
\beq
g(x_1, \tau | x_0,0 ) =  \frac {1  } {4m^2} \int_{0}^{\tau} dt_2
\int_0^{t_2} dt_1
\ \frac {\partial g_r} {\partial x} (x_1, \tau | x, t_2) \big|_{x=0} \ g (0,t_2| 0,t_1)
\ \frac {\partial g_r } { \partial x} (x,t_1| x_0,0) \big|_{x=0}.
\label{16a}
\eeq
This is clearly very useful for a step potential since the propagator is
decomposed in terms of propagation in $ x<0$ and in $x>0$, essentially
reducing the problem to that of computing the propagator along $x=0$,
$  g (0,t_2| 0,t_1) $. (See Figure \ref{pdxfig})

For paths with $x_0>0$ and $x_1 > 0 $, Eq.(\ref{PDX1}) is modified by the
addition of a term $g_r (x_1,t|x_0,0)$, corresponding to a sum over paths
which never cross $x=0$, so we have
\beq
g(x_1, \tau | x_0,0 ) = \frac {i } {2m} \int_{0}^{\tau} dt_1
\ g (x_1, \tau | 0, t_1) \frac {\partial g_r } { \partial x} (x,t_1| x_0,0) \big|_{x=0}
+ g_r (x_1,t|x_0,0).
\label{PDX2}
\eeq
Again a further decomposition involving the last crossing, as in Eq.(\ref{16a})
can also be included.
In general the usefulness of these expressions relies on the various partial propagators that occur in the PDX formulae being easier to compute than the full propagator. This depends crucially on the form of the potential, from now on we will assume that $f(x)=1$ in the above expressions, and thus that the potential is of simple step function form.

The various elements of these expressions are easily calculated
for a potential of simple step function form $ V(x) = V_0 \theta (-x)$.
The restricted propagator in $x>0$ is
given by the method of images expression
\beq
g_r (x_1, \tau |x_0,0) = \theta (x_1 ) \theta (x_0)
\left( g_f (x_1, \tau |x_0,0) - g_f (-x_1, \tau |x_0,0) \right),
\label{3.17}
\eeq
where $g_f $ denotes the free particle propagator
\beq
g_f (x_1, \tau |x_0,0) = \left( \frac {m} {2 \pi i \tau } \right)^{1/2}
\ \exp \left( \frac {i m (x_1 - x_0)^2 } { 2  \tau} \right).
\eeq
It follows that
\beq
\frac {\partial g_r } { \partial x} (x,t_1| x_0,0) \big|_{x=0} =
2 \frac {\partial g_f } { \partial x} (0,t_1| x_0,0) \theta (x_0).
\label{18}
\eeq
The restricted propagator in $x<0$ is given by an expression similar to Eq.(\ref{3.17}), multiplied by
$ \exp ( i V_0 \tau )$. 

Note that this means that Eqs.(\ref{PDX1}), (\ref{16}) and (\ref{16a}) can be written as,
\bea
\hspace{-10pt}\bra{x_{1}}e^{-i(H_{0}+V_{0}\theta(-\hat x))\t}\ket{x_{0}}&=&\frac{1}{m}\int_{0}^{\t}dt_{1}\bra{x_{1}}e^{-i(H_{0}+V_{0}\theta(-\hat x))(\t-t_{1})}\delta(\hat x)\hat p e^{-iH_{0}t_{1}}\ket{x_{0}}\label{PDX81}\\
&=& \frac{1}{m}\int_{0}^{\t}dt_{2}\bra{x_{1}}e^{-i(H_{0}+V_{0})(\t-t_{2})}\hat p \delta(\hat x) e^{-i(H_{0}+V_{0}\theta(-\hat x))t_{2}}\ket{x_{0}}\label{PDX81b}\\
&=&\frac{1}{m^{2}}\int_{0}^{\t}dt_{2}\int_{0}^{t_{2}}dt_{1}\bra{x_{1}}e^{-i(H_{0}+V_{0})(\t-t_{2})}\hat p\delta(\hat x) \nonumber\\
&&\times e^{-i(H_{0}+V_{0}\theta(-\hat x))(t_{2}-t_{1})}\delta(\hat x)\hat pe^{-iH_{0}t_{1}}\ket{x_{0}}\label{PDX81}
\eea
where $\delta(\hat x)=\ket{0}\bra{0}$ and $ \ket{0}$ denotes a
position eigenstate $ \ket{x}$ at $ x=0$. These operator forms of the PDX are the ones we shall use most often.

The only complicated propagator to calculate is the propagation from the origin to itself along the edge of the potential. We will show in the next chapter that in the case $ V(x) = V_0 \theta (-x)$ this is given by \cite{Ye1},
\beq
g(0,t | 0,0)
= \left( \frac {m} {2 \pi i t } \right)^{1/2}
\ \frac { \left(1 - \exp(- iV_0 t ) \right) } { iV_0 t}.
\label{27}
\eeq

However as well as allowing us to obtain exact results for the propagator in certain cases, the PDX also suggests some semiclassical approximations we could make to simplify calculation for a more general class of potentials. We will discuss this in Section \ref{semiclass}.

\section{Using the PDX: Scattering States for a Complex Step Potential}

In this Section we use the PDX to derive the standard
representations of the scattering solutions to the Schr\"odinger
equation with the simple complex potential $V(x)=-iV_{0}\theta(-x)$.
These are known results but this derivation confirms the validity of the PDX
method and also allows a certain heuristic path integral approximation to be tested.
The results will also be useful for the decoherent histories analysis
in Chapter 5.

The transmitted and reflected wave functions are defined by
\beq
\psi (x,\tau) = \theta  (-x) \psi_{tr} (x,\tau) + \theta (x) \left(
\psi_{ref} (x,\tau) + \psi_f (x, \tau) \right)
\label{28a}
\eeq
Here, $ \psi_{tr} $ is the transmitted wave function and is given by the propagation of the initial state $ \psi (x)$ using the PDX formulae Eq.(\ref{PDX1}) or Eq.(\ref{16a}). For large $\tau$, a freely evolving packet moves entirely into $x<0$ so that the free part, $ \psi_f (x, \tau) $ is zero, leaving just the transmitted and reflected parts.
Following the above definition, the transmitted wave function is given by
\bea
\psi_{tr} (x, \tau) &=& \frac {1} {m^2} \int_0^\tau ds \int_0^{\tau - s} dv
\ \langle x | \exp \left( - i H_0  s \right) \hat p | 0 \rangle \ e^{ - V_0 s }
\nonumber \\
& \times &
\langle 0 | \exp \left(  - i H v \right)  | 0 \rangle
\ \langle 0 | \hat p \exp \left(  - i H_0 ( \tau - v - s) \right) | \psi \rangle,
\label{4.2}
\eea
where $ | 0 \rangle $ denotes the position eigenstate $ | x \rangle $ at $ x = 0$.
Also, we have introduced $ s = \tau - t_1 $ and $ v = t_2 - t_1 $, and
$H = H_0 - i V_0 \theta (-x)  $ is the total Hamiltonian.
The scattering wave functions concern the regime of large $\tau$, so we let the upper
limit of the integration ranges extend to $\infty$.

Writing the initial state as a sum of momentum states $ | p \rangle$, and introducing
$ E = p^2 / 2m $, we have
\bea
\psi_{tr} (x, \tau) &=& \frac {1} {m^2} \int dp \int_0^{\infty} ds
\langle x | \exp \left( - i H_0  s \right) \hat p | 0 \rangle \ e^{ i (E+ i V_0) s }
\nonumber \\
& \times & \int_0^{\infty} dv \ \langle 0 | \exp \left(  - i H v \right)  | 0 \rangle \ e^{ i E v}
\ p \langle 0 | p \rangle \ \ e^{ - i E \tau} \psi (p).
\label{4.3}
\eea
To evaluate the $s$ integral, we use the formula \cite{Sch},
\beq
\int_0^{\infty} ds \ \left( \frac {m} {2 \pi i  s } \right)^{1/2}
\exp \left( i \left[ \lambda s + \frac { m x^2} {2 s} \right] \right)
= \left( \frac { m } { 2 \lambda } \right)^{\half} \ \exp \left( i | x | \sqrt {2 m \lambda} \right),
\label{4.4}
\eeq
from which it follows by differentiation with respect to $x$ and setting $\lambda = E+ i V_0$
that
\beq
\int_0^{\infty} ds
\langle x | \exp \left( - i H_0  s \right) \hat p  | 0 \rangle \ e^{ i (E+ i V_0) s }
= m \exp \left( i | x | [ 2m ( E + i V_0 ) ]^{\half} \right).
\eeq
The $v$ integral may be evaluated using the explicit
expression for the propagator along the edge of the potential, Eq.({\ref{27}),
together with the formula,
\beq
\left( \frac {m} {2 \pi i } \right)^{1/2}
\int_0^{\infty} dv \frac { (1 - e^{ - V_0 v}) } { V_0 v^{3/2} } \ e^{ i E v }
= \frac {\sqrt{2m} } { (E+ i V_0)^{\half} + E^{\half} }.
\label{4.5}
\eeq
We thus obtain the result,
\beq
\psi_{tr} (x, \tau) = \int \frac{ dp} { \sqrt{2 \pi} }
\exp \left( - i x [ 2m (E+ i V_0)]^{\half} - i E \tau \right) \ \psi_{tr} (p),
\label{4.6}
\eeq
where
\beq
\psi_{tr} (p) = \frac { 2 } { ( 1 + E^{- \half} ( E + i V_0 )^{\half} ) } \ \psi (p).
\label{4.7}
\eeq
Note that in this final result, it is possible to identify the specific
effects of the different sections of propagation: the propagation
along the edge of the potential corresponds to the coefficient in the
transmission amplitude Eq.(\ref{4.7}) (which is equal to $1$ when $V_0 = 0$),
and the propagation from final crossing to the final point produces
the $V_0$ dependence of the exponent. These observations will be useful below.

The reflected wave function $ \psi_{ref} $ is defined above using
the PDX Eq.(\ref{PDX2}) (rewritten using Eq.(\ref{16a}).
The first term in Eq.(\ref{PDX2}), the crossing part, is the same
as the transmitted case, Eq.(\ref{4.3}), except that $V_0 = 0$ in the last segment
of propagation, from $x=0$ to the final point, and also the sign of $x$ is reversed.
We must also add the effects of the reflection part of the restricted propagator,
and this simply
subtracts the reflection of the incoming wave packet.
The reflected
wave function is therefore given by,
\beq
\psi_{ref} (x,\tau) = \int \frac{ dp} { \sqrt{2 \pi} }
\exp \left( i x p  - i E \tau \right) \ \psi_{ref} (p),
\label{4.8}
\eeq
where
\bea
\psi_{ref} (p) &=& \psi_{tr} (p) - \psi (p)
\nonumber \\
&=& \frac { 1 -  E^{- \half} ( E + i V_0 )^{\half} ) } { ( 1 + E^{- \half} ( E + i V_0 )^{\half} ) } \ \psi (p).\label{4.7b}
\eea

We thus see that the PDX very readily gives the standard stationary wave
functions \cite{All}, without
having to use the usual (somewhat cumbersome) technique of matching eigenfunctions at $x=0$. In fact,
this procedure is in some sense already encoded in the PDX.

\section{A Semiclassical Approximation}\label{semiclass}\label{sec:2.4}

In this section we discuss a useful semiclassical approximation that allows us to obtain the approximate propagator for evolution under simple step potentials of the form $V(x)=V_{0}\theta(-x)$ for $V_{0}$ real or imaginary, under the condition that $|V_{0}|$ is small compared with the energy of the system.

Pictorially, the PDX equation Eq.(\ref{16a}) says that propagation from $x_{0}>0$ to $x_{1}<0$ may be considered in three parts, initial propagation to $x=0$, propagation along $x=0$ and propagation away from $x=0$, see Fig.(\ref{pdxfig}). Classically however, assuming we can neglect reflection,  we expect the propagator to be dominated by the single classical path. This path is a simple straight line that crosses $x=0$ only once. One might therefore expect that the propagator for the final segment from $x=0$ to $x_{1}<0$ in Eq.(\ref{PDX1}) is given approximately by
\beq
\langle x_{1} | \exp \left( - i H_{0} s-V_{0}\theta(-\hat x)s \right) | 0 \rangle \approx
\langle x_{1} | \exp \left( - i H_0 s \right) | 0 \rangle
  \exp \left( - V_0 s \right)
\label{3.20}
\eeq
and similarly that,
\beq
\langle 0 | \exp \left( - i H_{0} s-V_{0}\theta(-\hat x)s \right) | x_{0} \rangle \approx
\langle 0 | \exp \left( - i H_0 s \right) | x_{0} \rangle
\label{3.21}
\eeq
It is not entirely clear
that this is the case, however. On the one hand, the usual semiclassical approximation indicates
that paths close to the straight line paths dominate, but on the other hand, paths in
$x<0$ are suppressed, so maybe the wiggly paths that spend less time in $x<0$ make
a significant contribution. Since this approximation is potentially a useful one, it
is useful to compare with the exact result for the transmitted wave packet calculated
above.

We therefore evaluate the following approximate expression for the transmitted wave function,
\beq
\psi_{tr} (x, \tau) =
 \frac {1} {m} \int_0^{\tau} ds \ \langle x | e^{ - i H_0 s }  | 0 \rangle
\ e^{ - V_0 s }
\ \langle 0 | \hat p e^{ - i H_0 (\tau -s ) } | \psi \rangle
\label{4.12}
\eeq
This is the PDX, Eq.(\ref{PDX1}), in operator form with the semiclassical approximation described above
and we have set $ s = \tau - t_1 $. We now take $ \tau \rightarrow \infty$ in the integration
and evaluate. The key integral is,
\beq
\int_0^{\infty} ds \ \langle x | e^{ - i H_0 s }  | 0 \rangle
\ e^{ i (E  + i V_0 ) s}
=\left( \frac { m } { 2 (E + i V_0) } \right)^{\half} \ \exp \left( - i  x  [2 m (E+iV_0)]^{\half} \right)
\label{4.13}
\eeq
where we have used Eq.(\ref{4.4}) (and recall that $ x < 0 $).
The resulting expression for the transmitted wave function is of the form Eq.(\ref{4.6}), with
\beq
\psi_{tr} (p) = \frac {1} { E^{-\half} (E+ i V_0)^{\half} } \psi (p)
\eeq
This agrees with the exact expression for the transmission coefficient Eq.(\ref{4.7})
only when $V_0 = 0$, with the difference of order $ V_0 / E $ for small $V_0$.
This establishes that the approximation is valid for $ V_0 $ much less than
the energy scale of the initial state.

Crucial to the usefulness of this approximation, however, is that the {\it form} of the transmitted wavefunction is reproduced exactly, the discrepancy being in the prefactor. In the situations where we  make use of this semiclassical approximation we are always assuming refection may be neglected, and thus the prefactor is in fact irrelevant, since it can be obtained by normalization.  This makes this semiclassical approximation a very useful one.

We make some brief comments here about the significance of the semiclassical approximation in the arrival time problem. The classical phase space distribution representing the current,
\beq
J(p,x)=\frac{(-p)}{m}\delta(x)
\eeq
does not have a unique quantisation, because of the ambiguity in the operator ordering of $\hat p$ and $\delta(\hat x)$. The standard Schr\"odinger current comes from the symmetric choice
\beq
\hat J=\frac{(-1)}{2m}\left(\hat p\delta(\hat x) +\delta( \hat x)\hat p\right).
\eeq

Now in the PDX one can write the propagator from a point $x_{0}>0$ to a point $x_{1}<0$ in either of the forms
\bea
\bra{x_{1}}e^{-iHT}\ket{x_{0}}&=&\frac{1}{m}\int_{0}^{T}dt\bra{x_{1}}e^{-iH(T-t)}\delta(\hat x)\hat p e^{-iHt}\ket{x_{0}}\\
&=&\frac{1}{m}\int_{0}^{T}dt\bra{x_{1}}e^{-iH(T-t)}\hat p\delta(\hat x) e^{-iHt}\ket{x_{0}}.
\eea
These two integrals are equal, but the integrands are not. They differ in in terms of whether it is the incoming or outgoing paths which are restricted. However the semiclassical approximation says that the integrands, thought of as path integrals, are dominated by paths which cross only once and are therefore in effect ``restricted'' on both sides. This means we have,
\beq
\bra{x_{1}}e^{-iH(T-t)}\delta(\hat x)\hat p e^{-iHt}\ket{x_{0}}\approx\bra{x_{1}}e^{-iH(T-t)}\hat p\delta(\hat x) e^{-iHt}\ket{x_{0}}\nonumber.
\eeq
Because the times and positions are arbitrary this approximation means that there is essentially no difference between the terms $\hat p \delta(\hat x)$ and $\delta(\hat x)\hat p$. This means the semiclassically any result obtained by using the first crossing PDX, say, could equally well be obtained using the last crossing formula.

\chapter{The Propagator for the Step and Delta Function Potentials, using the PDX}\label{props}
\epigraph{The journey of a thousand miles begins with one step.}{Lao Tzu}

\section{Introduction}

One of the standard approaches to defining the time of entry to a given region of configuration space in quantum theory is to introduce a complex potential localised in the region of interest, and to define the arrival time probabilities in terms of the rate at which the wavefunction is absorbed by the potential \cite{All}. In the simplest case of the arrival time problem in 1D this means using a complex potential of the form $V(x)=-iV_{0}\theta(-\hat x)$. If we are to extract the arrival time probability from this set up we must be able to solve for the propagator in the presence of simple complex potentials. In fact, close inspection reveals that the propagator for a particle in the presence of a complex potential can generally be obtained by first considering the equivalent real potential, and then analytically continuing the result. This means our task is to consider propagators for a particle in the presence of a real step potential. Delta function potentials may be used in a similar way.

Calculations involving step and delta function potentials also occur in many other branches of physics. Step potentials can be used to represent ``hard wall'' boundary conditions, and are also involved in tunneling calculations. Delta function potentials can be used to model point interactions, especially in the low energy limit where the details of the process are largely independent of the form of the scattering potential \cite{deltapot}. 

We wish to compute the following propagator,
\begin{equation}
g(x_{1},T | x_{0},0)=\int_{x(0)=x_{0}}^{x(T)=x_{1}} \! \mathcal{D}x \; e^{iS}
\end{equation} where
\begin{equation}
S=\int_{0}^{T}dt\left(\frac{m\dot x^{2}}{2} - V(x)\right)
\end{equation}
and the potential $V(x)$ is either a step potential $\lambda\theta(-x)$, or a delta function potential, $\lambda\,\delta(x)$.  

In this Chapter we present a derivation of the propagators in the vicinity of these potentials by making use of the path decomposition expansion presented in the previous chapter.
The problem of calculating the full propagator therefore reduces to that of calculating the partial propagator for the interval where the path crosses the origin, $g(0,t_{2}|0,t_{1})$. We will show how these partial propagators may be derived using the Brownian motion definition of the path integral \cite{hartle, gert}. The full propagators may then be obtained with the help of Eq.(\ref{16a}).

For the step potential the full propagator has been derived in Refs.\cite{car, barut, che, crandall}, but we shall derive the partial propagator, and then direct the reader to Ref.\cite{car} for details of the use of the path decomposition expansion to recover the full propagator.
The partial propagator we will derive is given by
\begin{equation}
g(0,T|0,0)= -i\left(\frac{m}{2 \pi i}\right)^{1/2}\frac{(1-e^{-i\lambda T})}{\lambda T^{3/2}}.
\end{equation}
This result was used in Chapter 2 to compute the scattering states for a complex step potential.

For the delta function potential we will derive the full propagator and we quote the result from Ref.\cite{delta}
\begin{equation}
g(x_{1},T|x_{0},0)= g_{f}(x_{1},T|x_{0},0)-\lambda m\int_{0}^{\infty} \! du\; e^{-\lambda mu}g_{f}(|x_{1}|+|x_{0}|+u,T|0,0),
\end{equation}
where
\begin{equation}
g_{f}(x_{1},T|x_{0},0)=\left(\frac{m}{2\pi iT}\right)^{1/2}e^{im(x_{1}-x_{0})^{2}/2T}
\end{equation}
is the free propagator.
This chapter is based on Ref.\cite{Ye1}.

\section{The Brownian Motion Definition of the Propagator}

We begin with a review of some of the details of the Brownian motion approach to computing propagators. For more details see Refs.\cite{hartle, gert}. 
The first step is to switch to working with the Euclidean propagator $\overline g$ by means of a Wick rotation and we specialise immediately to the case of $x_{0}=x_{1}=0$.  That is, we wish to calculate
\begin{equation}
\overline g(0,T | 0,0)=\int_{x(0)=0}^{x(T)=0}\! \mathcal{D}x \;e^{-S_{E}}\label{3.1}
\end{equation} 
where $S_{E}$ is the Euclidean action given by
\begin{equation}
S_{E}=\int_{0}^{T}dt\left(\frac{m\dot x^{2}}{2} + V(x)\right)\label{3.2}.
\end{equation}
This propagator may be viewed as a conditional probability density for a random walk on the real line. The second step is then to make this integral over paths into a concrete object by defining it as the continuum limit of a discrete sum on a lattice.

To establish conventions and demonstrate the basic ideas we compute the case of a free particle, following closely the treatment in Ref.\cite{hartle}. We consider a rectangular lattice with spacing in the time direction of $\epsilon$, spacing in the $x$ direction of $\eta$ and we consider propagation for a time $T=2\epsilon n$, so we have $2n$ steps in our paths (the reason for this choice is that it simplifies a number of later expressions, and avoids clumsy factors of 1/2). The conditional probability $u(0,T|0,0)$ to start at (0,0) and end at (0,T) is given by the number of paths connecting the start and end points, divided by the total number of possible paths. The set of paths from $(0,0)$ to $(0,T)$ is bounded by the extremal paths that take $n$ steps to the left/right, followed by $n$ steps to the right/left, see Fig.(\ref{typical}). Since a path must have the same number of steps to the right as to the left to end up back at $x=0$ we find,
\begin{equation}
u(0,T|0,0) = \frac{1}{2^{2n}}{2n \choose n}
\end{equation}
The Euclidean propagator $\overline g$ is then defined as the continuum limit of $u/2\eta$ where we take $\epsilon,\eta \rightarrow 0$, $n\rightarrow \infty$, keeping $\epsilon/\eta^{2} = m$ and $T=2\epsilon n$ fixed. That is,
\begin{equation}
\overline g_{f}(0,T|0,0):=\lim_{\eta, \epsilon \to 0} (2\eta)^{-1}u(0,T|0,0) = \left(\frac{m}{2\pi T}\right)^{1/2}
\end{equation}
which is the expected result for the Euclidean free propagator.

\begin{figure}[h]
\begin{center}
\includegraphics[scale=0.5]{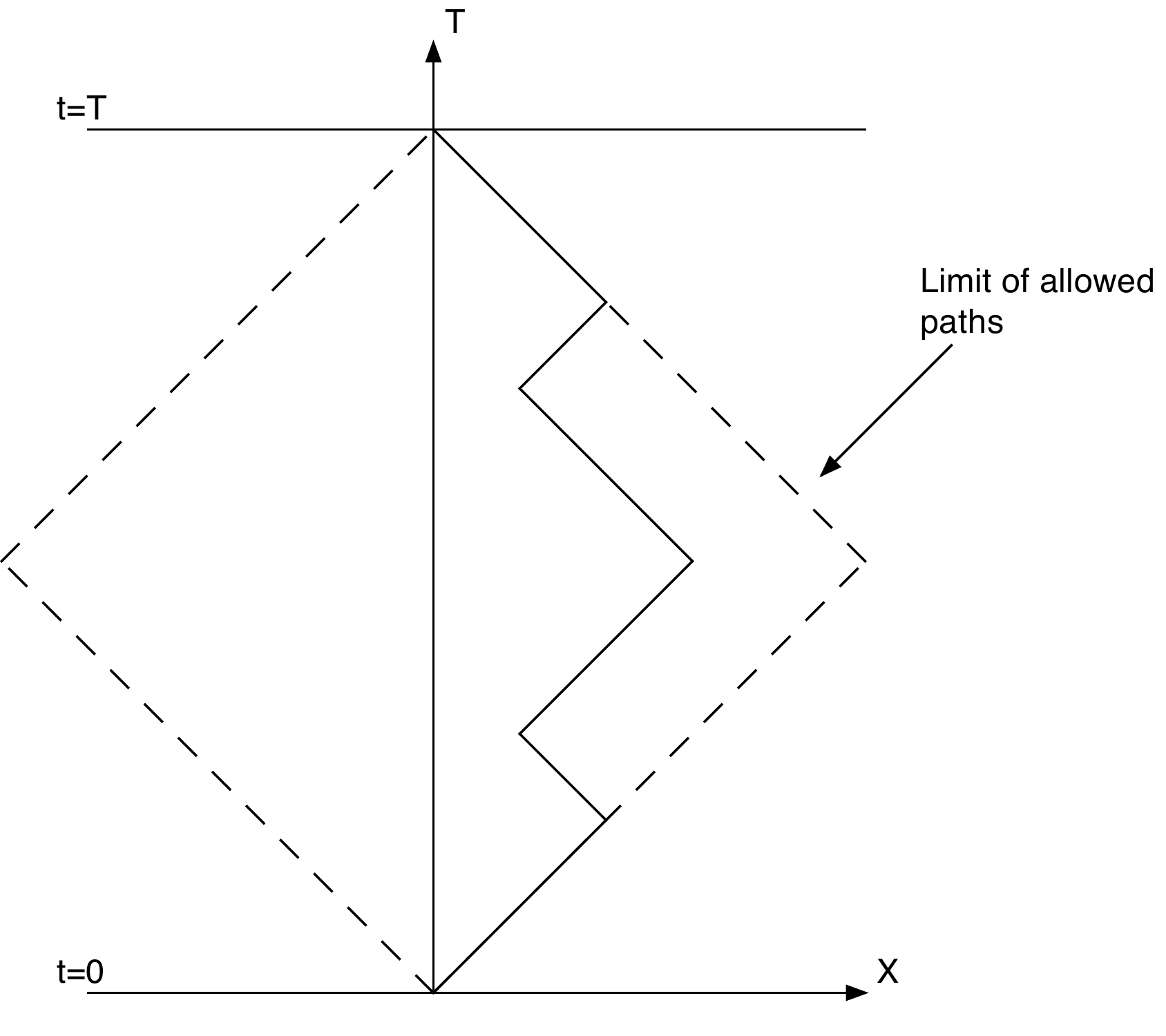}
\caption[A Typical Path From $x=0$ to $x=0$ On Our Lattice.]{{\em A typical path from $x=0$ to $x-0$. The dashed line shows the area within which all paths must remain.}}\label{typical}
\end{center}
\end{figure}

\section{The Step Potential}
The propagator along the edge of a step potential is given by Eqs.(\ref{3.1}) and (\ref{3.2}), with $V(x)=\lambda\theta(-x)$. We can write this as
\begin{equation}
\overline g(0,T|0,0) = \int \mathcal{D}x \exp \left(-\int_{0}^{T}dt \frac{m \dot x^{2}}{2}\right) \exp \left(-\lambda\int_{0}^{T}dt \theta(-x)\right)
\end{equation}
which is similar to the free particle case except that paths are weighted by a factor $e^{-\lambda\tau}$ where
\begin{equation}
\tau = \int_{0}^{T}dt \theta(-x)
\end{equation}
is the length of time spent in $x<0$. In the lattice case the corresponding conditional probability $u(0,T|0,0)$ is given by a sum of paths, each weighted by a similar factor.
This may be written as
\begin{equation}
u(0,T|0,0) = \frac{1}{2^{2n}}\sum_{k=0}^{n} \;n_{k}\; e^{-2k\epsilon \lambda}\label{exp},
\end{equation}
where $n_{k}$ is the number of paths spending a time $2k\epsilon\leq T$ in the region $x<0$. Expresions for these $n_{k}$ are known and are in fact independent of $k$ \cite{Cat}. They are equal to the Catalan numbers 
\begin{equation}
C_{n} = \frac{1}{n+1} {2n \choose n}.
\end{equation}
where $2n$ is the total number of time steps. We can see this as follows. First, note that the number of paths that never enter $x<0$ is given by $C_{n}$, this being one definition of the Catalan numbers. Next consider the following mapping on any path spending a time $2\epsilon k<T$ in $x<0$, Fig.(\ref{pathrule}).
\begin{enumerate}
\item Start from t=0, and follow the path until it first crosses $x=0$ (if it doesn't cross then stop, the path is in the set of non-crossing paths.)
\item Follow the path until it comes back to $x=0$ again, note the step at which this happens.
\item Swap the section of path \emph{before} this step with the section \emph{after} it.
\item The new path will now spend 2 fewer time steps in $x<0$.
\end{enumerate}
By repeated application of this mapping, any path can be transformed into one which never crosses $x=0$. The important point about this mapping however, is that it is bijective \cite{Cat}, which proves that the number of paths spending time $2k\epsilon$ in $x<0$ must equal the number of paths that never cross, for any value of $k$. This shows that $n_{k}=C_{n}$ for all $k\leq n$.
\begin{figure}[h] 
   \centering
   \includegraphics[width=3in]{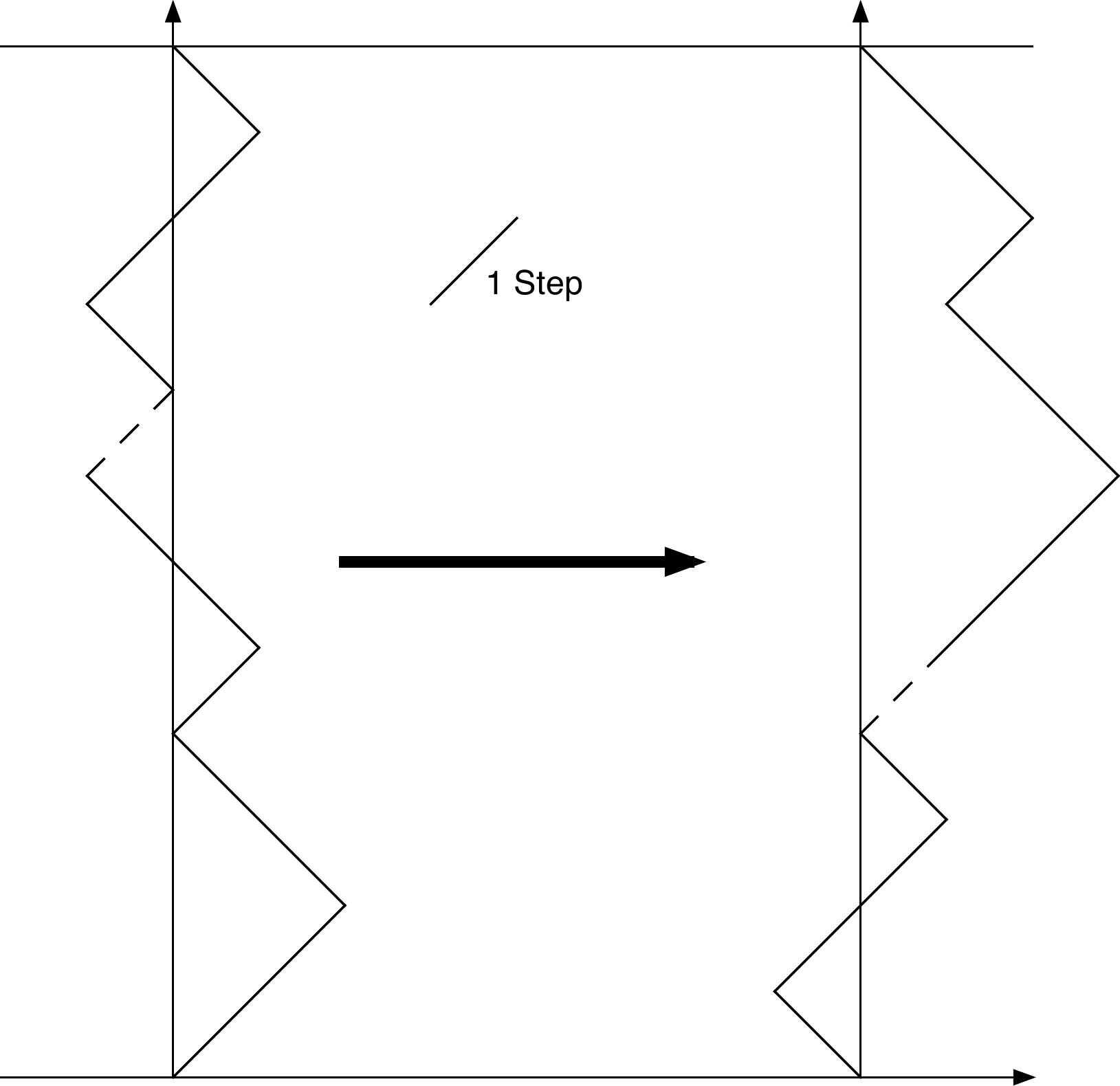} 
   \caption[The Bijection Between $n(k)$ and $C_{n}$.]{{\em Application of the rule to the path on the left produces the path on the right, with 2 timesteps less spent in $x<0$}}\label{pathrule}
\end{figure}
So we now have
\begin{eqnarray}
u(0,T|0,0) &=&\frac{1}{2^{2n}}\sum_{k=0}^{n}C_{n}e^{-2k\epsilon \l} = \frac{C_{n}}{2^{2n}}\sum_{k=0}^{n}e^{-2k\epsilon \l}\label{geo} \\ 
 &=&\frac{C_{n}}{2^{2n}}\left(\frac{1-e^{-2\epsilon(n+1)\l}}{1-e^{-2\epsilon \l}}\right). 
\end{eqnarray}
This coincides with the free particle case if $\l=0$.
Since we plan to take the continuum limit we can Taylor expand the exponential in the denominator to first order in $\epsilon$, and use the following useful assymptotic form for $C_{n}$ \cite{Cat}
\begin{displaymath}
C_{n} \approx \frac{4^{n}}{\sqrt{\pi}n^{3/2}}
\end{displaymath}
to get,
\begin{displaymath}
u(0,T|0,0) \approx  \frac{1}{\sqrt{\pi}2\epsilon n^{3/2}}\left(\frac{1-e^{-2\epsilon(n+1)\l}}{\l}\right).
\end{displaymath}
Now take the continuum limit as in Section 3.2 to obtain, after some simple algebra,
\begin{equation}
\overline g(0,T|0,0)= \left(\frac{m}{2 \pi}\right)^{1/2}\frac{(1-e^{-\l T})}{\l T^{3/2}}
\end{equation}
which implies 
\begin{equation}
g(0,T|0,0) = -i \left(\frac{m}{2 \pi i}\right)^{1/2}\frac{(1-e^{-i\l T})}{\l T^{3/2}}
\end{equation}
This is our first result, the propagator along the edge of a step potential. The  full propagator from $x_{0}$ to $x_{1}$ may now be obtained by making use of Eq.(\ref{16a}) \cite{car}. Note also that this result may be analytically continued in a simple way to imaginary potentials.

\section{The Delta Function Potential}
We now wish to evaluate Eqs.(\ref{3.1}) and (\ref{3.2}) with the  potential given by $V(x) = \l\delta(x)$. In a similar way as for the step potential, we can write this as,
\begin{equation}
\overline g(0,T|0,0) = \int \mathcal{D}x \exp \left(-\int_{0}^{T}dt \frac{m \dot x^{2}}{2}\right) \exp \left(-\l\int_{0}^{T}dt \delta(x(t))\right)
\end{equation}
which is again similar to the free particle case except that paths are weighted by a factor $e^{-\l\sigma}$ where
\begin{equation}
\sigma = \int_{0}^{T}dt \delta(x(t))
\end{equation}
is the number of times a given path crosses $x=0$. We model the delta function as a square potential of width $\eta$, and height $1/\eta$ so that every crossing of $x=0$ is weighted by a factor of $e^{-\l\epsilon/\eta}$. Since $\epsilon/\eta^{2}=m$, we can rewrite this as $e^{-\eta \l m}$. We take the potential to be located on the left of $x=0$ so that to start and end at $x=0$ involves an even number of crossings. Such choices are only made for convenience and have no significance in the continuum limit. 
The conditional probability density $u$ may be partitioned in a similar way to that for the step potential. However now the partitioning is with respect to the number of times a path crosses the square potential. Since the number of crossings will always be even we partition into classes of paths that cross $2l<2n$ times, so the conditional probability is
\begin{equation}
u(0,T|0,0)=\frac{1}{2^{2n}}\sum_{l=0}^{n}J(n,l)e^{-2lm\l\eta}\label{summ}
\end{equation}
Where $J(n,l)$ is the number of paths of $2n$ steps that cross the delta potential $2l$ times. It is known that these $J(n,l)$ are given by the $2l^{th}$ convolution of the Catalan numbers \cite{tri}, this can be demonstrated by writing a general path in terms of sums over non-crossing paths.
These convolutions form the diagonal elements in Catalan's triangle \cite{tri, tri2, tri3}. We need to know the $n^{th}$ element in the $2l^{th}$ diagonal from the right, which will give us $J(n,l)$.

From the formula for the elements of Catalan's triangle \cite{tri},
\begin{equation}
c(n,k) = \frac{(n+k)!(n-k+1)}{k!(n+1)!}
\end{equation}
it follows that 
\begin{equation}
J(n,l) =c(n+l,n-l)=\frac{(2n)! (2l+1)}{(n-l)!(n+l+1)!}={2n \choose n} \frac{n!n!(2l+1)}{(n-l)!(n+l+1)!}\label{inspect}
\end{equation}
where we have extracted the binomial factor for later convenience.
We cannot perform the summation in Eq.(\ref{summ}) directly as we did for the step function potential, so we take the continuum limit first to leave ourselves with an integral. In order to do this we need a simpler form for the $J$'s. We need the asymptotic form for $n\to \infty$, but plotting Eq.(\ref{inspect}) as a function of $l$ shows that the dominant contribution comes from taking $l\to \infty$ as well. It is possible to derive the following asymptotic form,
\begin{equation}
J(n,l)\sim {2n \choose n}\frac{2l}{n}e^{-l^{2}/n}\label{approx},
\end{equation}
and as this form shows, the maximum value of $J$ occurs when $l$ is of order $\sqrt{n}$. This is a consequence of the Brownian motion origin of the paths, which therefore have Hausdorff dimension 2 \cite{dimension}. The statement that a typical path crosses the origin an infinite number of times is a consequence of this fractal nature of a typical path. 

If we use the asymptotic form for the $J$'s, Eq. (\ref{approx}) and make the change of variable in the partitioning, Eq. (\ref{summ}), $u=2\eta l$ we can turn the summation into an integral,
\begin{equation}
u(0,T|0,0) = \frac{1}{2^{2n}}{2n \choose n} \int_{0}^{2\eta n} \! du \;\frac{u}{2\eta^{2}n}e^{-\frac{u^{2}}{4\eta^{2}n}}e^{-\l mu}.
\end{equation}
We now take the continuum limit as in Section 3.2 to obtain
\begin{eqnarray}
\overline g(0,T|0,0) &=&\left(\frac{m^{3}}{2\pi T^{3}}\right)^{1/2}\int_{0}^{\infty}\!du \;u e^{-\frac{mu^{2}}{2T}}e^{-\l mu}\nonumber \\
& =&\overline g_{f}(0,T|0,0)-\l m\int_{0}^{\infty}\!du \;e^{-\l mu}\overline g_{f}(u,T|0,0)\label{5}
\end{eqnarray}
which, after rotating back to real time, yields
\begin{equation}
g(0,T|0,0)= g_{f}(0,T|0,0)-\l m\int_{0}^{\infty}\!du \;e^{-\l mu}g_{f}(u,T|0,0).
\end{equation}
We can now use the path decompsition expansion to obtain the propagator for paths which start at $x_{0}$ and end at $x_{1}$. There are 4 cases, depending on the signs of $x_{0}$ and $x_{1}$, we shall present one case here, the others follow in a very similar fashion. 
First note that PDX formulae in Chapter 2 tell us that the free propagator obeys the following,
\begin{eqnarray}
g_{f}(x_{1},T|x_{0},0) &=& \frac{i}{m} \int_{0}^{T} dt_{1}\; g_{f}(x_{1},T|0,t_{1}) \left.\frac{\partial g_{f}}{\partial x}(x,t_{1}|x_{0},0)\right|_{x=0}\label{a}\\
&=& -\frac{i}{m} \int_{0}^{T} dt_{2}\; \left.\frac{\partial g_{f}}{\partial x}(x_{1},T|x,t_{2})\right|_{x=0} g_{f}(0,t_{2}|x_{0},0)\label{b}
\end{eqnarray}
provided $\e(x_{0})=\e(-x_{1})$. (where $g_{f}$ denotes the free propagator, and $\e(x)$ is the signum function.)
(If the start and end points have the same sign then there are paths between the two that never cross $x=0$, so the expression for the path decomposition expansion has to be modified as in Eq.(\ref{PDX2}) \cite{HaOr}.)

Secondly note the following identities,
\begin{equation}
g_{f}(a,t_{2}|b,t_{1})=g_{f}(-a,t_{2}|-b,t_{1})=g_{f}(0,t_{2}|b-a,t_{1})=g_{f}(0,t_{2}|a-b,t_{1})\label{point}.
\end{equation} 
which are just expressions of the symmetry of the free propagator.

Consider the case where $x_{0}<0$ and $x_{1}>0$, noting that $u\geq0$  we first use Eq. (\ref{a}) to attach the leg from $(x_{0},0)$,
\begin{eqnarray}
g(0,t_{2}|x_{0},0)&=&g_{f}(0,t_{2}|x_{0},0)-\l m\int_{0}^{\infty}du e^{-\l mu} g_{f}(u,t_{2}|x_{0},0)\nonumber\\
&=&g_{f}(0,t_{2}|x_{0},0)-\l m\int_{0}^{\infty}du e^{-\l mu} g_{f}(0,t_{2}|x_{0}-u,0)
\end{eqnarray}
where we have used Eq. (\ref{point}) to obtain the second line. Since $x_{0}<0$ we have that $x_{0}-u<0$, so we can attach a leg to $(x_{1},T)$ using Eq.(\ref{b}),
 \begin{eqnarray}
g(x_{1},T|x_{0},0)&=&g_{f}(x_{1},T|x_{0},0)-\l m\int_{0}^{\infty}du e^{-\l mu} g_{f}(x_{1},T|x_{0}-u,0)\nonumber\\
&=&g_{f}(x_{1},T|x_{0},0)-\l m\int_{0}^{\infty}du e^{-\l mu} g_{f}(|x_{1}|+|x_{0}|+u,T|0,0)\label{hurrah}
\end{eqnarray}
as expected.
Calculation in the other cases proceeds in a very similar manner, and confirms that Eq.(\ref{hurrah}) is valid in all cases.
 
\chapter{On the Relationship between Complex Potentials and Strings of Projection Operators}\label{cppo}
\epigraph{Time's a strange fellow;\\
\hspace{115pt}more he gives than takes\\ (and he takes all)nor any marvel finds\\ quite disappearance but some keener makes\\ losing,gaining\\
\hspace{70pt}-love!if a world ends\\
more than all worlds begin to(see?)begin}{E.E. Cummings}

\section{Introduction}

In Chapter 1 we argued that in decoherent histories the class operator describing histories which remain in some region defined by a projector $P$ is given by $C^{\e}_{nc}$, where,
\beq
C^{\e}_{nc}| \psi \rangle = e^{-iH \epsilon}
Pe^{-i H \epsilon}...P e^{-i H \epsilon} | \psi \rangle
\label{4.1.1}
\eeq
(Here, there are $n$ projection operators and $ \tau = (n+1) \epsilon$
and we use units in which $\hbar =1 $).
Such an object  also describes the evolution of an initial states under pulsed measurements.

We are interested in the specific case of a free particle with $P$ taken to be
the projection onto the positive $x$-axis, $P = \theta ( \hat x)$. For
sufficiently small $\epsilon$, Eq.(\ref{4.1.1}) is then a candidate for
the amplitude to remain in $x>0$ during the time interval $[0, \tau]$. In this Chapter we examine the object in Eq.(\ref{4.1.1}) without particular reference to it's relavence in the arrival time problem. Our aim will be to show that evolution under Eq.(\ref{4.1.1}) is equivalent to evolution under a particular form of continuous measurement. We will then use this result in Chapter 5 to address the arrival time problem.

It is of interest
to explore the properties of this Eq.(\ref{4.1.1}) for a range of values
of the time spacing $\epsilon$. It is known that as $\epsilon \rightarrow 0$,
we approach the Zeno limit, in which the state becomes entirely
confined to the Hilbert subspace
of states with support only in $x>0$, so that an incoming wave packet
from the right is totally reflected \cite{Zeno,Wall,Sch2}.
However, it is of greater physical
interest to explore the regime of non-zero $\epsilon$, in which the system
is monitored sufficiently well to get some idea of whether the particle is in $x>0$,
yet not monitored so much that an incoming state is significantly
reflected at $x=0$. An important question in this regime is to determine the
value of $ \epsilon $ for which reflection becomes significant. For an initial
state with energy width $\Delta H$, a timescale held to be significant
is the Zeno time,
\beq
t_Z = \frac {1} { \Delta H}
\eeq
which is the timescale on which the state becomes significantly different
from its initial value under unitary evolution \cite{Peres}.
For a wave
packet of momentum $p$ and spatial width $\sigma$, the Zeno time is of order $m \sigma / p$
which is the timescale on which the wave packet crosses the origin.
This indicates that the Zeno time for wave packets
is an essentially classical timescale and, in Eq.(\ref{4.1.1}), relates only
to the rate of removal of probability through projection. By contrast,
reflection in Eq.(\ref{4.1.1}) arises as a result of the increase in uncertainty in momentum
resulting from position projection, an obviously quantum process, so one would expect it to have a different
timescale, which could be much shorter than then Zeno time. It would be of
interest to compute this timescale. One reason it is important is that
there appears to be interesting physics very close to the Zeno limit \cite{Ech,Muga2,Hal4}.

A significant result in this area is due to Echanobe, del Campo and Muga, who claimed
that for finite $\epsilon$ the string of projection operators in Eq.(\ref{4.1.1})
is approximately equivalent to evolution in the presence of a complex potential
\cite{Ech}.
That is,
\beq
e^{ - i H \epsilon} Pe^{-i H \epsilon} \cdots P e^{-i H \epsilon}
\approx \exp \left( - i H \tau - V_0 \theta ( -x ) \tau \right)
\label{4.1.2}
\eeq
They assert that this result is valid if, for a given $\epsilon$, $V_0$ is chosen such that
two conditions
\bea
V_0 \epsilon  & \gg & 1
\label{4.1.3} \\
V_0 & \gg & \Delta H
\label{4.1.4}
\eea
are satisfied.
This is a very useful result since Eq.(\ref{4.1.1}) is not easy to evaluate analytically
but the Schr\"odinger equation with a complex step potential in Eq.(\ref{4.1.2}) can be
solved straightforwardly. Furthermore, such complex potentials have been studied extensively
in the literature and can often be linked to particular detection methods
\cite{complex}.

Given the connection Eq.(\ref{4.1.2}), one can determine the conditions under which
reflection becomes important. Known results on scattering with the complex
potential in Eq.(\ref{4.1.2}) show that, for an incoming state with energy scale $E$,
reflection becomes significant when $V_0 > E$ \cite{All, HaYe1}. Reflection is avoided,
therefore, when $ V_0 \ll E $, or equivalently, from Eq.(\ref{4.1.3}), when
\beq
\epsilon \gg \frac {1} {E}
\label{4.1.6}
\eeq
This is much less than the Zeno time for a state strongly peaked in energy.
There is therefore an interesting regime, namely
\beq
\frac{1} {E} \ \ll \ \epsilon \ \ll \ \frac {1} { \Delta H}
\label{4.1.7}
\eeq
in which the projections in Eq.(\ref{4.1.1}) are sufficiently frequent
to have a significant effect on the system, yet not so frequent that there is significant reflection.

Eq.(\ref{4.1.6}) is a very useful result, but it has been derived on the basis
of the claimed approximate
relationship Eq.(\ref{4.1.2}). The derivation of Eq.(\ref{4.1.2})
given by Echanobe et al
is very plausible (and was also hinted at by Allcock \cite{All}), but it
is rather heuristic, and the deduced connection
Eq.(\ref{4.1.3}) between $V_0$ and $\epsilon$ is rather loose.
There is therefore considerable scope for
a more detailed and substantial derivation.

The purpose of this chapter is to give a more substantial derivation of the equivalence
Eq.(\ref{4.1.2}) and to deduce a more precise relationship between $ V_0$ and $\epsilon$.
We will do this by computing the configuration space propagators associated
with each side of Eq.(\ref{4.1.2}) and show that they are approximately equal in
certain regimes. The propagator associated with the complex potential is in fact
known already, so the bulk of the work consists of an approximate evaluation of
the propagator associated with pulsed measurements, the left-hand side of Eq.(\ref{4.1.2}).

This is certainly not a rigorous mathematical proof of Eq.(\ref{4.1.2}), involving operator norms,
error bounds and the like, but a theoretical physicists style of proof involving
the approximate evaluation of propagators. A rigorous proof would certainly be of interest to construct
and the work described in this chapter may give some hints in that direction.

We begin in Section \ref{secl2} with a brief summary of the derivation of Echanobe et al,
with a small extension of it that turns out to be important and yields an equality
relating $V_0$ and $\eps$, thereby improving on Eq.(\ref{4.1.3}).
We then in Section 4.3 give a detailed formulation of the problem and how we solve it.
The key idea is to use the path decomposition expansion in which the propagators associated with
each side of Eq.(\ref{4.1.2}) are factored across the surface $x=0$
\cite{pdx, pdx2,HaOr,Hal3}. The problem of proving the equivalence Eq.(\ref{4.1.2}) thereby reduces to proving it for propagation between points at $x=0$ for different times. Since the propagator for the complex potential is known, the main work is to evaluate this propagator for pulsed measurements. This is actually rather difficult to do directly, but a good approximate analytic expression can be obtained for it
by attacking the problem from a number of different angles. This is described in
Sections 4.4, 4.5, 4.6 and 4.7. In Section 4.8
we give a detailed discussion of the timescales involved for the approximations
to be valid.
We summarize and conclude in Section 4.9. This chapter is based on Ref.\cite{HaYe3}.

\section{Review and Extension of Earlier Work}\label{secl2}

We first review and extend the derivation of Echanobe et al \cite{Ech}. They first note that
\beq
 \exp \left( -  V_0  \theta ( - \hat x ) \epsilon  \right)
 = P + e^{ - V_0 \eps } \bar P
\label{4.2.AA}
 \eeq
where, recall, $P = \theta ( \hat x ) $ and  $\bar P = 1 - P $. It follows that
\beq
P = \theta (\hat x ) \approx  \exp \left( -  V_0  \theta ( - \hat x ) \epsilon  \right),
\label{4.2.A}
\eeq
as long as the parameter
\beq
\alpha = V_ 0 \epsilon
\label{4.2.B}
\eeq
is sufficiently large that
\beq
e^{ - \alpha } \ll 1.
\label{4.2.BB}
\eeq
Eq.(\ref{4.1.2}) then follows from the approximate equivalence,
\beq
\exp \left( - i H \epsilon \right)
\ \exp \left( -  V_0  \theta ( - \hat x ) \epsilon  \right)
\approx
\exp \left( - i H \epsilon  -  V_0  \theta ( - \hat x ) \epsilon  \right),
\eeq
which will hold as long as
\beq
V_0 \epsilon^2 |\langle [H, \theta (-\hat x) ] \rangle| \ll 1,
\label{4.2.C}
\eeq
where the average is taken in the initial state.
Echanobe et al put an upper bound on the left-hand side using the Schr\"odinger-Robertson
inequality and Eq.(\ref{4.2.C}) may then be written in either of the two equivalent forms
\beq
\alpha^2 \Delta H \ll V_0
\label{4.2.D}
\eeq
or
\beq
\alpha \eps \ll \frac {1} { \Delta H}
\label{4.2.DD}
\eeq
which implies that the time between projections is much less than the Zeno time,
the typical timescale on which the state undergoes significant change \cite{Peres}.
(The conditions Eqs.(\ref{4.2.B}), (\ref{4.2.BB}), (\ref{4.2.D}) are a more precise version of the originally stated conditions
Eqs.(\ref{4.1.3}), (\ref{4.1.4})).

Since the parameter $\alpha$ need only satisfy an {\it inequality}, Eq.(\ref{4.2.BB}), the
relationship Eq.(\ref{4.2.B}) between $\eps$ and $V_0$ is not uniquely determined. However, it turns
out that the above derivation can be extended somewhat to give an {\it equality}
between $\eps$ and $V_0$. This turns out to be relevant to the more substantial derivation
given in the rest of this chapter.

The above result may be written,
\beq
\exp \left( - i H \epsilon  -  V_0  \theta ( - \hat x ) \epsilon  \right)
\approx \exp \left( - i H \epsilon \right) P
\label{4.2.E}
\eeq
under the conditions given above. Now suppose $\alpha $ is an integer and
write $ \eps = \alpha \eps'$, where
\beq
V_0 \eps' =1
\label{4.2.EE}
\eeq
Then, since
$P= P^2 $, we may approximate Eq.(\ref{4.2.E}) by
\bea
\exp \left( - i H \epsilon  -  V_0  \theta ( - \hat x ) \epsilon  \right)
& \approx &
e^{ - i H \eps'} Pe^{-i H \epsilon'} P \cdots e^{-i H \eps'} P
\nonumber \\
&=& \left( e^{ - i H \eps'} P \right)^{\alpha}
\label{4.2.F}
\eea
as long as the contribution from the commutator terms between $ e^{-iH \eps'}$ and $P$
is sufficiently
small. There will be of order $\alpha^2$ such terms, hence the error in this
approximation is of order $ \alpha^2 \eps' \Delta H $ and from Eq.(\ref{4.2.D}),
this error is much less than $1$. The key point here is that even the longer timescale
$\eps$ is still much less than the Zeno time and, since nothing changes on this
timescale, there is essentially no difference between Eqs.(\ref{4.2.E}) and (\ref{4.2.F}).

We therefore see that the desired result Eq.(\ref{4.1.2}) actually holds for much
smaller time steps, defined by the equality Eq.(\ref{4.2.EE}). This is
different to the original claim of Echanobe et al, since they require the
inequality Eq.(\ref{4.1.3}). However, the above argument shows that the restriction
Eq.(\ref{4.1.3}) is in fact stronger than necessary and in the following pages
our more detailed derivation will show that
Eq.(\ref{4.1.2}) does indeed hold with a timespacing of order $ 1 / V_0 $.

\section{Detailed Formulation of the Problem}

In this chapter, we will prove the relationship Eq.(\ref{4.1.2}) in a much more substantial
way by proving the approximate equivalence of the propagators
\bea
g_V ( x_1, \tau | x_0, 0 ) &=& \langle x_1 |   \exp \left( - i H \tau - V_0 \theta ( -x ) \tau \right)
| x_0 \rangle
\label{4.2.2}
\\
g_P ( x_1, \tau | x_0, 0 ) &=& \langle x_1 | e^{-iH \epsilon_{n}} Pe^{-i H \epsilon}
...e^{-i H \epsilon}P e^{-i H \eps_0} | x_0 \rangle
\label{4.2.1}
\eea
for some relationship between the parameters $\eps$ and $V_0$, to be determined.
Note that in Eq.(\ref{4.2.1}), we have chosen the initial and final time spacings
to be $\eps_0$ and $\eps_n $, with
\beq
\tau = (n-1) \eps + \eps_0 + \eps_n
\label{4.2.3}
\eeq
This turns
out to be necessary for the most general proof of Eq.(\ref{4.1.2}).
For the special case $ \eps_n = \eps = \eps_0 $, we may also write
\beq
g_P ( x_1, \tau | x_0, 0 ) =
\langle x_1 | e^{ - i H \tau} P( n \eps) \cdots P (2 \eps ) P ( \eps ) | x_0 \rangle
\label{4.T0}
\eeq

Each of the above propagators may be represented by a path integral,
\bea
g_V ( x_1, \tau | x_0, 0 ) &=& \int {\cal D}x \exp \left( i \int_0^\tau dt \left[ \half m \dot x^2 + i V_0
\theta (-x) \right] \right)
\label{4.2.5}
\\
g_P ( x_1, \tau | x_0, 0 ) &=& \int_P {\cal D}x \exp \left( i \int_0^\tau dt \half m \dot x^2 \right)
\label{4.2.4}
\eea
where in both cases the paths are from $x(0)=x_0$ to $x(\tau) = x_1$ and in the second
case, Eq.(\ref{4.2.4}), $P$ denotes that the paths are restricted to be in the positive $x$-axis
at times $t = \eps_0+ (k-1) \epsilon$, $k=1,2 \cdots n$.

A closely related object that will be important is the restricted
propagator,
\beq
g_r ( x_1, \tau | x_0, 0 ) = \int_{x(t)>0} {\cal
D}x \exp \left( i \int_0^\tau dt \half m \dot x^2 \right)
\label{4.2.6}
\eeq
where again the paths are from $x(0)=x_0$ to
$x(\tau) = x_1$ but with $x(t)>0$ for all times in $[0, \tau]$.
This is clearly equivalent to $g_P$ in Eq.(\ref{4.2.4}), in the limit
$ n \rightarrow \infty $, $ \eps \rightarrow 0 $
with $ \tau $ constant. If we take the same
limit in the equivalent expression for $g_P$, Eq.(\ref{4.2.1}),
one obtains the following convenient operator form of the restricted
propagator:
\beq
g_r (\tau,0) =  \exp ( - i P H P \tau )P
\label{4.2.7}
\eeq
The restricted propagator satisfies the Schr\"odinger equation in $x>0$
subject to the boundary conditions that it vanishes when either end
of the propagator sits on $x=0$.
For the free particle, considered here, one can easily solve for the
restricted propagator using the
method of images and the result is
\bea
g_r ( x_1, \tau | x_0, 0 ) &=& \left( \frac {m} { 2 \pi i \tau} \right)^{1/2}
\theta (x_1) \theta (x_0) \nonumber \\
& \times &
\left[
\exp \left(  \frac { i m (x_1 - x_0)^2 } { 2 \tau} \right)
- \exp \left(  \frac { i m (x_1 + x_0)^2 } { 2 \tau} \right) \right]
\label{4.2.8}
\eea
The restricted propagator, in any of the above forms, describes the regime
of ``Zeno dynamics'', in which all states are confined entirely to the Hilbert
subspace of states with support only in $x>0$ \cite{Sch2}.

The restricted propagator plays an important role here since not only does
$g_P \rightarrow g_r $
in the limit $ n \rightarrow \infty $, $ \eps \rightarrow 0 $
with $ \tau $ constant, but also $ g_V \rightarrow g_r $ as $V_0 \rightarrow \infty$.
It follows that $g_V$ and $g_P$ become arbitrarily close to each other for
sufficiently large $V_0$ and $n$, since they both tend to the same limit.
This is the underlying reason why we expect the approximate equivalence Eq.(\ref{4.1.2})
should hold.

The propagators $ g_P$ and $g_V$ may be decomposed using the path decomposition expansion (PDX), which we discussed in detail in Chapter 2. There are two cases to consider.

\begin{figure}[htp] 
   \centering
   \includegraphics[width=10cm]{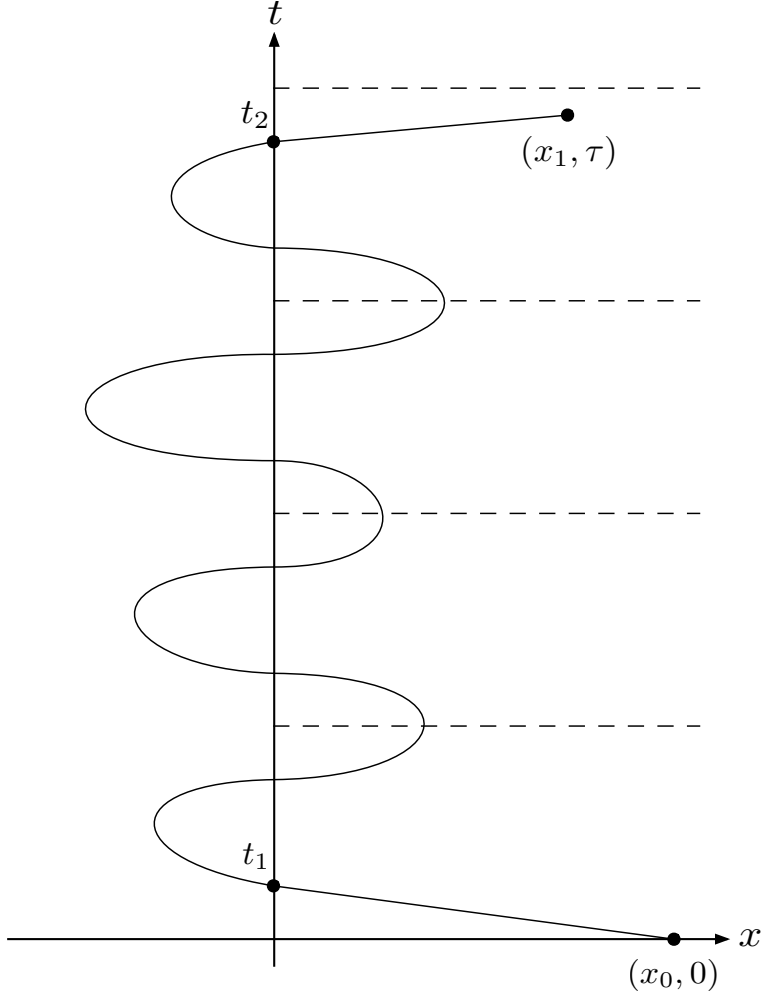} 
   \caption[The Path Decomposition Expansion for $g_{P}$.]{{\em The path decomposition expansion for $g_P$.
Any path from $ x_0 > 0 $ at $t=0$ to a final point $x_1 > 0 $  at $t=\tau$ which crosses $x=0$ has a first crossing of $x=0$ at $t_1$ and a last crossing at $t_2$. The crossing part of the propagator from $(x_0,0)$ to $(x_1, \tau)$ may therefore be decomposed into three parts: (A) restricted propagation entirely in $x>0$, (B) propagation starting and ending on $x=0$ with the restriction that the paths are in $x>0$ at a discrete set of times, and (C) restricted propagation entirely in $x>0$. The corresponding path decomposition expansion formula is given in Eq.(\ref{4.2.9}). (The $g_r$ term corresponds to paths which never cross $x=0$).}} \label{fig:cpfigure1}
\end{figure}

We consider first the case in which the initial and final points are on the same side of the surface in $x>0$. Recall from Chapter 2 that the corresponding PDX has the form
\bea
g(x_1, \tau | x_0,0 ) &=&
\frac {1  } {4m^2} \int_{0}^{\tau} dt_2
\int_0^{t_2} dt_1
\ \frac {\partial g_{r}} {\partial x} (x_1, \tau | x, t_2) \big|_{x=0} \ g (0,t_2| 0,t_1)
\ \frac {\partial g_{r} } { \partial x} (x,t_1| x_0,0) \big|_{x=0}
\nonumber \\
&+&g_r ( x_1, \tau | x_0, 0 )
\label{4.2.9}
\eea
This is depicted in Figure \ref{fig:cpfigure1}.

Eq.(\ref{4.2.9}) holds for both $g_P$ and
$g_V$, but the restricted part $g_r $ is the same in each case and equal to
Eq.(\ref{4.2.8}) above, since the restriction on paths defining $ g_P$ and
the presence of the complex potential in
$g_V$
in Eqs.(\ref{4.2.5}), (\ref{4.2.4}) are both redundant if $x(t)>0$.

It follows from the above that
$g_P$ and $g_V$ could differ only in terms of their propagation
along $x=0$, hence to prove the approximate equivalence of the propagators
Eqs.(\ref{4.2.2}), (\ref{4.2.1}), we need to prove that
\beq
g_P (0, t_2 | 0, t_1 ) \approx g_V (0, t_2| 0, t_1 )
\label{4.2.11}
\eeq
Note that, unlike $g_V (0,t_2|0,t_1)$, $ g_P (0, t_2 | 0, t_1 ) $ is
not covariant under time translation, that is,
\beq
g_P (0, t_2 | 0, t_1 ) \ne g_P (0, t_2 -t_1 | 0, 0 )
\label{4.cov}
\eeq
although it has an approximate covariance on timescales much greater than $\eps$,
as we will see below.

The second case is that in which the initial
and final points are on opposite sides of the surface, $x_1 < 0 $ and $x_0 > 0 $.
In this case we have for $g_V$,
\bea
g_V(x_1, \tau | x_0,0 ) &=&
\frac {1  } {4m^2} \int_{0}^{\tau} dt_2 \ \ e^{ - V_0 ( \tau - t_2) }
\ \int_0^{t_2} dt_1
\ \frac {\partial g_r} {\partial x} (x_1, \tau | x, t_2) \big|_{x=0}
\nonumber \\
& \times & \ g_V (0,t_2| 0,t_1)
\ \frac {\partial g_r } { \partial x} (x,t_1| x_0,0) \big|_{x=0}
\label{4.2.13}
\eea
since the last section of the paths is in $x<0$ where the complex potential acts.
For $g_P$, if $x_1 < 0$, since the paths must be in $x>0$ at the given discrete
set of times,
the last crossing time $t_2$ cannot be less than the last time $ \tau - \eps_n $
at which the projectors act, hence
\bea
g_P(x_1, \tau | x_0,0 ) &=&
\frac {1  } {4m^2} \int_{\tau - \eps_n}^{\tau} dt_2
\int_0^{t_2} dt_1
\ \frac {\partial g_r} {\partial x} (x_1, \tau | x, t_2) \big|_{x=0}
\nonumber \\
& \times &
\ g_P (0,t_2| 0,t_1)
\ \frac {\partial g_r } { \partial x} (x,t_1| x_0,0) \big|_{x=0}
\label{4.2.14}
\eea
Again we will have to prove that Eq.(\ref{4.2.11}) holds, but these two expressions
Eqs.(\ref{4.2.13}), (\ref{4.2.14}) also differ in the form of the $t_2$ integral.
However, since $ V_0 \eps \approx 1 $, the form of the exponential in
Eq.(\ref{4.2.13}) effectively squeezes $t_2$ to lie approximately
within $\eps $ of $\tau$, so we have
\beq
\int_{0}^{\tau} dt_2 \ \ e^{ - V_0 ( \tau - t_2) } \approx
\int_{\tau - \eps_n}^{\tau} dt_2
\eeq
and Eqs.(\ref{4.2.13}) and (\ref{4.2.14}) are approximately the same.

We see that both cases reduce to proving Eq.(\ref{4.2.11}).
In Chapter 3 we showed that the propagator along the boundary of the complex potential is given by,
\bea
g_{V}(0,t|0,0)&=&\left(\frac{m}{2\pi i
t}\right)^{1/2}\frac{(1-e^{-V_{0}t })}{V_{0}t }
\nonumber \\
&:=& \left(\frac{m}{2\pi i t}\right)^{1/2} f_V (t)
\label{4.CP}.
\eea
The main purpose of the remainder of this chapter is to calculate
the propagator with projection operators along the boundary,
\beq
g_P ( 0, \tau | 0, 0 ) = \langle 0 | e^{-iH \epsilon_{n}} Pe^{-i H \epsilon}
\cdots e^{-i H \epsilon}P e^{-i H \eps_0} | 0 \rangle
\label{4.GP}
\eeq
and show that
the approximation Eq.(\ref{4.2.11}) holds, under conditions to be determined.
(Here, $ |0 \rangle $ denotes a position eigenstate $ | x \rangle $ at $x=0$).

In what follows, our main result is to show that, to a good approximation,
\beq
g_P ( 0, t | 0, 0 ) \approx \left(\frac{m}{2\pi i t}\right)^{1/2} f_P (t)
\label{4.FP1}
\eeq
where $f_P (t)$ is a kind of saw-tooth function -- a piecewise linear function
with peaks immediately followed by troughs at $t_k = \eps_0 + (k-1) \eps $,
$k= 1,2,\cdots$.
Approaching $t_k$ from below, there is a peak of value
\beq
f_P (t_k) = \frac {1} {k+1}
\label{4.peaks}
\eeq
and approaching $t_k$ from above there is a trough of half that size.
That is
\beq
f_P (t) = \frac {(t - t_{k-1})} {(k+1) \eps} + \frac { (t_{k} - t) } {2 k \eps}
\ \ \ \ {\rm for } \ \ \ \ t_{k-1} \le t < t_{k}, \ \ k = 2,3 \cdots
\label{4.FP2}
\eeq
and
\beq
f_P(t) = 1 \ \ \ \ {\rm for} \ \ \ \ 0 \le t < \eps_0
\eeq
The functions $f_P (t)$ and $f_V (t) $  are shown in Figure \ref{fig:cpfig2}.
We see that $f_P (t)$
oscillates with period $\eps$ about $ f_V (t)$, as long as
we choose $V_0$ so that $f_V (t) $ lies between the peaks and troughs of $f_P (t)$.
That is,
for large $k$, we require that
\beq
\frac {1} {2k} < \frac{1} { V_0 t} < \frac {1} {k}
\eeq
Since $ t \approx k \eps $, $f_V (t)$ will
lie approximately midway between the peaks and troughs of $f_P (t)$ if
\beq
V_0 \eps \approx \frac {4} {3}
\eeq
Recalling that the propagator is attached through the PDX Eq.(\ref{4.2.9})
to an initial state, the oscillations, and hence the
differences between $g_P$ and $g_V$, will be smoothed out as long
as $\eps $ is chosen to be smaller than the timescale of variation
of the initial state. With some qualifications (discussed further in Section 4.8),
this timescale
is the Zeno time, $ 1 / \Delta H$.
The desired approximation Eq.(\ref{4.2.11}) will therefore hold in a time-averaged
sense, and
we thus have significant agreement
with the extended version of the original argument of Echanobe et al
described in Section 4.2.

\begin{figure}[htp] 
\centering
\includegraphics[width=15cm]{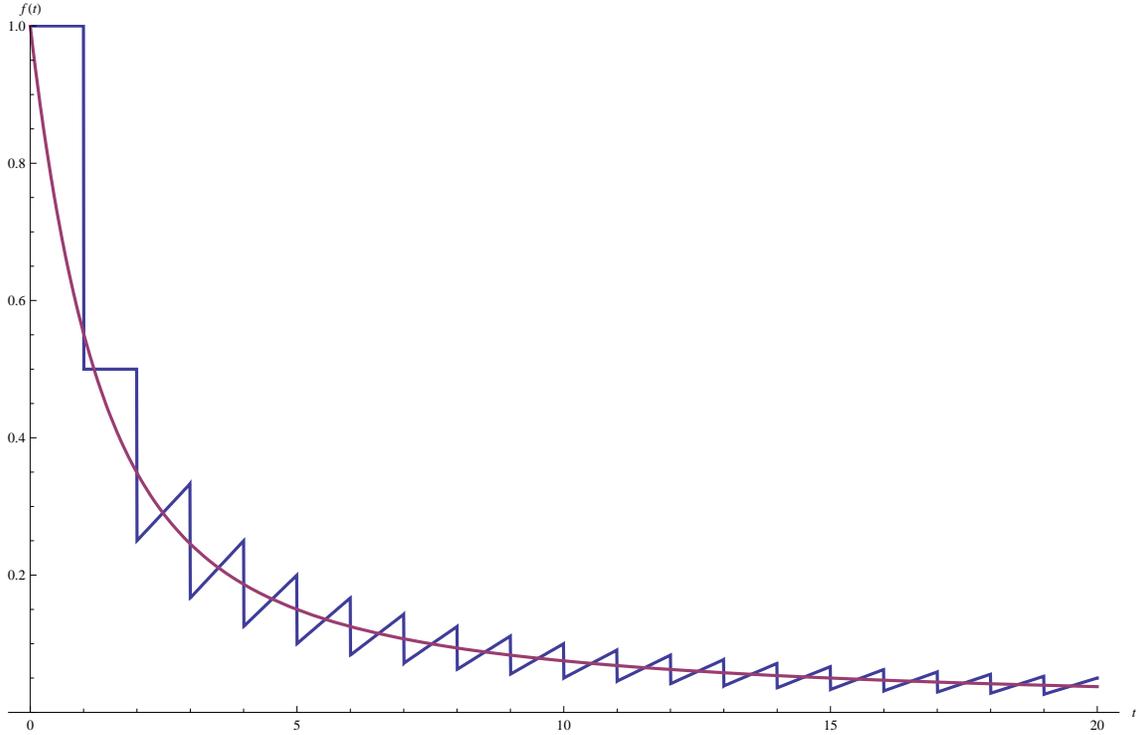} 
  \caption[A Plot Comparing the Functions $f_{V}(t)$ and $f_{P}(t)$.]{{\em A plot of the functions $f_V(t)$ (smooth curve) and $ f_P (t)$ (saw-tooth curve), defined for the complex potential by Eq.(\ref{4.CP}) with $ V_0 \eps = 4/3$, and for intermittent projectors by Eqs.(\ref{4.FP1}), (\ref{4.FP2}). (The time scaling is such that $\eps = 1$ in the plot, so that the peaks occur at integer values of $t$). We see that $f_P (t)$ oscillates around $f_V (t)$ with period $\eps$ so the two functions are equal in a time-averaged sense, when integrated against functions which are slowly varying on a timescale $\eps$.}}
 \label{fig:cpfig2}
\end{figure}

In the following sections, we evaluate Eq.(\ref{4.GP}) and confirm the
form Eq.(\ref{4.FP2}) of $f_P(t)$.
In Section 4.4 we evaluate Eq.(\ref{4.GP}) exactly for the cases of one, two and
three projections.
We also show why in general the troughs of $f_P(t)$ are exactly half the size of the
peaks by considering the limit $\eps_n \rightarrow 0 $ of Eq.(\ref{4.GP}). In Section 4.5
we use a lattice method to derive the magnitude of the peaks of
$f_P (t)$ for large $k$.
These results are substantiated
in Section 4.6, where we use an S-matrix expansion to derive some exact
results for a certain time-averaged version of Eq.(\ref{4.GP}).
In Section 4.7,
we fill in some of the gaps in these
regimes and approximations by computing Eq.(\ref{4.GP}) using
numerical methods.

We will find in the numerical and analytic solutions
that the function interpolating between the troughs
and peaks of $f_P(t)$ is not in fact a linear function in general. It is
interesting however, that, because of the slow time variation of the initial
state in comparison to $\eps$,
the precise form of this function turns out to be unimportant, and
this is what makes the problem more tractable than one might expect.
All that is important is the location of the peaks and troughs
of the saw-tooth function and the period of their oscillation.
It is this separation of timescales that also restores an approximate
time translation covariance to $g_P$, even though it does not hold exactly
in general, Eq.(\ref{4.cov}).

\section{Exact Analytic Results}

In this section we carry out an exact evaluation of Eq.(\ref{4.GP}) for the case
of one, two and three projections. For simplicity, we take the initial time interval
$\eps_0$ to be $\eps$ and for the case of three projections we are able to carry
out the calculation only in when the final time $\eps_n = \eps$.
We thus compute the objects
\beq
g_P (0,t|0,0) =
\begin{cases}
\langle 0 | e^{ - i H t} | 0 \rangle, & \mbox{if} \ 0\le t < \eps \\
\langle 0 | e^{ - i H (t-\eps)} P e^{ - i H \eps} | 0 \rangle, &\mbox{if}  \ \eps \le t < 2 \eps \\
\langle 0 | e^{ - i H (t-2 \eps)} P e^{ - i H \eps} P e^{ - i H \eps} | 0 \rangle, &\mbox{if}  \ 2 \eps \le t < 3 \eps \\
\langle 0 | e^{ - i H \eps }P e^{ - i H \eps} P e^{ - i H \eps} P e^{ - i H \eps} | 0 \rangle, &\mbox{if} \ t = 4 \eps
\end{cases}
\label{4.4E.1}
\eeq
We will show that
\beq
g_P (0,t|0,0) =
\begin{cases}
\left( \frac {m} {2 \pi i t} \right)^{\half} &\mbox{if}  \ 0\le t <  \eps \\
\half \left( \frac {m} {2 \pi i t} \right)^{\half} &\mbox{if}  \ \eps \le t < 2 \eps \\
\frac {1} {4} \left( 1 + \frac {2} {\pi} {\rm ArcTan} [(t-2 \eps) /t ]^{\half}  \right) \left( \frac {m} {2 \pi i t} \right)^{\half}
&\mbox{if}  \ 2 \eps \le t \ < 3 \eps \\
\frac {1} {4} \left( \frac {m} {2 \pi i t} \right)^{\half}
&\mbox{if}  \ t = 4 \eps
\end{cases}
\label{4.4E.2}
\eeq
thereby confirming the approximate form of the saw-tooth function, Eq.(\ref{4.FP2}), for the
first few projections. These expressions all have the property that the value
of $g_P(0,t|0,0)$ drops to half its value immediately after a projection and we show
why this is true in general at the end of this section.

To compute Eq.(\ref{4.4E.1}) it turns out to be
useful to consider more general objects, in which the projections may
be $P$, $ \bar P$ or the identity. We therefore consider the object
\bea
g_{P,C}(0,t |0,0) &=&\int_{C}dx_{1}...dx_{n}\left(\frac{m}{2\pi i \e}\right)^{n/2} \left(\frac{m}{2\pi i (t - n \eps)}\right)^{1/2}
\nonumber \\
& \times & \exp\left(\frac{imx_{1}^{2}}{2\e}+\frac{im}{2\e}\sum_{k=1}^{n-1}(x_{k}-x_{k+1})^{2}+\frac{imx_{n}^{2}}{2(t - n \eps)}\right)
\eea
Here $C=\{+,-,0...\}$ symbolically stands for the integration ranges of the $x_{k}$, eg $C=\{+,-,0...\}$
means $0<x_{1}<\infty, -\infty<x_{2}<0, -\infty<x_{3}<\infty$ etc.
Performing a Wick rotation, $\e\to-i\e$ and changing variables yields
\bea
g_{P,C}(0,-i t|0,0)&=&\left(\frac{m}{2\pi }\right)^{1/2}\int_{C}\frac { dy_{1}...dy_{n} } { (\pi \eps)^{n/2} (t- n \eps)^{1/2} }
\nonumber \\
& \times &
\exp\left(-\frac {y_{1}^{2}} {\eps} -\sum_{k=1}^{n-1}\frac {(y_{k}-y_{k+1})^{2}} {\eps} -\frac {y_{n}^{2}} { (t - n \eps)}
\right)\\
\nonumber \\
& := & \left(\frac{m}{2\pi }\right)^{1/2} T_C ( \eps, t)
\label{4.tc}
\eea

The cases of no projection and one projection are trivially evaluated, and we easily
obtain the first two equations in Eq.(\ref{4.4E.2}).
For the case of two projections, we need to evaluate the object
\beq
T_{++}(\e_{1},\e_{2},\e_{3})=\int_{0}^{\infty}\frac{dy_{1}dy_{2}}{\pi \sqrt{\e_{1} \e_{2} \e_{3}}}
\exp\left(-\frac{y_{1}^{2}}{\e_{1}}-\frac{(y_{1}-y_{2})^{2}}{\e_{2}}-\frac{y_{2}^{2}}{\e_{3}}\right)
\eeq
(a slight generalization of the $T_C$ defined in Eq.(\ref{4.tc})).
Changing variables to $r=y_{1}$, $y=y_{2}/y_{1}$ gives,
\bea
T_{++}(\e_{1},\e_{2},\e_{3})&=&\int_{0}^{\infty}\frac{dr dy}{\pi\sqrt{\e_{1}\e_{2}\e_{3}}}r
\exp\left(-r^{2}\left\{\frac{1}{\e_{1}}+\frac{(1-y)^{2}}{\e_{2}}+\frac{y^{2}}{\e_{3}}\right\}\right)\nonumber\\
&=&\int_{0}^{\infty}\frac{dy}{2\pi}\frac{1}{(\e_{2}\e_{3}+\e_{1}\e_{3}(1-y)^{2}+\e_{1}\e_{2}y^{2})}\nonumber\\
&=&\int_{0}^{\infty}\frac{dy}{2\pi}\frac{1}{a+by+cy^{2}}
\eea
Where $a=\e_{3}(\e_{1}+\e_{2})$, $b=-2\e_{1}\e_{3}$, $c=\e_{1}(\e_{2}+\e_{3})$.
Now use the change of variables,
\beq
u=\frac{b}{2\sqrt{c}}+\sqrt{c}y
\eeq
to obtain
\beq
T_{++}(\e_{1},\e_{2},\e_{3})=\frac{1}{2\pi}\sqrt{\frac{\e_{2}}{\e_{1}\e_{3}c}}
\int_{\frac{b}{2\sqrt{c}}}^{\infty}du\frac{1}{u^{2}+\frac{4ac-b^{2}}{4c}}
\eeq
Noting that $4ac-b^{2}>0$, and using the standard integral \cite{GrRy}
\beq
\int du \frac{1}{u^2+\a^{2}}=\frac{1}{\a}{\rm ArcTan}\left(\frac{u}{\a}\right)
\eeq
we find finally
\bea
T_{++}(\e_{1},\e_{2},\e_{3})
&=&\frac{1}{4\pi\sqrt{(\e_{1}+\e_{2}+\e_{3})}}\left(\pi+2{\rm ArcTan}
\left(\sqrt{\frac{\e_{1}\e_{3}}{\e_{2}(\e_{1}+\e_{2}+\e_{3})}}\right)\right)
\label{4.ex1.7}
\eea
Setting $\e_1 = \e_2 = \eps $ and $\e_3 = t - 2 \e $ it follows that
\beq
g_P (0, t |0, 0) =
\frac {1} {4} \left( 1 + \frac {2} {\pi} {\rm ArcTan} [(t-2 \eps) /t ]^{\half}  \right)
\left( \frac {m} {2 \pi i t} \right)^{\half}
\eeq
for $ 2 \eps \le t < 3 \eps $,
so we confirm the third equation in Eq.(\ref{4.4E.2}).

It will be useful for the three projection case below to record the following related results.
A similar analysis to that above yields
\bea
T_{+-}(\e_{1},\e_{2},\e_{3})&=&\int_{0}^{\infty}dy_{1}\int_{-\infty}^{0}dy_{2}
\frac{1}{\pi \sqrt{\e_{1} \e_{2} \e_{3}}}\exp\left(-\frac{y_{1}^{2}}{\e_{1}}-
\frac{(y_{1}-y_{2})^{2}}{\e_{2}}-\frac{y_{2}^{2}}{\e_{3}}\right)\nonumber\\
&&=\frac{1}{4\pi\sqrt{(\e_{1}+\e_{2}+\e_{3})}}\left(\pi-2{\rm ArcTan}
\left(\sqrt{\frac{\e_{1}\e_{3}}{\e_{2}(\e_{1}+\e_{2}+\e_{3})}}\right)\right)
\label{4.ex1.8}
\eea
In particular we then have
\bea
T_{++}&=& T_{++}(\e,\e,\e)=\frac{1}{3\sqrt{3 \e}}
\\
T_{+-}&=&\frac{1}{6\sqrt{3\e}}
\\
T_{+0}&=&\frac{1}{2\sqrt{3\e}}= \half T_{00}
\eea

We are able to evaluate the three projection case only in the situation where all time intervals
are equal, so we set $ t - n \eps = \eps $ in the definition of $T_C$ in  Eq.(\ref{4.tc}).
In this case, $T_{C}$ possesses a number of helpful symmetries. The first is ``reflection'' symmetry:
if we define $-C$ by the string obtained by letting $(+\to -), (-\to+) ,(0\to 0)$ then
\beq
T_{-C}=T_{C}\label{4.prop1}
\eeq
The second symmetry is ``time reversal'': if we define $\tilde C$ by the string
obtained by reversing the order of $C$, then we have
\beq
T_{\tilde C}=T_{C}\label{4.prop2}
\eeq
In addition we have the simple property that
\beq
T_{...0...}=T_{...+...}+T_{...-...}\label{4.prop3}
\eeq
These properties imply the following for the three projection case.
On the one hand we have,
\beq
T_{+++}=T_{++0}-T_{++-}=T_{++0}-T_{0+-}+T_{-+-}
\eeq
but we also have,
\beq
T_{+++}=T_{+0+}-T_{+-+}=T_{+0+}-T_{-+-}
\eeq
Combining these expressions gives,
\beq
2T_{+++}=T_{+0+}+T_{++0}-T_{0+-}
\eeq
For each of the objects on the right, we may carry out the full range integral,
to leave ourselves with an object of the form $T_{C}(\e_{1},\e_{2},\e_{3})$ computed above.
For example we have that $T_{+0+}=T_{++}(\e,2\e,\e)$.
Each of these objects may then be evaluated using Eq.(\ref{4.ex1.7}) and Eq.(\ref{4.ex1.8}),
and combined to give,
\beq
T_{+++}=\frac{1}{4\sqrt{4\e}}
\eeq
It follows that
\beq
g_P (0, t |0,0 ) = \frac {1} {4} \left( \frac {m} {2 \pi i t} \right)^{\half}
\eeq
when $t= 4 \eps $, thus confirming the fourth equation in Eq.(\ref{4.4E.2}).

To end this section, we confirm our claim in Section 4.3 that the function $g_P (0,t|0,0)$
drops to half its value immediately after a projection.
On the face of it, this involves
interpreting the expression $ \langle 0 | P $ which is ambiguous, so must be defined
by a limiting procedure (where, recall $ \langle 0 |$ denotes $ \langle x |$
at $x=0$).
We thus consider the limit
$ \eps_n \rightarrow 0 $ in the expression
\bea
g_P ( 0, \tau | 0, 0 ) &=& \langle 0 | e^{-iH \epsilon_{n}} Pe^{-i H \epsilon} P
\cdots e^{-i H \epsilon}P e^{-i H \eps_0} | 0 \rangle
\nonumber \\
&=& \int dy \ \langle 0 | e^{-iH \epsilon_{n}} P e^{-i H \epsilon} | y \rangle \langle y |  P
\cdots e^{-i H \epsilon}P e^{-i H \eps_0} | 0 \rangle
\label{4.GP4}
\eea
We seek to show that the result obtained by this limit
is half the result obtained when the final projection is absent.
We write
\beq
P = \half ( P + \bar P)  + \half ( P - \bar P)
\eeq
so that
\beq
\langle 0 | e^{-iH \epsilon_{n}} P e^{-i H \epsilon} | y \rangle
= \half \langle 0 | e^{-iH \epsilon_{n}} e^{-i H \epsilon} | y \rangle +
\half \langle 0 | e^{-iH \epsilon_{n}} (P - \bar P)  e^{-i H \epsilon} | y \rangle
\eeq
The first term on the right-hand side yields the sought after factor of a half as $\eps_n \rightarrow 0 $,
so it remains to show that the second term is zero. We have
\bea
\langle 0 | e^{-iH \epsilon_{n}} (P - \bar P)  e^{-i H \epsilon} | y \rangle
&=& \int_0^{\infty} dx \ \langle 0 | e^{-iH \epsilon_{n}} | x \rangle
\left( \langle x | e^{ - i H \eps } | y \rangle -  \langle - x | e^{ - i H \eps } | y \rangle \right)
\label{4.4E.23}
\eea
where we have used the fact that
\beq
\langle 0 | e^{-iH \epsilon_{n}} | x \rangle = \langle 0 | e^{-iH \epsilon_{n}} | -x \rangle
\eeq
(for the free particle, considered here).
As $\eps_n \rightarrow 0 $, $ \langle 0 | e^{-iH \epsilon_{n}} | x \rangle \rightarrow \delta (x)$
and the pair of terms in brackets on the right-hand side cancel in Eq.(\ref{4.4E.23}),
so we obtain zero as required.

\section{Asymptotic Limit}

We now consider the asymptotic evaluation of $g_P (0, \tau | 0, 0 )$ for large $n$
in the case $\eps_n = \eps = \eps_0$. This will give the values of the peaks
of $f_P(t)$, Eq.(\ref{4.peaks}).

Recall that in the limit $ n \rightarrow \infty $ and $ \eps \rightarrow 0 $
with $ \tau = (n+1) \eps $ fixed, $g_P (x_1, \tau | x_0, 0 )$
tends to the restricted propagator $g_r (x_1, \tau | x_0, 0) $.
To obtain the asymptotic form of $g_P (0, \tau | 0, 0 )$ for large $n$ we therefore
need to determine the lowest non-trivial correction to this result. It is clear
from the definition Eq.(\ref{4.2.1}) of $g_P$ that, close to the limit,
we have the general form
\beq
g_P (x_1, \tau | x_0, 0 ) =  g_r (x_1, \tau | x_0, 0 ) + \eps g_1 (x_1, \tau | x_0, 0 )
+ O (\eps^2)
\label{4.5.1}
\eeq
for some function $g_1 (x_1, \tau | x_0, 0)$.
Since the restricted propagator vanishes if either $x_1 = 0 $ or $ x_0 = 0 $, the object
we need to calculate is
\beq
g_1 (0,\tau|0,0) = \lim_{\eps \rightarrow 0, n \rightarrow \infty} \frac { g_P (0, \tau | 0, 0 ) } {\eps}
\label{4.limit}
\eeq

A standard and convenient way to do this is to rotate to imaginary time $\tilde \tau $
(we use tilde to denote Euclideanized time)
and then write $g_P (x_1, \tilde \tau | x_0, 0 )$  as a
Euclidean path integral. This is
then defined in terms of the continuum limit of probabilities of random walks on a space time lattice
of temporal spacing $ \Delta \tilde \tau $ and spatial spacing $\eta $.
The details of this construction are very conveniently
given by Hartle \cite{Har}
so we will give only the briefest account here.
Using this language,
$g_P (0, \tilde \tau | 0, 0 )$ is then the continuum limit of the object
$ (2 \eta)^{-1} u_P  ( 0, \tilde \tau | 0, 0) $, where $ u_P  ( 0, \tilde \tau | 0, 0)  $ is
the probability for a random walk on the lattice starting at the origin and
ending at
the origin at time $ \tilde \tau$, with the restriction that the walker is
in the positive $x$-axis at the intermediate times $ \tilde \eps, 2 \tilde \eps, 3 \tilde \eps
\cdots$ (where $ \tilde \eps \ge \Delta \tilde \tau $).

The calculation of $u_P$ defined in this way, for values of $\tilde \eps $ generally
greater than the lattice spacing $\Delta \tilde \tau$,
is in fact a known problem in combinatorics called the tennis ball problem. It
appears to have  a formal solution, but this solution is too implicit for us to
extract a useful result \cite{tennis}.

Fortunately, however, for the purposes of calculating
the limit Eq.(\ref{4.limit}) the results of Ref.\cite{Har}
are sufficient.
For this case, we set
$\tilde \eps = \Delta \tilde \tau$
and $ u_P( 0, \tilde \tau | 0, 0) $ is then
the probability for a random walk from the origin to itself, with the restriction
that the walker is in the positive $x$-axis at {\it every} intermediate step.
On a finite lattice Hartle's calculations give the result,
\beq
\frac {1} {2 \eta} u_P( 0, \tilde \tau | 0, 0) =
\left( \frac{ m } { 2 \pi \tilde \tau} \right)^{1/2} \ \frac { \tilde \eps} {\tilde \tau}
\eeq
to leading order for small $\tilde \eps$, $\eta$ \cite{Har}. We may use this result
to compute the limit Eq.(\ref{4.limit}), which, continued back to real time, is
\beq
g_1 (0, \tau | 0, 0 ) = \left(\frac{m}{2\pi i
\tau}\right)^{1/2} \frac { 1} {\tau}
\eeq
Through Eq.(\ref{4.5.1}), this confirms Eq.(\ref{4.peaks})
for large $k$.

\section{A Time Averaged Result}

We have argued that at the peak values of $ g_P (0,\tau|0,0) $ with $n$
projections, we have the approximate result
\beq
\langle 0 | e^{-iH \tau } P (n \eps) \cdots P (2 \eps ) P ( \eps )
| 0 \rangle
\approx \left(\frac{m}{2\pi i \tau}\right)^{1/2} \frac {1} {n+1}
\label{4.TA}
\eeq
We showed above that this is exact for $n=1,2,3$ and true asymptotically
for large $n$.
Some interesting exact
results for any $n$ may be obtained by considering a more general version of this object
in which the projections are not restricted to act at the given set of
times, $ \eps, 2\eps, 3 \eps \cdots$.

On the one hand, Eq.(\ref{4.CP}) may be expanded as a power series in powers of
$V_0$:
\beq
g_{V}(0,\tau|0,0)=\left(\frac{m}{2\pi i
\tau}\right)^{1/2}\sum_{n=0}^{\infty} \frac{ (-1)^n }{ (n+1)! } V_0^n \tau^n
\label{4.T1}
\eeq
On the other hand, the evolution operator with complex potential may be expanded
in the usual S-matrix expansion,
\bea
e^{-i(H-iV)\tau} &=& e^{-iH\t} \ T \exp \left( -\int_{0}^{\t}dt V(t)\right)
\nonumber \\
&=& e^{-iH \tau}\sum_{n=0}^{\infty}\frac{(-1)^{n}}{n!}\int_{0}^{\tau}dt_{n}...\int_{0}^{\tau}dt_{1}
T [V(t_{n})...V(t_{1})]
\label{4.T2}
\eea
where $T$ denotes time ordering,  $V (t) = V_0 \bar P (t) $ and $\bar P(t) = e^{ i H t } \bar P e^{ - i H t} $.
It follows that
\bea
g_{V}(0,\tau|0,0) &=& \langle 0 | \exp \left(-i(H-iV)\tau \right) | 0 \rangle
\nonumber \\
&=& \sum_{n=0}^{\infty}\frac{(-1)^{n}}{n!} V_0^n \int_{0}^{\tau}dt_{n}...\int_{0}^{\tau}dt_{1}
\ \langle 0 | e^{-iH\t} T [\bar P(t_{n})...\bar P(t_{1})] | 0 \rangle
\label{4.T3}
\eea
Equating powers of $V_0$ in Eqs.(\ref{4.T1}) and (\ref{4.T3}) and writing out the time ordering
explicitly, we deduce that
\beq
\frac{n!} {\tau^n}  \int_{0}^{\tau}dt_{n} \int_0^{t_n} dt_{n-1}...\int_{0}^{t_2}dt_{1}
\ \langle 0 | e^{-iH\t}  \bar P(t_{n})... \bar P(t_{1}) | 0 \rangle
= \left(\frac{m}{2\pi i
\tau}\right)^{1/2} \frac {1} {n+1}
\label{4.T4}
\eeq
We would like to write this in a form involving $P$, instead of $\bar P$. We introduce
the reflection operator
\beq
R = \int dx \ | x \rangle \langle -x |
\eeq
and note that $ \bar P = R P R $, the Hamiltonian $H$ commutes with $R$, and $ R | 0 \rangle = | 0 \rangle $
(where, recall, $ | 0 \rangle $ denotes $ | x \rangle $ at $x=0$). It follows that
\beq
\langle 0 | e^{-iH\t}  \bar P(t_{n})... \bar P(t_{1}) | 0 \rangle
= \langle 0 | e^{-iH\t}  P(t_{n})... P(t_{1}) | 0 \rangle
\eeq
so that Eq.(\ref{4.T4}) holds with all the $\bar P$'s replaced with $P$'s.
Noting that
\beq
\frac{n!} {\tau^n}  \int_{0}^{\tau}dt_{n} \int_0^{t_n} dt_{n-1}...\int_{0}^{t_2}dt_{1} = 1
\eeq
we see that Eq.(\ref{4.T4}) is of the desired general form, Eq.(\ref{4.TA}), but time-averaged
over the times of the projections.

The question is now to what extent the time-averaged expression on the left-hand
side
of Eq.(\ref{4.T4}) is close to Eq.(\ref{4.TA}). We may take this further in two different
ways.

First, note that for a real-valued function $f$ of $n$ variables, we have
the mean value theorem
\beq
\frac{n!} {\tau^n}  \int_{0}^{\tau}dt_{n} \int_0^{t_n} dt_{n-1}...\int_{0}^{t_2}dt_{1}
f(t_n, t_{n-1} \cdots t_1) =  f( \xi_n, \xi_{n-1} \cdots \xi_1 )
\label{4.T5}
\eeq
for some set of numbers $\xi_n \ge \xi_{n-1} \ge \cdots \ge \xi_1 $ in the interval
$[0,\tau]$. The integral in Eq.(\ref{4.T4}) is easily made into a real integral
over a real-valued function by analytic continuation, and
we therefore deduce the exact result
\beq
\langle 0 | e^{-iH\t}  P(\xi_{n})...P(\xi_{1}) | 0 \rangle
= \left(\frac{m}{2\pi i
\tau}\right)^{1/2} \frac {1} {n+1}
\eeq
for {\it some} set of projection times $\xi_n \ge \xi_{n-1} \ge \cdots \ge \xi_1 $
in the interval $[0, \tau]$.

Second, we may expand the integrand in the left-hand side of Eq.(\ref{4.T4}) about
the values $ k \eps $, to get some insight into why these particular values
have any special significance. We have
\bea
\langle 0 | e^{-iH\t}  P(t_{n})...P(t_{1}) | 0 \rangle
&=& \langle 0 | e^{-iH\t}  P(n \eps)...P(\eps ) | 0 \rangle
\nonumber \\
&+&
\sum_{k=1}^n \ (t_k - k \eps ) \  \frac {\partial} { \partial t_k} \langle 0 | e^{-iH\t}
P(t_{n})...P(t_{1}) | 0 \rangle \big|_{t_k = k \eps}
\nonumber \\
&+& \cdots
\label{4.5.8}
\eea
Inserting in Eq.(\ref{4.T4}), and noting that
\beq
\frac{n!} {\tau^n}  \int_{0}^{\tau}dt_{n} \int_0^{t_n} dt_{n-1}...\int_{0}^{t_2}dt_{1} \ t_k = k \eps
\eeq
(where, recall, $ \tau = (n+1) \eps $)
we see the first order term in the expansion Eq.(\ref{4.5.8}) averages to zero in
Eq.(\ref{4.T4}). This is clearly only
true for expansion about the special values $t_k = k \eps $. We therefore obtain
the desired result Eq.(\ref{4.TA}) up to second order corrections. This suggests that
the values $t_k = k \eps $ are significant because they
give the best approximation to the average in Eq.(\ref{4.T4}).
These results give evidence that Eq.(\ref{4.peaks}) holds approximately
for all $n$, including the intermediate values not covered in the previous
two sections.


\section{Numerical Results}

To support the analytic results described in the previous sections
we evaluate Eq.(\ref{4.GP}) numerically for up to $20$ projections and confirm
the conjectured form Eq.(\ref{4.FP1}), (\ref{4.FP2}).
It is convenient to define a sequence of functions
$F_n (t,x)$ defined by
\beq
F_0 (t,x) = \langle x | e^{ - i H t } | 0 \rangle
\eeq
for $ 0 \le t < \eps $ and
\beq
F_n (t,x) = \langle x | e^{ - i H ( t - n \eps ) } \left( P e^{ - i H \eps} \right)^n | 0 \rangle
\eeq
for $ n \eps \le t \le (n+1) \eps $ where $n=1,2,3 \cdots $. The desired object Eq.(\ref{4.GP})
(with, for convenience, $\eps_0 = \eps$), is then given by
\beq
g_P (0,t|0,0) = F_n (t,0)
\eeq
The sequence $F_n (t,x)$ may be calculated using the recursion relation,
\beq
F_n (t,x) = \int_0^\infty dy \ \langle x | e^{ - i H (t - n \eps ) } | y \rangle \ F_{n-1} (n \eps, y)
\eeq
Using a new time coordinate $s$ defined by $ t = s \eps $, rotating to imaginary time,
and defining $\tilde F_n ( s, x) = F_n ( - i t, x ) $,
we have
\beq
\tilde F_n (s, x ) = \left( \frac{m} { 2 \pi ( s - n ) } \right)^{\half}
\int_0^\infty dy \ \exp \left( - \frac { m(x-y)^2 } { 2(s-n) } \right)\ \tilde F_{n-1} (n, y)
\label{4.7.5}
\eeq
for $ n \le s \le n+1 $.

We have attempted to find an approximate analytic solution to Eq.(\ref{4.7.5}) for large $n$,
but without success. However, a numerical solution is straightforward and yields
all the information we require.
This was done using a simple mid-point rule. The lattice
size was chosen to be of order $10^{-3}$, and the results were checked
for robustness against changes in lattice size.

The numerical result for $f_P (t) $ (defined in Eq.(\ref{4.FP1}) in terms of $g_P (0,t|0,0)$)
is plotted in Figure \ref{fig:cpfigure3}, along with our claimed approximate analytic expression
for $f_P (t)$, Eq.(\ref{4.FP2}). We see that there is excellent agreement.
The values at the peaks and troughs appear to agree perfectly.
The only small
discrepancy is that the interpolating functions between the peaks and troughs
are not exactly linear. This discrepancy is only noticeable for intermediate
values of $n$ and in any event since, as argued, the curve is effectively
averaged over time in the PDX, this discrepancy is insignificant. We therefore
find substantial numerical confirmation for our our main result, Eqs.(\ref{4.FP1}), (\ref{4.FP2}).
\begin{figure}[htp] 
   \centering
   \includegraphics[width=15cm]{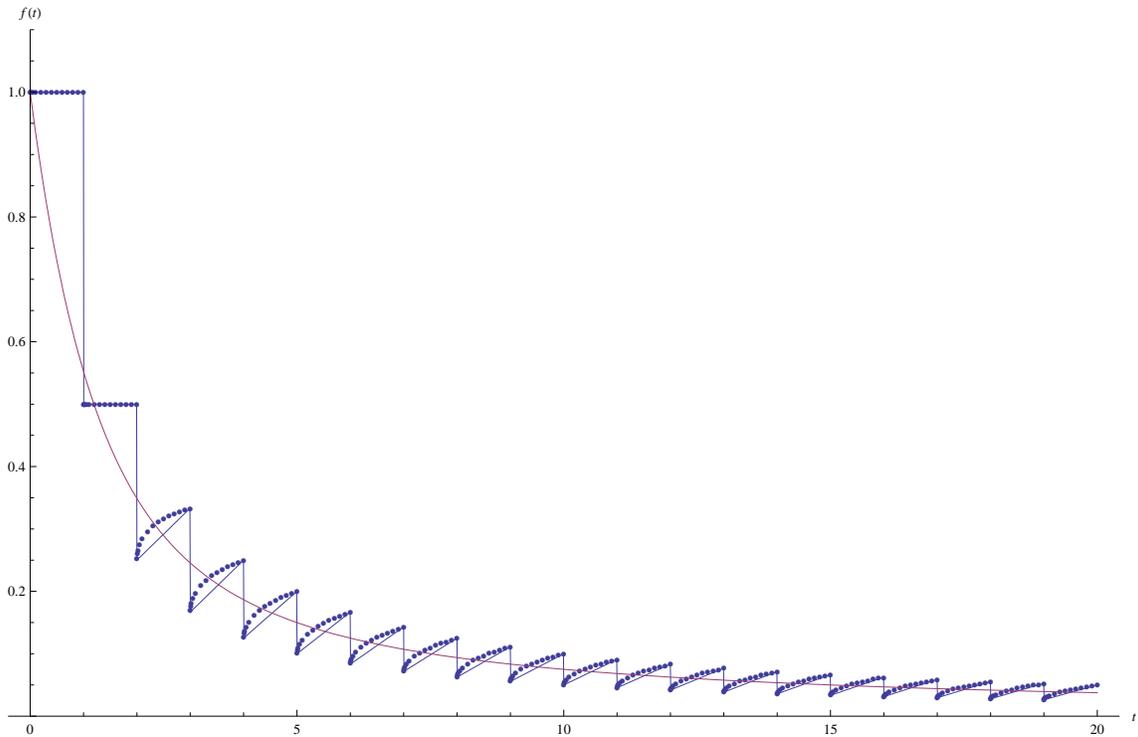} 
   \caption[Plot of the Exact Value of $f_{P}(t)$ Compared With an Analytic Approximation.]{ {\em A plot of the numerical calculation of $f_P (t)$
(the set of dotted lines),  and the conjectured analytic expression for
$f_P(t)$, Eqs.(\ref{4.FP1}), (\ref{4.FP2}) (the set of straight lines).
(Again we use a time scaling such that $\eps = 1$).
There is excellent agreement between analytic and numerical results at the
peaks and troughs of the functions. The numerical result
shows some departure from exact linear behaviour between the peaks and troughs
but this is insignificant as argued in the main text.
A plot of $f_V (t)$, the smooth curve, is also given for reference.}}
   \label{fig:cpfigure3}
\end{figure}

\begin{figure}[htp] 
   \centering
   \includegraphics[width=15cm]{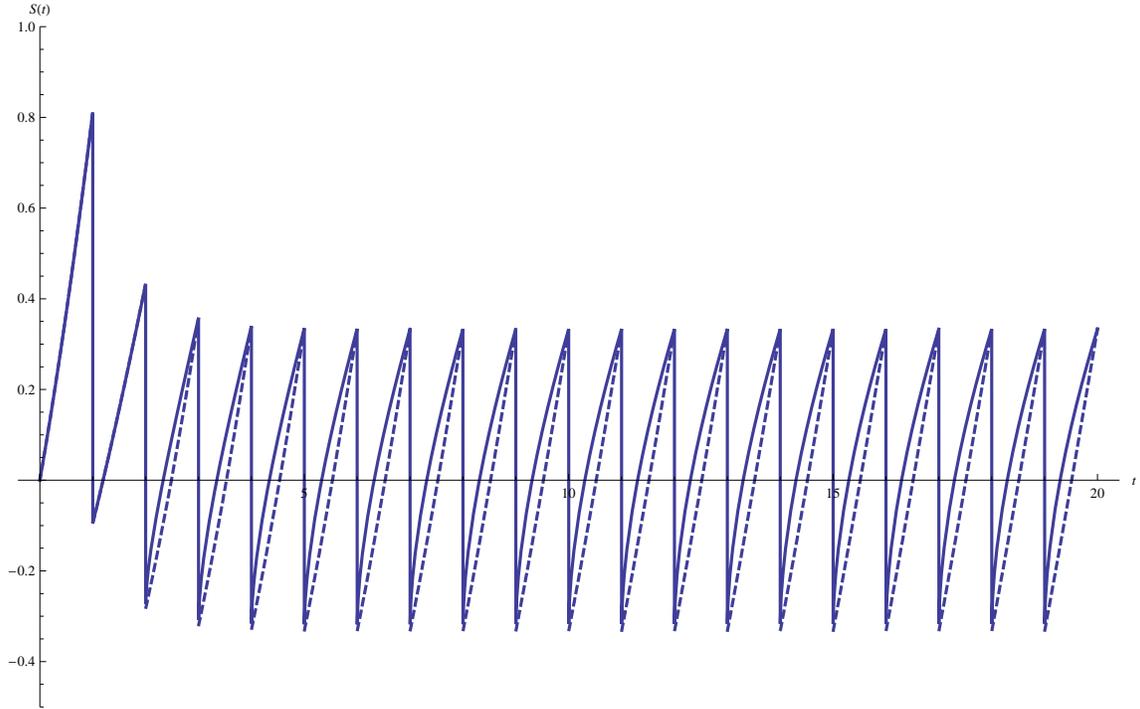} 
   \caption[A Plot of the Function $S(t)$.]{ {\em A plot of the function $S(t)$ defined in Eq.(\ref{4.7.6}),
describing the oscillations of $g_P (0,t|0,0)$ around $g_V (0,t|0,0)$. The bold line
represents the numerical calculation and the dashed line the analytic result
(from Eqs.(\ref{4.CP}), (\ref{4.FP2})). After the first few oscillations, it
oscillates between $ \pm 1/3$ (approximately).}}
   \label{fig:cpfigure4}
\end{figure}

The apparently perfect agreement of numerical results with the approximate analytic expression
Eq.(\ref{4.FP2}) at the peaks and troughs is striking. We wonder if the approximate
analytic expression is in fact exact at these points, but we have not been able
to prove this, except for the cases $n=1,2,3$ and for large $n$.

A useful way of seeing even more precisely the relationship between $f_P
(t)$ and $ f_V (t)$ is to define the function
\beq
S(t) = \frac { f_P (t) } { f_V (t) } - 1
\label{4.7.6}
\eeq
so that
\beq
f_P (t) = ( 1 + S(t) ) f_V (t)
\eeq
This is plotted in Figure \ref{fig:cpfigure4}.
It is a simple function oscillating around
zero between $ \pm 1/3 $ with period $\eps$. In terms of $S(t)$, the relationship between $g_P$ and $g_V$ then has the particularly
simple form
\beq
g_P (0,t|0,0) = ( 1 + S(t) ) g_V (0,t|0,0)
\label{4.7.7}
\eeq
This relationship is perhaps the most concise summary of the sought-after connection between the propagators.

\section{Timescales}

We now give a more detailed explanation as to the timescales
involved in proving the approximate equivalence of $g_V$ and $g_P$.
We broadly expect that the appropriate timescale is the
Zeno time of the initial state, $ t_Z = 1 / \Delta H$. However, we have derived a very precise
connection between evolution in the presence of a complex potential
and evolution with projection operators so we are in a position
to investigate the specific way in which an initial state may discriminate
between these two types of evolution.

We have argued that the equivalence boils down to proving
the equivalence of propagation along the boundary, Eq.(\ref{4.2.11}),
and we have shown that $ g_P (0,t|0,0) $ oscillates around $ g_V (0,t|0,0)$
with period $\eps$. Suppose we have an initial state $\psi (x,0)$.
Let us consider the change in the wave function
arising from replacing $ g_V $ with $g_P$
using
the PDX, Eq.(\ref{4.2.9}). It
is
\beq
\delta \psi ( x_1, \tau) = \frac {1  } {m^2} \int_{0}^{\tau} dt_2
\int_0^{t_2} dt_1
\ \frac {\partial g_f} {\partial x} (x_1, \tau | x, t_2) \big|_{x=0} \ \delta g (0,t_2| 0,t_1)
\ \frac {\partial \psi} { \partial x } (0,t_1)
\label{4.8.1}
\eeq
where
\beq
\delta  g (0,t_2| 0,t_1) =  g_V (0,t_2| 0,t_1) -  g_P (0,t_2| 0,t_1).
\eeq
The important part of this expression is the $t_1$ integral, which we
expect will be small if the initial state is sufficiently slowly varying
in time.

The results
of Section 4.3 and 4.8 show that $ \delta g (0,t|0,0)$ oscillates around zero
with period $\eps$. In particular, Eq.(\ref{4.7.7}) shows that
\beq
\delta g (0,t|0,0) = S(t) g_V (0,t|0,0)
\eeq
The explicit form of $S(t)$ is given in Figure \ref{fig:cpfigure4}, but its important
qualitative feature is its oscillation with period $\eps$, so
for simplicity, $S(t)$ may
be loosely modelled by the function $ \sin (2 \pi t/\eps)$.
To be definite
we take the initial state $\psi$ to be a Gaussian wave packet, so we have
\beq
\psi (x,t) = N \exp \left( - \frac { ( x - q - pt/m)^2 } {4 \sigma^2} + i p x - i E t \right)
\eeq
where $E = p^2 / 2m$, $N$ is a normalization factor, and we have ignored wave packet spreading
effects in evolving the state. For such a state the Zeno time is $t_Z = m \sigma / p $.
In Eq.(\ref{4.8.1}) the differentiation of $ \psi $ produces a prefactor which does not contribute
to the leading order evaluation of the time integral, and if the range of integration is much
greater than $\eps$, we obtain the order of magnitude result
\beq
\int dt_1 \ \delta g (0,t_2| 0,t_1)
\ \frac {\partial \psi} { \partial x } (0,t_1) \sim \exp \left( - \frac {t_Z^2} {\eps^2} (E \eps - 1)^2 \right)
\label{4.8.4}
\eeq
If $ E \eps \ll 1 $, the right-hand side is clearly very small if $ \eps \ll t_Z $.
If $ E \eps > 2 $, the right-hand side if bounded from above by $ \exp ( -  {t_Z^2}/ {\eps^2} ) $
so again will be small if $ \eps \ll t_Z $. Hence the Zeno time of the initial state
is indeed the timescale controlling the validity of the approximation, as expected.
Note also that Eq.(\ref{4.8.4}) goes to zero as $\eps \rightarrow 0$, as it must, since
$g_P$ and $g_V$ become exactly equal (and equal to $g_r$) in this limit.

The only problematic case is that in which the initial state has energy $ E \sim 1 / \eps$. In this
case, the right-hand side of  Eq.(\ref{4.8.4}) is not necessarily small and the approximation
may fail. This is not surprising since it is the case in which the oscillations in time of the
incoming state are comparable to the temporal spacing of the projections, so that the state
can ``see'' the difference between the complex potential and projections at a discrete
set of times.

Hence, apart from the above exception, Eq.(\ref{4.2.11}) holds for $ \eps \ll t_Z$.
As outlined in the Introduction to this chapter, reflection from the complex potential
is negligible if $V_0 \ll E $ which is equivalent to  $ \eps \gg 1/E $.
Therefore there is an interesting regime, namely
\beq
\frac {1}  {E} \ll \eps \ll \frac {1} {\Delta H}
\eeq
in which the approximate
equivalence Eq.(\ref{4.2.11}) holds, yet there is negligible reflection. This
regime is important in, for example, study of the arrival time problem using
complex potentials \cite{HaYe1}, and we will make use of it in Chapter 5 of this thesis.
(See also Ref.\cite{Sch} for an interesting
discussion of timescales in the Zeno effect.)

\section{Summary and Discussion}

This chapter was physically motivated by the desire
to understand the effect of periodically acting projections
onto the positive $x$-axis for a free particle, Eq.(\ref{4.1.1}).
A valuable way to proceed is to use the conjectured
relationship Eq.(\ref{4.1.2}) with a complex potential first put forward by Echanobe et al.
This connection, together with known results on scattering,
establishes the timescale under which significant reflection occurs
in Eq.(\ref{4.1.1}).
We noted that the arguments for the relationship Eq.(\ref{4.1.2}) are only heuristic and
there is scope for a more substantial proof.

We proved Eq.(\ref{4.1.2}) by considering the associated propagators Eqs.(\ref{4.2.2}), (\ref{4.2.1}).
We noted that an approximate equivalence between these propagators
is expected since both propagators tend to the restricted
propagator $g_r (x_1, \tau | x_0,0)$
in the limits
$ \eps \rightarrow 0$, $ n \rightarrow \infty $ and $ V_0 \rightarrow \infty$.
The path decomposition expansion reduced the proof of equivalence
of these propagators to the simpler case of propagation between points
lying on the origin, Eq.(\ref{4.2.11}). The propagator along
the origin for the complex potential $g_V (0,t|0,0)$ is already known, Eq.(\ref{4.CP}), so the bulk of the proof
was to derive the analogous result for the propagator with
projections, $g_P( 0 , \tau | 0, 0 )$. Our main result was to prove
that this propagator
has the approximate form Eqs.(\ref{4.FP1}), (\ref{4.FP2}), which we proved
using a variety of analytic and numerical methods. In effect, the main achievement
of this chapter has therefore been to obtain a good approximate analytic expression
for the propagator $g_P$ that appears in Eq.(\ref{4.1.1}).

We found that $g_P(0,t|0,0)$ oscillates with period $\eps$
around $g_V (0,t|0,0)$ as
long as $V_0 \eps \approx 4/3 $, a result most concisely summarized in Eq.(\ref{4.7.7}).
The approximate equivalence Eq.(\ref{4.2.11})
of these propagators then holds in a time-averaged sense as long as the
timescale $\eps$ between projections is much smaller than the Zeno
time of the initial state, $ 1 / \Delta H $ (but may fail in the special
case when the incoming state is peaked about energy $ E \sim 1/ \eps$).
These conditions agree in essence with those of an extended version
of the results of Echanobe et al, with an advantage over their results
that a definite relationship between $V_0$ and $\eps$ is obtained.
We noted that their restriction Eq.(\ref{4.1.3}) relating
$ V_0$ and $\eps$ is in fact stronger than required and the equality
$ V_0 \eps \approx 4/3 $ derived here gives the best approximate
equivalence between $g_P $ and $g_V$.


Let us make some comments here on the possible generality of the connection Eq.(\ref{4.1.2}).
We first note that it may in fact be written
\beq
e^{ - i H \eps} Pe^{-i H \epsilon} \cdots P e^{-i H \eps}
\approx \exp \left( - i H \tau - V_0 \bar P \tau \right)
\label{4.9.1}
\eeq
where $\bar P = 1 - P$. We have proved Eq.(\ref{4.9.1}) for the case
in which the projections are onto the positive $x$-axis, but it seems
reasonably clear that the relationship will hold for projections onto
any region $[a,b]$ of the $x$-axis (as long as it is not too small) and indeed
for regions of configuration space in a many-dimensional model.
Such a potential has been used recently in the decoherent histories analysis
of quantum cosmological models \cite{jjhqc}.
Moreover, although the proof of Eq.(\ref{4.9.1}) given in this chapter
relied heavily on the fact that $P$ projects onto position,
the form Eq.(\ref{4.9.1}) and the heuristic
argument for it in Section 4.2 do not rely on the particular form of $P$.
We therefore conjecture that Eq.(\ref{4.9.1}) may hold for a wider
variety of projection operators, not just projectors onto position.
This will be pursued elsewhere.

Finally we note that the left hand side of Eq.(\ref{4.9.1}) is a class operator in the decoherent histories (DH) approach to quantum theory outlined in Chapter 1. The equivalence expressed in Eq.(\ref{4.9.1}) means we can also use the right hand side as a class operator. The practical value of this is that it opens the way to dealing with time observables in DH by replacing strings of projectors, which may be hard to work with, with complex potentials which are much simpler. However on a more fundamental level, it also suggest new ways to construct class operators which would not have been obvious starting from the exposition in Chapter 1. We will pursue this in the next chapter.


\chapter{Arrival Times, Complex Potentials and Decoherent Histories}\label{atp}
\epigraph{Tomorrow, and tomorrow, and tomorrow,
Creeps in this petty pace from day to day}
{Shakespeare}

\section{Introduction}

We saw in the Introduction, Chapter 1, that there are many different approaches to the arrival time problem in quantum theory, (see Fig.(\ref{fig:singlewave_fig1}) for a recap of conventions).
\begin{figure}[htbp] 
   \centering
   \includegraphics[width=10cm]{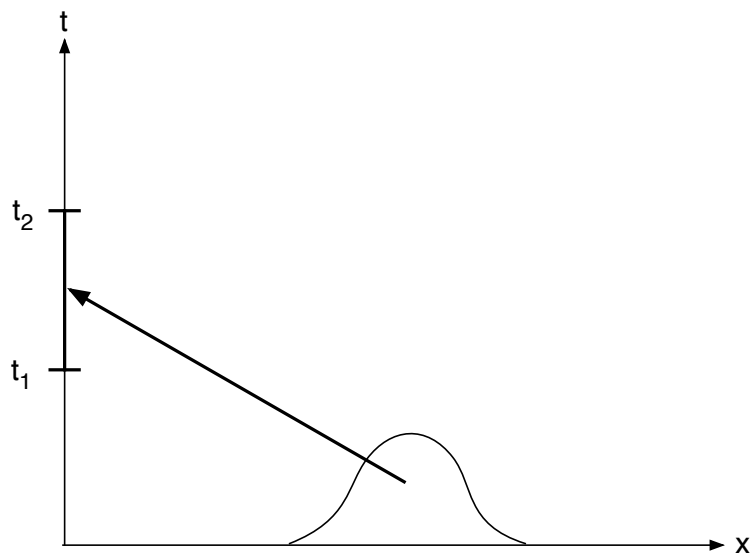} 
   \caption[Recap of the Arrival Time Problem.]{{\em The quantum arrival time problem. We prepare an initial state
localized entirely in $x>0$ and consisting entirely of negative momenta. What is
the probability that the particle crosses the origin during the time interval $[t_1,t_2]$?}}
   \label{fig:singlewave_fig1}
\end{figure} 
The approach we wish to pursue in the next two chapters is based on the decoherent histories approach to quantum theory, as outlined in Sections \ref{sec:1.4} and \ref{sec:1.5}. We saw in Section \ref{sec:1.5} that the key objects for this analysis are the class operators for crossing and not crossing the origin during a given time interval. The class operator for not crossing was determined to be,
\beq
C_{nc}(\t)=Pe^{-iH\e}P...e^{-iH\e}P\label{5.1}
\eeq
where $P=\theta(\hat x)$, there are $N+1$ projections and $\t=N\e$. In order to avoid the Zeno effect we need to leave $\e$ finite. However in Section \ref{sec:1.5} we concluded that this object was somewhat problematic, firstly because we could not determine the required timescale $\e$, and secondly because this object is difficult to work with analytically. 

However, armed now with the results of Chapter 4 we are able to make progress. The results of Chapter 4 tell us that evolution with the class operator above,  Eq.(\ref{5.1}), is equivalent to evolution in the presence of a complex potential,
\beq
C_{nc}(\t)=Pe^{-iH\e}P...e^{-iH\e}P\approx e^{-iH\t-V_{0}\bar P\t}\label{5.2},
\eeq
where $\bar P=\theta(-\hat x)$, $\t=N\e$ and $V_{0}\sim1/\e$. Eq.(\ref{5.2}) holds provided
\beq
\e<<\frac{1}{\Delta H}.
\eeq
This means we can compute the probabilities and the decoherence functional by replacing class operators of the form Eq.(\ref{5.1}) with evolution under the complex potential, and this simplifies the calculations enormously. The equivalence in Eq.(\ref{5.2}) also tells us that we must have $E\e>>1$, where $E$ is the energy of the incoming state, if we are to avoid reflection.

Of course the results of Chapter 4 as they stand only give us the class operator for not crossing during a given time interval. We also require the class operator for crossing. We shall compute this in this chapter, as well as calculating the probabilities and decoherence functional for questions such as, ``Did the particle cross during the time interval $[t_{1},t_{2}]$?'' Our end result will be a complete analysis of the arrival time problem in the language of decoherent histories. Before we do this, however, it is useful to explore some aspects of the arrival time distribution defined using complex potentials. This will help us to understand the key features of our final result.

The first aim of this chapter is to look in some detail at the calculation and properties of the arrival time distribution defined using a complex potential. In particular, we will use path integral methods, which in some ways are more concise and transparent than previous derivations. 
The second aim of this chapter is to carry out a decoherent histories analysis of the
arrival time problem, using the equivalence between strings of projection operators and complex potentials demonstrated in Chapter \ref{cppo}. 

Our main result is that for intervals of size $ \Delta \gg 1 / V_0 $, these class
operators are given approximately by
\bea
C_c^k &=&  e^{ - i H_0 \tau} \int_{t_k}^{t_{k+1}} dt \ \ \frac {(-1)} {2m}
\ \left( \hat p \ \delta (\hat x_t)  + \delta (\hat x_t ) \ \hat p \right)
\nonumber \\
&=&  e^{ - i H_0 \tau} \left( \theta ( \hat x (t_k) ) - \theta ( \hat x (t_{k+1} ) ) \right)
\eea
Significantly, the dependence on the complex potential has dropped out entirely. We will
show that there is decoherence for an interesting class of states, and, for such states
the probabilities
are then given by 
\beq
q(t_k, t_{k+1} ) = \int_{t_k}^{t_{k+1}} dt \ J(t)
\label{5.9}
\eeq
This, we will show, coincides with the expected result Eq.(\ref{1.7}), when $\Pi (t)$ is
integrated over a range of time much greater than $ 1/V_0$. Hence there is complete
agreement with standard results on the arrival time distribution at sufficiently
coarse-grained timescales. Furthermore, our results
shed some light on the problem of backflow -- we find that the situations when
Eq.(\ref{5.9}) is negative are those in which there is no decoherence in which case
probabilities cannot be assigned.

In the next Chapter, we compute the crossing class operators in a simpler but more heuristic
way, by exploring a semiclassical approximation to Eq.(\ref{5.1}) keeping $\epsilon$ finite. The results are essentially the same.

The rest of this Chapter is arranged as follows: In Section \ref{sec:5.2} we consider the classical arrival time defined via a complex
potential. This well help us understand how the probabilities we obtain depend on the choice of $V_{0}$, or equivalently $\e$, in Eq.(\ref{5.2}). In Section \ref{sec:5.3} we then repeat this analysis for the quantum case,  using the PDX to calculate the arrival time distribution function. We make heavy use in this section of the results of Chapter 2.
In Sections \ref{sec:5.4} and \ref{sec:5.5}, we use the results of the previous sections to carry out the decoherent histories analysis. We summarize and conclude in Section \ref{sec:5.6}. This chapter is based on Ref.\cite{HaYe1}.

\section{The Classical Arrival Time Problem via a Complex Potential}\label{sec:5.2}

Before looking at the quantum arrival time problem it is
enlightening to look at the corresponding classical arrival
time problem defined using an absorbing potential. This gives
some understanding of the expected form of the result
in the quantum case. In particular since the complex potential is a simple measurement model we expect the probabilities obtained in this way to be of the form Eq.(\ref{1.8.best}), which include an ``apparatus response function.'' We will see in this section that exactly the same thing happens in the classical case, and in fact the response functions are the same in the quantum and classical cases. This means the complex potential is acting like a classical measuring device, and all dependence on it can be removed by coarse-graining, in the manner of Eq.(\ref{1.8.cg}).

We consider a classical phase space distribution $w_t (p,q)$, with initial
value $w_0 (p,q)$ concentrated entirely in $q>0$ with only negative momenta. This distribution evolves in the presence of a complex potential $V(q)=-iV_{0}\theta(-q)$.
The equation of motion is,
\beq
\frac{ \partial w} { \partial t} = - \frac {p} {m} \frac {\partial w} { \partial
q} - 2 V (q) w.
\label{5.2.1}
\eeq
This form may be deduced, for example, by computing the evolution equation of the
Wigner function with this complex potential and dropping the higher order terms (involving
powers of $\hbar$). Eq.(\ref{5.2.1}) is readily solved and has solution
\beq
w_\tau (p,q) = \exp \left( - 2 \int_0^\tau ds \ V( q - p s / m ) \right)
\ w_0 (p, q - p \tau / m )
\label{5.2.2}
\eeq

The survival probability, that is the probability of not having been absorbed by the complex potential by a time $\t$, is
\beq
N( \tau ) = \int_{-\infty}^{\infty} dp \int_{-\infty}^{\infty} dq \ w_\tau (p,q).
\eeq
The arrival time distribution is therefore
\beq
\Pi (\tau ) = - \frac { d N} {d \tau}
= 2 V_0 \int_{-\infty}^{\infty} dp \int_{-\infty}^0 dq \ w_\tau (p,q)
\eeq
where we have made use of Eq.(\ref{5.2.1}). This expression is conveniently
rewritten by noting that, again using Eq.(\ref{5.2.1}), $ \Pi (\tau) $ obeys
the equation
\beq
\frac { d \Pi} { d \tau} + 2 V_0 \Pi =
- 2 V_0 \int_{-\infty}^{\infty} dp \ \frac {p} {m} w_\tau (p,0)
\eeq
This may be solved to yield
\beq
\Pi (\tau) = - 2 V_0 \int_0^\tau dt \ e^{ - 2V_0 (\tau - t) }
\ \int_{-\infty}^{\infty} dp \ \frac {p} {m} w_t (p,0)
\eeq
From Eq.(\ref{5.2.2}), we see that
\beq
w_t (p,0) = \exp \left( - 2 V_0 \int_0^t ds \ \theta ( p s / m ) \right)
\ w_0 (p,  - p t / m )
\eeq
but since the momenta are all negative the exponential factor makes
no contribution. We thus obtain
\beq
\Pi (\tau) =  2 V_0 \int_0^\tau dt \ e^{ - 2V_0 (\tau - t) } \ J(t)
\label{5.2.8}
\eeq
where
\beq
J(t) = - \int_{-\infty}^{\infty} dp \ \frac {p} {m} w_0 (p,  - p t / m)
\eeq
The current $J(t)$ is the usual classical arrival time distribution that we would have expected in the absence of the absorbing potential.

Eq.(\ref{5.2.8}) has the same form as the expressions for the quantum arrival time we expect to obtain from a measurement model, Eq.(\ref{1.8.best}). The response function is given by the quantity,
\beq
R(\t,s)=2V_{0}\exp(-2V_{0}(\t-s)).
\eeq
In Chapter 1 we noted that $R(\t,s)$ is related to a sort of
coarse graining in time,  we now demonstrate this for this specific case. Eq.(\ref{5.2.8}) gives the probability
$ \Pi (\tau ) d \tau $
for arriving during the infinitesimal time interval $ [\tau, \tau + d \tau]$. Supposer we consider the probability for
arriving during a finite time interval, $[ \tau_1, \tau_2]$. This is given by
\bea
p (\tau_2, \tau_1) &=& \int_{\tau_1}^{\tau_2} d \tau \ \Pi (\tau)
\nonumber \\
&=& \int_{\tau_1}^{\tau_2} d \tau \int_0^\tau dt \ 2 V_0  \ e^{ - 2V_0 (\tau - t) } \ J(t)
\label{5.2.10}
\eea
Rearranging the order of integration and integrating over $\tau$ yields,
\bea
p(\tau_2, \tau_1) &=& \int_0^{\tau_1} dt \int_{\tau_1}^{\tau_2} d \tau \ 2 V_0  \ e^{ - 2V_0 (\tau - t) } \ J(t)
\nonumber \\
&& \ \ \ \ + \int_{\tau_1}^{\tau_2} dt \int_{t}^{\tau_2} d \tau \ 2 V_0  \ e^{ - 2V_0 (\tau - t) } \ J(t)
\\
&=& \int_0^{\tau_1} dt \left( e^{- 2V_0 (\tau_1 - t)} -  e^{- 2V_0 (\tau_2 - t)} \right) \ J(t)
\nonumber \\
&&\ \ \ \ + \int_{\tau_1}^{\tau_2} dt \left( 1 -  e^{- 2V_0 (\tau_2 - t)} \right) \ J(t)
\label{5.2.12}
\eea
$1/V_0$ plays a role as a fundamental short timescale in the problem, so now suppose we assume that $\tau_1$, $\tau_2$ and $ (\tau_2 -\tau_1) $ are all much greater
than $ 1 / V_0 $. It follows that all the exponential terms may be dropped in Eq.(\ref{5.2.12})
and we obtain the very simple result,
\beq
p (\tau_2, \tau_1) \approx \int_{\tau_1}^{\tau_2} dt \ J(t)
\label{5.2.13}
\eeq
That is, all dependence on the resolution function $R(\t,s)$ and the complex potential
parameter $V_0$ completely drops out when we look at probabilities defined on timescales
much greater than $1/V_0$. This result is very relevant to the decoherent histories analysis
considered later where it is natural to look at the arrival time during a finite time interval much greater than $\e$.

\section{Calculation of the Arrival Time Distribution via a Complex Potential}\label{sec:5.3}

We now repeat the analysis of the previous section for the quantum case. The analogue of Eq.(\ref{5.2.1}) is the inclusion of the complex potential,
\beq
V(x) =
- i V_0 \theta (-x)
\label{5.3}
\eeq
in the Schr\"odinger equation. With such a potential, the state at time $\tau$ is
\beq
| \psi (\tau) \rangle = \exp \left( - i H_0 \tau -  V_0  \theta (-x) \tau \right) | \psi \rangle
\label{5.4}
\eeq
where $H_0$ is the free Hamiltonian. In a similar way to the classical case the idea is that the part of the wave packet that reaches the origin during the time interval $[0, \tau]$ should be absorbed, so that
\beq N( \tau )
= \langle \psi (\tau) | \psi (\tau) \rangle
\label{5.5}
\eeq is the probability of not crossing $x=0$ during the time interval $[0,\tau]$. The probability of crossing between $ \tau$ and $\tau + d \tau $ is then
\beq
\Pi (\tau) = - \frac { d N } { d \tau}
\label{5.6}
\eeq
Complex potentials such as Eq.(\ref{5.3}) were originally considered by Allcock in his seminal work on arrival time \cite{All} and have subsequently appeared in detector models \cite{Hal1, meas}. (See also Ref.\cite{complex, HSM} for further work with complex potentials). From the results of Chapter \ref{cppo} we see that evolution according to Eq.(\ref{5.4}) is essentially the same as evolution under pulsed measurements.

For large $V_0$, the wave function defined by the evolution Eq.(\ref{5.4})
undergoes significant reflection, with total reflection
in the ``Zeno limit'', $V_0 \rightarrow \infty$ \cite{Zeno}. Here, we are interested
in the opposite case of small $V_0$, where there is small reflection
and Eq.({\ref{5.6}) can give reasonable expressions for the arrival time
distribution. A number of different authors \cite{All,meas, HSM} indicate that the
resulting distribution is of the form
\beq
\Pi (\tau) = \int_{-\infty}^{\infty} dt \ R (V_0, \tau - t) J (t)
\label{5.7}
\eeq
where $J(t)$ is the current,
\bea
J(t)&=&\bra{\psi_{0}}\frac{(-1)}{2m}\left\{\hat p \delta(\hat x)+\delta(\hat x)\hat p\right\}\ket{\psi_{0}}\nonumber\\
&=&\frac{i}{2m}\left(\psi^{*}(0,\t)\frac{\partial \psi(0,\t)}{\partial x}-\frac{\partial \psi^{*}(0,\t)}{\partial x}\psi(0,\t)\right) \label{5.7a}
\eea
and,
\beq
R (V_0, t) = 2 V_0 \theta (t) \exp \left( - 2 V_0 t \right)
\label{5.8}
\eeq
It is therefore closely related to the current $J(t)$ but also includes the influence of the complex potential via the ``apparatus resolution function'' $ R(V_0,t)$.  Our aim in this section is to prove this claim.

With the complex potential Eq.(\ref{5.3}), the arrival time distribution
Eq.(\ref{5.6}) is given by
\beq
\Pi (\tau)
= 2  V_0 \langle \psi_{\tau} | \theta ( - \hat x ) | \psi_{\tau} \rangle
\label{5.11}
\eeq
where
\bea
| \psi_\tau \rangle &=& \exp \left( - i H \tau \right) | \psi \rangle
\nonumber \\
&=& \exp \left( - i H_0 \tau  - V_0 \theta (-\hat x)  \tau  \right) | \psi \rangle
\eea
(so we use $H = H_0 - i V_0 \theta ( - \hat x )  $ to denote the total
non-Hermitian Hamiltonian).
We are interested in calculating this expression for the case in which $V_0$
is much smaller than the energy scale of the initial state. (The very different limit,
of $V_0 \rightarrow \infty $, the Zeno limit, has been explored elsewhere
\cite{Hal4}).

One way to evaluate Eq.(\ref{5.11}) is to use the transmitted wave functions,
Eq.(\ref{4.3}). However, we give here instead a different and more direct
derivation using the PDX.
We use the first crossing PDX, written in operator form, Eq.(\ref{PDX81}),
\beq
\langle  x |  \exp ( - i H \tau ) | \psi \rangle
=  -  \frac {1} {m} \int_0^{\tau} dt \ \langle x | \exp ( - i H (\tau -t) )
\ \delta  ( \hat x ) \hat p \ \exp \left( - i H_0 t \right)  | \psi \rangle
\label{5.19}
\eeq
Now note that the operator $\delta (\hat x) = | 0 \rangle \langle 0 | $ has the simple
property that for any operator $A$
\beq
\delta (\hat x) A \delta (\hat x) = \delta ( \hat x ) \langle 0 | A | 0 \rangle
\label{5.20}
\eeq
(where, recall, $ | 0 \rangle $ denotes the position eigenstate $ | x \rangle$
at $x=0$).
Inserting Eq.(\ref{5.19}) in Eq.(\ref{5.11}), together with the property Eq.(\ref{5.20})
and the change of variables $ s= \tau - t $, $ s' = \tau - t'$,
yields
\bea
\Pi (\tau) &=& \frac {2 V_0 } {m^2}
\int_0^\tau ds' \int_0^{\tau} ds \int_{-\infty}^0 dx
\nonumber \\
& \times & \langle 0 | \exp \left(  i  H^{\dag} s' \right) | x \rangle \langle x | \exp \left(
- i H s \right) | 0 \rangle
\nonumber \\
& \times & \langle \psi | \exp \left( i H_0 (\tau - s') \right)  \hat p \ \delta (\hat x) \ \hat p \ \exp \left(
 - i H_0 (\tau - s)  \right) | \psi \rangle
\label{5.5.5}
\eea
We are aiming to show that this coincides with Eq.(\ref{5.7}) with the current
Eq.(\ref{5.7a}), and the main challenge is to show how the
$ \hat p  \delta (\hat x)  \hat p $ combination turns into the
combination $ \hat p \delta (\hat x )  + \delta (\hat x ) \hat p $ in the
current Eq.(\ref{5.7a}).

Consider first the $x$ integral. Since we are assuming small $V_0$,
we may use the semiclassical approximation Eq.(\ref{3.20}), which
reads
\beq
\langle x | \exp \left( - i H s \right) | 0 \rangle
\ \approx\ \left( \frac {m} {2 \pi i  s } \right)^{1/2}
\ \exp \left( i \frac { m x^2 } { 2  s} - V_0 s \right)\label{5.5.6}
\eeq
The $x$ integral may now be carried out, with the result,
\bea
\Pi (\tau) &=& \frac { V_0 } {m^2}
\int_0^\tau ds' \int_0^{\tau} ds \ \left( \frac {m} {2 \pi i   } \right)^{1/2}
\ \frac{ e^{ - V_0 (s+s') } } { (s-s')^{\half} }
\nonumber \\
& \times & \langle \psi_{\tau} | \exp \left( -i H_0 s' \right)
\hat p \ \delta (\hat x) \ \hat p \ \exp \left(
i H_0 s  \right) | \psi_\tau  \rangle
\label{5.5.7}
\eea
where $ | \psi_{\tau} \rangle $ denotes the free evolution of the
initial state.

We now carry out one of the time integrals. Note that,
\beq
\int_0^\tau ds' \int_0^\tau ds =
\int_0^\tau  ds' \int_{s'}^\tau ds +
\int_0^\tau ds \int_s^\tau ds'
\eeq
In the first integral, we set $u=s'$, $ v = s-s'$, and in the second
integral we set $ u=s$, $v=s'-s$. We thus obtain
\bea
\Pi (\tau) &=& \frac { V_0 } {m^2} \left( \frac {m} {2 \pi    } \right)^{1/2}
\int_0^\tau du \ e^{ - 2V_0 u } \ \int_0^{\tau - u} dv \ \frac {e^{- V_0 v}} {v^{\half}}
\nonumber \\
& \times &
\left[ \ \frac{1} {i^\half} \langle \psi_{\tau} | \exp \left( -i H_0 u \right) \hat p \ \delta (\hat x) \ \hat p \ \exp \left(
i H_0 (u+v)  \right) | \psi_\tau  \rangle \right.
\nonumber \\
&+&  \left.  \frac{1} {(-i)^\half}\langle \psi_{\tau} | \exp \left( -i H_0 (u+v) \right) \hat p \ \delta (\hat x) \ \hat p \ \exp \left(
i H_0 u  \right) | \psi_\tau  \rangle  \right]
\label{5.5.9}
\eea
The factors of $ 1/ ( \pm i)^{\half} $ in front of each term come from a careful consideration
of the square root in the free propagator prefactor (and must have this form because
$ \Pi (\tau) $ is real).

We will assume that $ V_0 \tau \gg 1 $, which means that the integrals are concentrated around
$ u = v = 0$. This means that we may take the upper limit of the $v$ integral to be $\infty$,
and it may be carried out, to yield,
\bea
\Pi (\tau) &=& 2 V_0 \int_0^\tau du \ e^{ - 2V_0 u }
\nonumber \\
& \times &
\frac {1} {2m} \langle \psi_{\tau - u} |
\ \hat p \ \delta (\hat x)  \   \Sigma (\hat p)  +  \Sigma^{\dag} ( \hat p)
\ \delta (\hat x ) \ \hat p
\ | \psi_{\tau - u} \rangle
\label{5.5.10}
\eea
where the operator $\Sigma (\hat p) $ is given by
\beq
\Sigma (\hat p ) = \frac { \hat p } { [ 2 m ( H_0 + i V_0)]^\half}
\eeq
For $V_0$ much less than the energy scale,
\beq
\Sigma (\hat p ) \ \approx \ \frac{ \hat p } { | \hat p | }
\eeq
so $\Sigma (\hat p)$ is simply the sign function of the momentum, which is $ - 1$ in this case,
since the initial state consists entirely of negative momenta.
Finally, writing $ u = \tau - t$, we obtain
\bea
\Pi (\tau) &=&  2 V_0 \int_{0}^\tau dt \ e^{ - 2V_0 (\tau - t) }
\ \frac {(-1)} {2m} \langle \psi_t |
\left( \hat p \ \delta (\hat x)  \    +
\ \delta (\hat x ) \ \hat p
\right) | \psi_t \rangle
\nonumber \\
&=&  2 V_0 \int_{0}^\tau dt \ e^{ - 2V_0 (\tau - t) } \ J(t)
\label{5.5.13}
\eea
We therefore have precise confirmation of the classical result Eq.(\ref{5.2.8}),
and also agreement with the expected quantum result, Eq.(\ref{5.7}).

Some comments are in order concerning the positivity of
the result for $\Pi (\tau)$. The expression (\ref{5.5.10}) is positive because it was derived from the manifestly positive
expression Eq.(\ref{5.5.5}). (Two approximations were used: the semiclassical approximation
Eq.(\ref{5.5.6}), and
the condition $V_0 \tau \gg 1$, neither of which affect the positivity of the
result.)

However, to obtain the final result Eq.(\ref{5.5.13}) we took the limit $V_0 \rightarrow 0 $
in the current part only of Eq.(\ref{5.5.10}), leaving behind the $V_0$-dependent
term in the exponential part, and the resulting expression is not guaranteed to be
positive. In Eq.(\ref{5.5.13}}), $J(t)$ is not always positive due
to the backflow effect \cite{cur,BrMe}
and integration
over time does not necessarily remedy the situation. (See the Appendix for a more thorough discussion of this point). The lack of positivity
for a $\Pi (\tau) $ obtained in this way is not surprising since
taking the limit $V_0 \rightarrow 0 $ in one part of the expression Eq.(\ref{5.5.10})
only but not the other will not necessarily preserve its property of positivity.
The violation of positivity is generally small, however, so Eq.(\ref{5.5.13}) may still be a good
approximation to the manifestly positive expression Eq.(\ref{5.5.10}).

It should also be added that it would be misleading to explore the first
order corrections in $V_0$ in going from Eq.(\ref{5.5.10}) to Eq.(\ref{5.5.13}),
since comparable correction terms have already been dropped in using
the semiclassical approximation Eq.(\ref{5.5.6}).



\section{Decoherent Histories Analysis for a Single Large Time Interval}\label{sec:5.4}

We turn now to the decoherent histories analysis of the arrival time problem. The important question we wish to answer is; What is the probability of arriving during some general time interval $[t_{1},t_{2}]$? We answer this question in detail in the next section, here we address a simpler question; What are
the probabilities for crossing or not crossing
during the large time interval $ [0, \tau] $? The reason for tackling this question first is that it gives us useful insight into the more general case and also suggests an interesting way of thinking about the situations in which we might have decoherence.

The class operators for not crossing and crossing during this time interval  are
\bea
C_{nc} &=& \exp \left( - i H_0\tau - V_{0}\theta(-x) \tau \right)
\label{5.6.1}\\
C_c &=& \exp\left( - i H_0 \tau \right) - \exp \left( - i H_0 \tau - V_{0}\theta(-x) \tau \right)
\label{5.6.2}
\eea
and they satisfy
\beq
C_{nc} + C_c = e^{- i H_0 \tau }
\label{5.6.3}
\eeq
We are interested in the probabilities for not crossing and crossing,
\bea
p_{nc} (\tau)  &=& {\rm Tr} \left( C_{nc} \rho C_{nc}^{\dag} \right)
\\
p_c (\tau) &=& {\rm Tr} \left( C_{c} \rho C_{c}^{\dag} \right)
\eea
and the off-diagonal term of the decoherence functional,
\bea
D_{c, nc} &=& {\rm Tr} \left( C_{nc} \rho C_{c}^{\dag} \right)
\nonumber \\
&=& {\rm Tr} \left( C_{nc} \rho e^{i H_0 \tau} \right) - p_{nc}
\label{5.6.6}
\eea
These quantities obey the relation
\beq
p_{nc} + p_c + D_{c,nc} + D^*_{c,nc} = 1
\label{5.6.5}
\eeq
We look for situations where there is decoherence,
\beq
D_{c, nc} = 0
\eeq
(which is usually only approximate), in which case the probabilities
then sum to $1$,
\beq
p_c (\tau ) + p_{nc} (\tau)  = 1
\label{5.6.7}
\eeq

It is useful to relate some of these expressions
to the standard expressions for arrival time $ \Pi (t)$ defined in Eqs.(\ref{5.5}), (\ref{5.6}).
To do this, note that $p_{nc}$ is in fact the same as the survival probability,
$N(\tau)$ defined in Eq.(\ref{5.5}), and that $p_{nc}$ obeys the trivial identity
\beq
p_{nc} (\tau)  = 1 + \int_0^\tau dt \ \frac { d p_{nc} } {dt}
\label{5.6.10}
\eeq
since $p_{nc} (0) = 1 $. It follows that
\beq
p_{nc} (\tau)  = 1 - \int_0^\tau \ dt \ \Pi (t)
\label{5.6.11}
\eeq
When there is decoherence, Eq.(\ref{5.6.7}) holds and we may deduce that
\beq
p_c (\tau) = \int_0^\tau \ dt \ \Pi (t)
\eeq
Hence the decoherent histories analysis is compatible with the standard result, but only
when there is decoherence.

We turn now to the calculation of the decoherence functional. Following the general pattern described in
Eqs.(\ref{1.16}), (\ref{1.17}),
consider the quantity,
\beq
q_{nc} (\tau)  = {\rm Tr} \left( C_{nc} \rho e^{i H_0 \tau} \right)
\label{5.6.18}
\eeq
From Eq.(\ref{5.6.6}), we see that the decoherence functional may be written,
\beq
D_{c,nc} = q_{nc} (\tau) - p_{nc} (\tau)
\eeq
This means that $q_{nc} = p_{nc}$ when there is decoherence. Or to put it the
other way round, decoherence of histories may be checked by comparing $q_{nc}$
with $p_{nc}$ and this is what we now do. Recall that $p_{nc}$ is given
by Eqs.(\ref{5.6.10}), (\ref{5.6.11}) (which hold in the absence of decoherence).
We may write $q_{nc}$ in a similar form:
\beq
q_{nc} (\tau) = 1 + \int_0^\tau dt \ \frac { d q_{nc} } {dt}
\eeq
The integrand is similar to $\Pi (t)$ defined in Eq.(\ref{5.11}), so we define
\beq
\tilde \Pi (t) \equiv - \frac { d q_{nc} } {dt}
\eeq
We now have that
\beq
q_{nc} (\tau)  = 1 - \int_0^\tau \ dt \ \tilde \Pi (t)
\eeq
It then follows that the decoherence functional is
\beq
D_{c,nc} =  \int_0^\tau \ dt \left ( \Pi (t) - \tilde \Pi (t) \right)
\eeq

To compute the decoherence functional we need to calculate $\tilde \Pi (t)$, which
is given by
\beq
\tilde \Pi (t) =  V_0 \langle \psi | \ \exp \left( i H_0 t \right)
\ \theta (- \hat x)\  \exp \left( - i H_0 t - V_0 \theta ( - \hat x ) t \right)\  | \psi \rangle
\label{5.6.21}
\eeq
This is almost the same as $ \Pi (t) $ except
that the exponential on the left involves only $H_0$ and not the complex potential
(and also an overall factor of $2$). We therefore follow the calculation of $\Pi (t)$ in Section 5.3
with small modifications. With a little care, one may see that the
final result is the same as that for $\Pi (t)$, Eq.(\ref{5.5.13}),
except that $2 V_0$ is replaced with $V_0$, that is,
\beq
\tilde \Pi (t) =  V_0 \int_{0}^t ds \ e^{ - V_0 (\tau - s) } \ J(s)\label{5.4.99}
\eeq
This result holds for timescales greater than $ 1/ V_0$ and under the semiclassical
approximation (which required $ E \gg V_0$ so is equivalent to the
requirement of negligible reflection encountered above).
Finally, a calculation similar to that of Eqs.(\ref{5.2.10})-(\ref{5.2.13}) implies that
\beq
\int_0^{\tau} dt \ \Pi (t ) \ \approx \ \int_0^{\tau} dt \ J (t)
\eeq
as long as $ V_0 \tau \gg 1 $. Since this result is independent of $V_0$,
$ \tilde \Pi (t)$ will satisfy the same relation. We thus deduce that
\beq
D_{c,nc} \approx 0
\eeq
hence there is decoherence, under the above conditions. 

This analysis gives us an interesting way of thinking about decoherence. Decoherence is achieved if Eq.(\ref{5.4.99}) is equal to Eq.(\ref{5.5.13}). This will be the case if the current varies on a timescale that is long compared with $1/V_{0}>>1/E$, where $E$ is the energy of the wavepacket. The timescale on which the current varies is given, for a gaussian wavepacket, as $1/\Delta H$. In Chapter 4 we noted that whereas the timescale given by $1/E$ is genuinely quantum, $1/\Delta H$ is in fact a classical timescale. This means we will have decoherence if there is a large separation between the quantum and classical timescales associated with the incoming wavepacket, a satisfying result.

\section{Decoherent Histories Analysis for an Arbitrary Set of Time Intervals}\label{sec:5.5}

We now turn to the more complicated question of much more refined histories,
that may cross the origin at any one of a large number of times, during the time interval
$ [ 0, \tau] $. This corresponds more directly to the standard crossing probability,
$ \Pi (t) dt $, the probability that the particle crosses during an infinitesimal
time interval $ [t, t+ dt] $.

\subsection{Class Operators}

We have defined class operators Eqs.(\ref{5.6.1}), (\ref{5.6.2})
describing crossing or not crossing during a time interval $[0,\tau]$.
We now split this time interval into $n$ equal parts of size $\e$,
so $ \tau = n \e $ and we seek class operators describing crossing
or not crossing during any one of the $n$ intervals. We first note
that
\beq
e^{ - i H_0 \e } = C_{nc} (\e ) + C_c ({\e} )
\eeq
where $C_{nc} (\e) $ and $C_c (\e )$ are defined as in Eqs.(\ref{5.6.1}), (\ref{5.6.2})
except that here they are for a time interval $ [0,\e]$. We now use this to decompose $ e^{ - i H_0 \tau}$
into the desired class operators. We have
\bea
e^{ - i H_0 \tau } &=& \left( e^{ - i H_0 \e } \right)^n
\nonumber \\
&=& \left( e^{ - i H_0 \e } \right)^{n-1} \ \left( C_{nc} (\e ) + C_c ({\e}) \right)
\nonumber \\
&=&
\left( e^{ - i H_0 \e } \right)^{n-1} C_{nc} (\e) + e^{ - i H_0 (\tau - \e) } C_c (\e )
\eea
Repeating the same steps on the first term, this yields,
\beq
e^{ - i H_0 \tau }  = \left( e^{ - i H_0 \e } \right)^{n-2} C_{nc} (2 \e)
+ e^{ - i H_0 (\tau - 2\e) } C_c (\e ) C_{nc} (\e)
+ e^{ - i H_0 (\tau - \e) } C_c (\e )
\eeq
Repeating more times eventually yields,
\beq
e^{- i H_0 \tau} = C_{nc} (\tau ) + \sum_{k=0}^{n-1} \ e^{ - i H_0 (\tau - (k+1) \e ) }
C_c (\e) C_{nc} (k \e )
\label{5.7.4}
\eeq
From this expression, we see that the class operator for crossing $x=0$ for the first
time during the time interval $ [k \e, (k+1) \e ] $ is given by the summand of the second term,
\beq
C_c ( (k+1) \e, k \e) = e^{ - i H_0 (\tau - (k+1) \e ) }
C_c (\e) C_{nc} (k \e )
\label{5.7.5}
\eeq

We will not in fact work with the class operator Eq.(\ref{5.7.5}),
since a more useful similar but alternative expression can also be found. Taking the continuum
limit of Eq.(\ref{5.7.4}) (and inserting the explicit expression for $C_{nc})$, we obtain,
\beq
e^{- i H_0 \tau} = e^{- i H_0 \tau - V \tau }
+ \int_0^\tau dt \ e^{- i H_0 (\tau - t)} V e^{- i H_0 t - V t}
\label{5.7.6}
\eeq
(where, recall, $V = V_0 \theta ( - \hat x ) $). This indicates that the class operator
for first crossing during the infinitesimal time interval $ [t, t+dt]$, is
\beq
C_c (t) = e^{- i H_0 (\tau - t)} V e^{- i H_0 t - V t}
\label{5.7.7}
\eeq
We do not, however, expect histories characterized by such precise crossing time to be decoherent,
so it is natural to consider coarser-grained class operators,
\beq
C_c^k  = \int_{t_k}^{t_{k+1}} dt \ C_c (t)
\label{5.7.8}
\eeq
which represents crossing during one of the $N$ time intervals $ [t_k, t_{k+1}]$ of size $\Delta$, where
$ t_k = k \Delta $ with $ k = 0,1 \cdots N-1 $ and $ \tau = N \Delta $.
The complete set of class operators $C_{\a}$ for crossing
and not crossing is the set of $N+1$ operators
\beq
C_{\a} = \{ C_{nc}, C_c^k \}
\eeq
and Eq.(\ref{5.7.6}) implies that they satisfy
\beq
e^{- i H_0 \tau} = C_{nc} + \sum_{k=0}^{N-1} C_c^k
\label{5.7.10}
\eeq
To check for decoherence of histories we need to calculate two types of decoherence functional
\bea
D_{k k'} &=& {\rm Tr} \left( C_c^k \rho ( C_c^{k'} )^{\dag} \right)
\label{5.7.11} \\
D_{ k, nc} &=& {\rm Tr} \left( C_c^k \rho ( C_{nc} )^{\dag} \right)
\label{5.7.12}
\eea
and this will be carried out below.


\subsection{An Important Simplification of the Class Operator}

There turns out to be a very useful simplification in the class operator Eq.(\ref{5.7.7}). Consider
the amplitude
\beq
\langle x | e^{ i H_0 \tau} C_c (t) | \psi \rangle = V_0 \langle x | e^{- i H_0 t } \theta (- \hat x)
e^{- i H_0 t - V t} | \psi \rangle
\label{5.7.13}
\eeq
for any $ x $. The right-hand side is very similar to Eq.(\ref{5.11}), except that there is no complex potential
in one of the exponential terms and also the ``final" state is $ | x \rangle $ not $ | \psi \rangle$.
(And there is also an overall factor of $2$ different). Despite these differences, we may once again
make use of the details of the calculation of Section 5.3,
and we deduce from the analogous result Eq.(\ref{5.5.13}),  that
\beq
\langle x | e^{ i H_0 \tau} C_c (t) | \psi \rangle =
V_0 \int_{0}^t ds \ e^{ - V_0 (t - s) }
\ \frac {(-1)} {2m} \langle x |
\left( \hat p \ \delta (\hat x_s)  \    +
\ \delta (\hat x_s ) \ \hat p
\right) | \psi \rangle
\eeq
Like the derivation of Eq.(\ref{5.5.13}), this is valid under the conditions that all energy
scales are much greater than $V_0$ and all time scales much greater than $ 1 / V_0 $.
Now we integrate this over time to obtain the coarse-grained crossing time operator, Eq.(\ref{5.7.8}),
and again use approximations of the form Eqs.(\ref{5.2.10})-(\ref{5.2.13}) (again using the assumption
of timescales much greater than $1/V_0$), to yield, the remarkably
simple and appealing form,
\beq
e^{ i H_0 \tau} C_c^k = \int_{t_k}^{t_{k+1}} dt \ \ \frac {(-1)} {2m}
\ \left( \hat p \ \delta (\hat x_t)  \    +
\ \delta (\hat x_t ) \ \hat p \right)
\label{5.7.15}
\eeq
This may also be written even more simply
\beq
e^{ i H_0 \tau} C_c^k = \theta ( \hat x (t_k) ) - \theta ( \hat x (t_{k+1} ) )
\label{5.7.16}
\eeq

\subsection{Probabilities for Crossing}

The above expressions for the crossing time class operator are the most important results of this Chapter
and provide an immediate connection to the standard expression for the arrival time distribution.
Supposing for the moment that there is decoherence of histories, we may assign
probabilities to the histories.
The probability for crossing during the time interval $ [t_k, t_{k+1}]$ is
\beq
p (t_k, t_{k+1} ) = {\rm Tr} \left( C_c^k \rho (C_c^k)^{\dag} \right)
\eeq
However, as noted in Eqs.(\ref{1.16}), (\ref{1.17}) when there is decoherence of histories, this expression for
the probabilities for histories is equal to the simpler expression,
\bea
q (t_k, t_{k+1} )  &=& {\rm Tr} \left( C_c^k \rho e^{i H_0 \tau } \right)
\nonumber \\
&=&
\int_{t_k}^{t_{k+1}} dt \ \ \frac {(-1)} {2m}
\ \langle \psi |  \left( \hat p \ \delta (\hat x_t)  \    +
\ \delta (\hat x_t ) \ \hat p \right) | \psi \rangle
\nonumber \\
&=& \int_{t_k}^{t_{k+1}} dt \ J(t),
\label{5.7.18}
\eea
which is precisely the standard result! The expression for the probability $ q (t_k, t_{k+1} )$ is not
positive in general (although is real in this case, as it happens), but when there is decoherence,
it is equal to $ p (t_k, t_{k+1} ) $, which {\it is} positive. Hence the decoherent histories result
coincides with the standard result under the somewhat special conditions of decoherence of histories.

\subsection{Decoherence of Histories and the Backflow Problem}\label{sec:5.5.4}

There is an interesting connection between decoherence of histories and backflow.
To see this, consider the following simple case. We consider histories which either
cross or do not cross the origin during the time interval $[t_1, t_2]$. So the crossing
and not crossing class operators are $C$ and $ 1 - C $, where
\beq
C =  \theta ( \hat x_1 ) - \theta ( \hat x_2 )
\eeq
where we have adopted the notation $ \hat x_k = \hat x (t_k) $
(and for convenience we have dropped the exponential factor which is just a matter
of definition and drops out of all expression of interest).
The decoherence functional is
\bea
D &=& \langle C ( 1 - C) \rangle
\nonumber \\
&=& \langle C \rangle - \langle C^2 \rangle
\label{5.7.25}
\eea
This may also be written
\beq
D =  - \langle \left( \theta ( - \hat x_1 ) \theta ( \hat x_2 ) + \theta ( \hat x_2 )
\theta ( - \hat x_1 ) \right) \rangle
\label{5.7.21}
\eeq
a form we will use below to check decoherence.
When there is decoherence, $D = 0 $ and the probability for crossing is
\beq
p (t_1, t_2) = \langle C^2 \rangle = \langle C \rangle
\eeq
As noted above, $ \langle C \rangle $ is the standard result, Eq.(\ref{5.7.18}),
for the probability of crossing.

There is an interesting connection here between backflow and decoherence. If there is
decoherence, $D$ is zero so
$ \langle C \rangle$ must cancel  $ \langle C^2 \rangle$ in Eq.(\ref{5.7.25}),
which means that $ \langle C \rangle \ge 0 $, so there is no backflow. Or we
may make a logically equivalent statement:
if there is backflow, $ \langle C \rangle < 0 $, then there cannot be decoherence, since $|D|$ is then
greater than the probability $ \langle C^2 \rangle $. Hence, states with backflow do not permit
decoherence of histories.
(Note that absence of backflow, $ \langle C \rangle \ge 0 $,
is not itself enough to guarantee decoherence -- the stronger condition $D = 0$
must be satisfied).

This is an important result. The quantity $ \langle C \rangle $ is regarded as the
``standard'' result for the arrival time probability and its possible negativity is
disturbing. Here, the decoherent histories approach sheds new light on this issue.
In the decoherence histories apporach,
the true probability for crossing is the manifestly positive quantity $ \langle C^2 \rangle$
and this is equal to $ \langle C \rangle $ only when there is decoherence. In particular,
when there is significant backflow, there cannot be decoherence, so probabilities cannot
be assigned and $ \langle C^2 \rangle$ is not equal to $ \langle C \rangle $.

\subsection{The Decoherence Conditions}

The crossing probabilities described above are only valid when all components of the
decoherence functional, Eqs.(\ref{5.7.11}), (\ref{5.7.12}), are zero. We therefore
address the issue of finding those states for which these conditions hold.

The most important decoherence condition is that $D_{kk'}$ defined  in Eq.(\ref{5.7.11})
vanishes, so we focus on that.
We write the class operator Eq.(\ref{5.7.16}) for crossing during the time
interval $ [t_k, t_{k+1}]$ as
\beq
C_c^k = e^{ - i H_0 \tau} \left( \theta ( \hat x_k ) - \theta ( \hat x_{k+1}  ) \right)
\eeq
For an initial state $ | \psi \rangle$, the quantity
$ C_c^k | \psi \rangle $ is a quantum state representing the property of
crossing of the origin in the time interval $ [t_k, t_{k+1}]$. The decoherence
condition $D_{kk'} = 0$ is simply the condition that the ``crossing states''
$ C_c^k | \psi \rangle $ for different
time intervals have negligible interference.
The states
$ C_c^k | \psi \rangle $ consist of an initial state
which has been localized to a range of time at $x=0$.
This is closely related to the interesting question
of diffraction in time \cite{diff} and this connection will be explored in more
detail elsewhere \cite{HaYe}.

The decoherence functional is given by
\bea
D_{kj} &=& \langle \left( \theta ( \hat x_k ) - \theta ( \hat x_{k+1}  ) \right)
\left( \theta ( \hat x_{j} ) - \theta ( \hat x_{j+1}  ) \right) \rangle
\nonumber \\
&=& \langle \left( \theta ( \hat x_k ) - \theta ( \hat x_{k+1}  ) \right)
\left( \theta ( -\hat x_{j+1} ) - \theta (- \hat x_{j}  ) \right) \rangle \label{5.7.26}
\eea
where without loss of generality we take $ t_{j+1} < t_k $.
It is a sum of terms each of the form,
\beq
d_{kj} =  \langle \theta ( - \hat x_k ) \theta ( \hat x_j )  \rangle
\eeq
where $t_k < t_j $. Note that
\beq
| d_{kj} |^2 \le  d^2_m = \langle  \theta ( - \hat x_k ) \theta ( \hat x_j )  \theta ( - \hat x_k ) \rangle
\label{5.7.30}
\eeq
The key thing is that $d_m^2$ has the form of a probability -- it is the probability to
find the particle in $x<0$ at $t_k$ and then in $x>0$ at $t_j $. Semiclassical
expectations suggest that this is small in general for the states considered here,
which are left-moving wave packets, and indeed the classical limit of this probability
is zero. So this is a useful object to calculate in terms
of checking decoherence. (Although it may not be small for states with
backflow). Note that it also implies that the other parts of the decoherence functional,
Eq.(\ref{5.7.26}) will also be small. In detailed calculations, the upper bound
Eq.(\ref{5.7.30}) must be compared with the probabilities, as in Eq.(\ref{DA}).

\subsection{Checking the Decoherence Condition for Wavepackets}

Decoherence starts to become lost as the the size $ \Delta $ of the time intervals
$ [t_k, t_{k+1}]$ is reduced to close to $1/E$. This is because
the ``localisation in time'' produces a corresponding spreading in energy, and thus momentum, and for sufficiently small time intervals this is enough to give rise to positive momenta with significant probability.
We will see in a specific example how this works.

For simplicity we work with the simple wavepacket initial states, more general initial states require an environment to produce decoherence and will be considered in Chapter 7.
We take the initial state to be,
\beq
\psi (x) = \frac{1} {(2 \pi \sigma^2)^{1/4} } \exp \left( - \frac {(x-q_0)^2} {4 \sigma^2 } + i p_0 x \right)
\label{5.7.33}
\eeq
where $q_0 > 0 $, $ p_0 < 0 $ and $|p_{0}|\sigma>>1$. 
We note that the decoherence functional Eq.(\ref{5.7.21})
satisfies
\beq
| D |^2 \le 2 d_m^2
\eeq
with $d_m^2$ given by Eq.(\ref{5.7.30}) with $k=1$, $j=2$.
We need some probabilities
to compare this with. We have that
\beq
| D |^2 \le \langle C^2 \rangle \langle (1 - C)^2 \rangle
\eeq
The interesting case is that in which the crossing probability $ \langle C^2 \rangle$
is somewhat less than $1$, less than about $1/2$, say, in which case the non-crossing
probability $ \langle (1-C)^2 \rangle $ will be of order $1$. It is therefore
sufficient to compare $d_m^2$ with $ \langle C^2 \rangle $. Now note that
\bea
\langle C^2 \rangle &=& \langle \left( \theta ( \hat x_1 ) - \theta ( \hat x_2 ) \right)^2 \rangle
\nonumber \\
&=& \langle \theta ( \hat x_1 ) \theta ( -\hat x_2 )  \theta (  \hat x_1 ) \rangle
+ \langle \theta ( \hat x_1 ) \theta ( -\hat x_2 )  \theta (  - \hat x_1 ) \rangle
+ \langle \theta ( \hat x_2 )  \theta (  - \hat x_1 ) \rangle.
\eea
This means that $\langle C^2 \rangle $ is in fact equal to the probability,
\beq
p_{12} = \langle \theta ( \hat x_1 ) \theta ( -\hat x_2 )  \theta (  \hat x_1 ) \rangle,
\label{5.7.39b}
\eeq
up to terms which vanish when $D = 0$. This is useful since it is now identical in form
to the expression for $d_m^2$ and our goal is to show that
\beq
d_m^2 \ll p_{12}.
\eeq

It is useful to work in the Wigner representation \cite{Wig}, defined, for a state $\rho(x,y)$
by
\beq
W(p,q) = { 1 \over 2 \pi } \int d \xi \ e^{- i  p \xi}
\ \rho( q + \half \xi, q - \half \xi).
\eeq
The probabilities $p_{12}$ and $d_m^2 $ are then given by
\bea
p_{12} &=& 2 \pi \int dp dq \ W_{P} (p,q) \ W_0 (p,q,t_1),
\label{5.7.39}
\\
d_m^2 &=& 2 \pi \int dp dq \ W_ D (p,q) \ W_0 (p,q,t_1 ).
\label{5.7.40}
\eea
Here, $  W_0 (p,q,t_1 ) $ is the Wigner function of the initial state,
evolved in time to $t_1$,
\beq
 W_0 (p,q,t_1 ) = \frac {1} {\pi} \exp \left( - \frac { (q - q_0 - p_0 t_1 / m)^2 } { 2 \sigma^2}
 - 2 {\sigma^2}   ( p - p_0 )^2 \right).
\label{5.7.41}
\eeq
The objects $W_{P}$ and $W_D$ are the Wigner transforms of the $\theta$-function
combinations appearing in Eqs.(\ref{5.7.30}), (\ref{5.7.39b}) and are given by
\bea
W_{P} (p,q) &=& \frac {1} {2 \pi^2} \theta (q) \int_{u(p,q)}^\infty dy \ \frac { \sin y } {y},
\label{5.7.42}
\\
W_D (p,q) &=& \frac {1} {2 \pi^2} \theta (- q)  \int_{u(p,q)}^\infty dy \ \frac { \sin y } {y},
\label{5.7.43}
\eea
where
\beq
u(p,q) = 2 q \left( p + \frac {m q } {(t_2 - t_1)} \right).
\eeq
We see that the only difference between the expressions for $p_{12}$
and $d_m^2$ is in the
sign in the $\theta$-functions.

The integral
\beq
f(u) = \int_{u}^\infty dy \ \frac { \sin y } {y}
\eeq
may be expressed in terms of the Sine integral function ${\rm Si} (x)$,
\beq
f(u) = \frac {\pi } {2} - {\rm Si} (u)
\eeq
but its properties are not hard to see directly. For large negative $u$,
$ f(u) \approx \pi $, at $u=0$, $f(0) = \pi / 2 $, and for large
positive $u$, $f(u)$ goes to zero, oscillating around $ 1/u$.
(See Figure \ref{fig:fu_fig4}.)

\begin{figure}[htbp] 
   \centering
   \includegraphics[width=12cm]{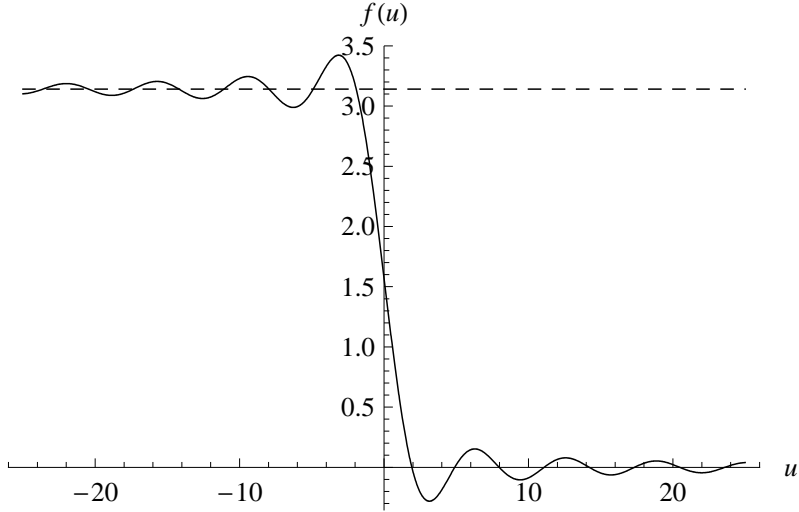} 
   \caption[Plot of the Function $f(u)$.]{{\em The function $f(u)$. It oscillates around zero for $u \gg 0$
and oscillates around $ \pi $ for $ u \ll 0 $. As a function of $q$, $f(u(p_0,q))$
differs from $0$ or $ \pi$ only in a region of size $1/|p_0|$ around $q=0$.}}
   \label{fig:fu_fig4}
\end{figure}

We now compare the size of $p_{12}$ and $d_m^2$.We assume that the wave packet
is spatially broad, so $\sigma$ is large and the Wigner function Eq.(\ref{5.7.41})
is therefore concentrated strongly about $ p = p_0 < 0 $. We therefore
integrate out $p$ and set $ p = p_0$ throughout.
The most important case to check is that in which the wave packet is reasonably
evenly divided between $ x>0$ and $x<0$ at time $t_1$ so that both $p_{12}$
and $d_m^2$ have a chance of being reasonably large. This means that
$ q_0 + p_0 t_1 / m $ should be close to zero (to within a few widths $ \sigma$),
so for simplicity we take it to be exactly zero.

With these simplifications, we have
\beq
d_m^2 = \frac {1 } {(2 \pi^3 \sigma^2)^{1/2}} \int_{-\infty}^0 dq \ \exp \left( - \frac {q^2 } {2 \sigma^2} \right)
\ f ( u (p_0, q) ).
\label{5.7.47}
\eeq
Here, since $q \le 0$ and $p_0 < 0 $, we have $ u(p_0, q) \ge 0 $. We can evaluate
this expression by examining the comparative effects of  $f(u)$ and the exponential
term. From the plot
of $f(u)$ (see Figure \ref{fig:fu_fig4}), we see that it drops to zero at $ u=u_0$ (which is of order $1$)
and oscillates rapidly
around zero for $u>u_0$, so we expect the integral to be dominated by values of
$q$ for which $ 0 \le u \le u_0$. The value $u=u_0$ corresponds
to $q=q_0$ where
\beq
q_0 = - \frac { | p_0 | \Delta } { 2 m } \left( \left[ 1 + \frac {u_0} { E_0 \Delta } \right]^{1/2}
 -1 \right),
\eeq
where $E_0 = p_0^2 / 2m $. In the complex potential calculations, we have assumed
that $ E_0 \gg V_0 $ and we also assumed that all timescales are much greater than $ 1 / V_0 $,
and these together imply that $ E_0 \Delta \gg 1 $. We may therefore expand the square root
to leading order and obtain
\beq
q_0 \approx - \frac { u_0 } { 2 |p_0 |}.
\eeq
We have assumed that the wave packet is sufficiently broad that $ \sigma p_0 \gg 1 $,
and this means that
\beq
|q_0| \ll \sigma.
\eeq
This in turn means that $f(u)$ is significantly different from zero
only in the range $ | q | \ll \sigma $, and most importantly, in this range, the exponential term
in Eq.(\ref{5.7.47}) is approximately constant. We may therefore evaluate Eq.(\ref{5.7.47})
by ignoring the exponential term, integrating from $ 0 $ to $ q_0$ and
approximating $f(u)$ as
\beq
f(u) \approx \frac { \pi } {2} - u + O (u^3).
\eeq
We thus obtain the simple result,
\beq
d_m^2 \approx \frac {1 } {(2 \pi^3 )^{1/2}} \ \frac {1} { |p_0| \sigma} \ll 1.
\eeq

In the expression for $p_{12}$, there is a key difference in that
$q>0$ which means that $ u(p_0,q)$ can be positive or negative.
Introducing
\beq
q_z  = \frac {| p_0 |} {m} \Delta  = \sigma \frac { \Delta  } {t_z}
\eeq
(where $t_z$ is the Zeno time) we see that $ u < 0 $ for $ q< q_z $ and $ u> 0 $ for $ q > q_z $.
We therefore have
\bea
p_{12} &=& \frac {1 } {(2 \pi^3 \sigma^2)^{1/2}} \int_0^{q_z} dq
\ \exp \left( - \frac { q^2 } { 2 \sigma^2}
\right) \ f ( u (p_0, q) )
\nonumber \\
& &  \ \ \ \ \ + \frac {1 } {(2 \pi^3 \sigma^2)^{1/2}} \int_{q_z}^{\infty} dq \ \exp \left( - \frac { q^2 } { 2 \sigma^2}
\right) \ f ( u (p_0, q) ).
\label{5.7.49}
\eea
Here, $ f(u(p_0,q)) \approx \pi $ in the first term, differing from this value only in
a region of size $1/p_0$ close to $ q=0$. In the second term $f(u)$ will
tend to be small except for a small region of size $ 1 / p_0 $ around the origin.

If $q_z \gg \sigma $, (that is, $ \Delta \gg t_z$), the second term in $p_{12}$
is exponentially suppressed and in the first
term the integration range is effectively $ 0$ to $ \infty $, so we obtain
\beq
p_{12} \approx \frac {1 } {2}.
\eeq
This is the expected result, since under the above assumptions on the wave
packet, half of it will cross $x=0$ if the time interval is sufficiently
large. Clearly $ p_{12} \gg\ d_m^2 $ in this case so there is decoherence.

If $q_z \ll \sigma $, the first term in $p_{12}$ is of order $ q_z / \sigma $ and the second
of order $ 1 / ( |p_0| \sigma ) $, the same order of magnitude as $d_m^2$. Hence
in this case we have decoherence if
\beq
\frac { \Delta } { t_z } \gg \frac { 1 } { |p_0| \sigma},
\eeq
which is equivalent to,
\beq
 E \Delta\gg1 \label{5.Ee}.
\eeq
Crucially this is equivalent to the condition that there be negligible reflection from the complex potential.  

In brief, we therefore get decoherence of histories for a single
wave packet under a wide variety of circumstances. It is easily seen that similar results also hold for orthogonal superpositions of gaussian wavepackets, see Fig.(\ref{fig:superposition_fig3}). More general states will require an environment to ensure decoherence. This will be addressed in Chapter 7.
\begin{figure}[htbp] 
   \centering
   \includegraphics[width=10cm]{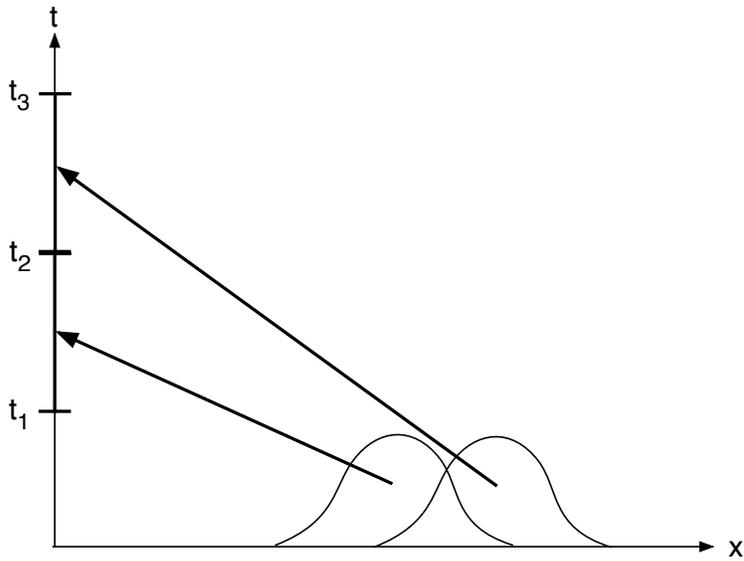} 
   \caption[The Arrival Time of a Superposition of Two Wavepackets.]{{\em An initial state consisting of a superposition of wave packets
may have significant crossings in at least two different time intervals. If the initial
packets are orthogonal, and the time intervals are sufficiently large (greater than $1/E$), the packets will remain orthogonal after passing through the time intervals
and the corresponding histories will be decoherent.}}
   \label{fig:superposition_fig3}
\end{figure}

The condition Eq.(\ref{5.Ee}) has previously been obtained in studies of the accuracy with which a quantum system can be used as a clock \cite{clock, clock2} (see also Chapter 8.) In these studies it was proposed that $1/E$ is a fundamental limitation of the accuracy with which times can be {\em defined} in quantum theory. One weakness of those studies, however, is that it is always possible that the limitation is just an artifact of the particular choice of clock model. It is therefore very interesting that the same condition arises as a decoherence condition in the study of the arrival time problem presented here. The fact that what is, on the face of it, a very different analysis of the problem nevertheless arrives at the same limitation on the accuracy of the definition of arrival times  adds weight to the idea that Eq.(\ref{5.Ee}) really is a {\em fundamental} limit in quantum theory.

\section{Summary and Conclusions}\label{sec:5.6}

We considered the arrival time problem using a complex potential to kill paths entering
$x<0$. In Section 5.2 we gave a classical analysis of the problem. We derived a
result of the expected form exposing the resolution function as an essentially
classical effect summarizing the role of the complex potential. We also showed
that coarse graining over time scales much greater than $1/ V_0$ produces a formula
for the arrival time of expected form and which is independent of the complex potential.
This is an important result for the rest of the paper.

In Section 5.3, we used the
PDX to rederive the standard form of the arrival time distribution with a complex potential,
in the limit of weak potential. The form of this calculation turned out to be useful for
the subsequent work on the decoherent histories approach.

In Section 5.4, we considered the decoherent histories analysis for the simple case
of a particle crossing or not crossing $x=0$ during a large time interval $[0,\tau]$.
We found the simple and expected result that the histories are decoherent as long
as reflection by the complex potential is negligible. The resultant probabilities
are consistent with the standard result for the arrival time.

The main part of the decoherent histories analysis was given in Section 5.5, where we first
derived the class operators described crossing $x=0$ for an arbitrary set of small
time intervals. Here we obtained our most important result: the crossing class operator
Eq.(\ref{5.7.15}) for timescales much greater than $1/V_0$. This form of the class operator
gives an immediate connection with the standard result for probabilities when there is
decoherence. Indeed, one may have {\it guessed} the form of the class operator from
the standard form of the probabilities, and this is pursued in the next chapter.
However, it is also gratifying that it can be derived in some detail using the complex
potential approach used here.

To assign probabilities, the decoherence functional must be diagonal and we considered
this condition. We found a variety of states for which there is decoherence, under
certain more detailed conditions, which we discussed.

We also noted an interesting and important relationship between decoherence and backflow:
If there is decoherence, the probabilities for crossing must be positive so there cannot
be any backflow. If there is no decoherence, the integrated current may still be positive,
but one can say that if there is backflow there will definitely be no decoherence.
This means that the decoherent histories approach brings something genuinely new to
the arrival time problem: it establishes the conditions under which probabilities can be
assigned and in particular forbids the assignment of probabilities in cases where there
is backflow.

\chapter{Arrival Times, Crossing Times and a Semiclassical Approximation}
\epigraph{Some historians hold that history ... is just one damned thing after another.}{Arnold Toynbee}

\section{Introduction}

We have seen in the previous chapter that arrival time probabilities can indeed be defined via decoherent histories, with the class operators defined in terms of complex potentials. However probably the single most important result was that the class operators corresponding to arriving in a time interval $[t_{1},t_{2}]$ can be written as,
\beq
C(t_{2},t_{1})=P(t_{1})-P(t_{2})\label{6.1}
\eeq
with $P(t)=\theta(\hat x_{t})$, provided $E(t_{2}-t_{1})>>1$ where $E$ is the energy around which the state is strongly peaked.
 In this limit, all dependence on the complex potential has dropped out and the class operator is exceedingly simple. This strongly suggests that there is some simple approximation that can lead us to these class operators, without having to resort to complex potentials. The calculations in the previous chapter would then be important only to establish the conditions under which this approximation holds. One aim of this chapter is to present such a derivation.

Eq.(\ref{6.1}) was derived for the special case of a wavepacket incoming from the right, with entirely negative momentum. This analysis can be changed in a simple way to allow for states consisting entirely of positive momenta, but it is less clear how to handle states with momenta of both signs. 

This is important for the following reason. Consider the case where the initial state is a superposition of two wave packets, identical except that one is the mirror image of the other, ie $x\to-x,p\to-p$. The symmetry here implies that the current at the origin is always zero, even though we would like to say that both wave packets cross the origin. This suggests that crossing probabilities cannot in general be simply related to the current, and it is of interest to know how they may be defined instead.

\begin{figure}[htbp] 
   \centering
   \includegraphics[width=12cm]{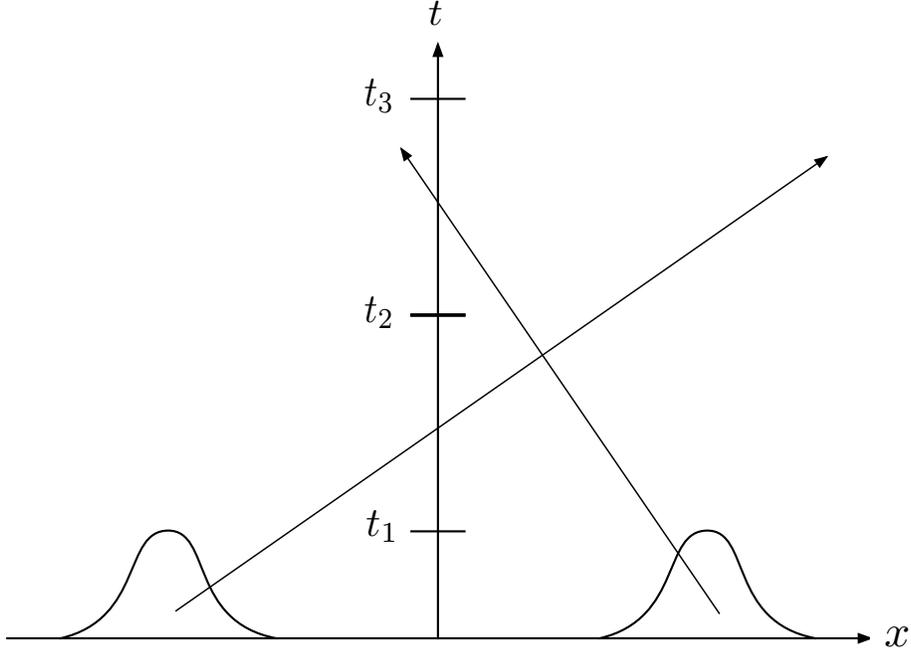} 
   \caption[Crossing Times]{{\em Crossing times. Consider an initial state which is a superposition of a left and a right moving wavepacket. If we can define the probability of crossing for the left and right moving wavepackets separately, it seems reasonable to expect that we can define the crossing time probability for the superposition.}}
   \label{fig:fig6.1}
\end{figure}

Useful insight can be gained by looking at the classical case. There if one has momenta of both signs one can seperate the classical trajectories into ones that cross the origin from the left and ones that cross from the right \cite{MugLev, MugUnpub}. The crossing time distribution, $\Pi(t)$ can therefore be written as
\beq
\Pi(t)=\Pi_{+}(t)+\Pi_{-}(t)=J_{+}(0,t)-J_{-}(0,t)\label{6.1.11},
\eeq
where the current associated with the positive momentum components of the system is given by
\beq
J_{+}(0,t)=\int_{0}^{\infty}dp\frac{p}{m}w_{t}(p,0)
\eeq
and similarly for $J_{-}(0,t)$. Here $w_{t}(p,q)$ is the classical phase space distribution function. This can be written in the alternative form
\beq
\Pi(t)=\int_{-\infty}^{\infty}dp \frac{|p|}{m}w_{t}(p,0)=\int_{-\infty}^{\infty}dp \frac{p\e(p)}{m}w_{t}(p,0)\label{6.1.2}
\eeq
where $\e(p)$ is the signum function.

In the quantum case, however, we do not have a description in terms of trajectories. This means that we cannot justify the procedure used in the classical case. In fact it turns out that even if we could, we would not arrive at an acceptable arrival time distribution. This is because the quantum analogue of Eq.(\ref{6.1.2}) would be
\beq
\Pi_{qm}(t)\stackrel{?}{=}\int dp \frac{|p|}{m}W_{t}(p,0)
\eeq
where $W_{t}(p,0)$ is the Wigner function of the system. However unlike the classical phase space distribution function, the Wigner function need not be positive. Therefore in general this expression will not be positive, so cannot be a probability distribution. It is important to note that this negativity is not due to any ``trajectories moving backwards'' since the modulus of the momentum has been taken. Rather this negativity is the result of quantum interference between wave packets moving in different directions. The second aim of this chapter therefore, is to find the correct class operators to describe arrival times where the initial state is a superposition of left and right moving wavepackets, to compute the crossing probabilities thus obtained, and to check the conditions under which we have decoherence. Our aim will be to demonstrate that if we have a left and a right moving wavepacket for which the arrival time histories would individually decohere according to the analysis of Chapter 5, then the crossing time histories of a superposition of such wavepackets will also be decoherent.

Our analysis follows from the simple observation that although in standard quantum theory there is no notion of trajectory, in decoherent histories the concept of a history is very similar to that of a trajectory. We will use this insight to show how an analysis similiar to the classical case can lead us to an expression for the object representing the probability of crossing a surface from either side.

The rest of this chapter is arranged as follows. In Section \ref{sec:6.2} we present a semiclassical derivation of the crossing class operators previously derived via complex potentials in Chapter 5. In Section \ref{sec:6.3} we address some issues relating to the differences between arrival times and crossing times that are relevant for the rest of this chapter. In Section \ref{sec:6.4} we introduce the crossing time class operators, and then simplify them using a similar semiclassical aproximation to that used in Section \ref{sec:6.2}. In Section \ref{sec:6.5} we compute the crossing probabilities and decoherence conditions using these crossing class operators. We conclude in Section \ref{sec:6.6}. The first part of this chapter, Section \ref{sec:6.2}, is based on Ref.\cite{HaYe2}.

\section{Semiclassical Derivation of the Arrival Time Class Operators}\label{sec:6.2}

In this section we give a semiclassical derivation of Eq.(\ref{6.1}) following Ref.\cite{HaYe2}. We wish to see if there is a simpler way of arriving at the main results of Chapter 5, without recourse to the machinery of complex potentials. Instead we see if some semiclassical approximation can lead to the same conclusions.

We consider an initial state at $t=0$ with $p<0$ concentrated in $x>0$. Suppose the state crosses the origin during a large time interval $[0, \tau]$. It is useful to introduce a discrete set of times
$t_k = k \epsilon $, where $ k = 0,1 \cdots n $ and $ \tau = n \epsilon $.
We also define the projection operators
\bea
P &=& \theta ( \hat x )
\\
\bar P &=& 1 - P = \theta ( - \hat x )
\eea

Consider first the class operator for remaining in $x>0$ (i.e., not crossing $x=0$) during
the time interval $ [ 0, \tau ]$.
We assert that the appropriate class operator is
\beq
C_{nc} = P (t_n) \cdots P(t_2) P(t_1)
\label{6.17}
\eeq
This corresponds to the statement that the particle is in $x>0$ at the discrete set of
times $t_1, t_2, \cdots t_n$,
but its location is unspecified at intermediate times. We know from our studies of the Zeno effect that we must have $\e>>1/E$ to avoid reflection.


We construct the first crossing class operator by partitioning the
histories according to whether they are in $x<0$ or $ x>0$ at the
discrete set of times $t_k$ and noting that the class operators
must sum to the identity. We write
\bea
1 &=& \bar P(t_1) + P(t_1)
\nonumber \\
&=& \bar P(t_1) + \bar P (t_2) P(t_1) + P (t_2) P(t_1).
\eea
Repeating inductively, we obtain
\bea
1 &=& \bar P (t_1) + \sum_{k=2}^n \bar P ( t_{k} ) P (t_{k-1}) \cdots P(t_2) P(t_1)
\nonumber \\
&+& P (t_n) \cdots P(t_2) P(t_1).
\label{6.23}
\eea
We thus identify the first crossing class operator as
\beq
C_k =  \bar P ( t_{k} ) P (t_{k-1}) \cdots P(t_2) P(t_1)
\label{6.24}
\eeq
for $ k \ge 2 $, with $C_1 = \bar P (t_1)$. This clearly describes histories
which are in $ x>0$ at times $ t_1, t_2, \cdots t_{k-1}$ and in $x<0$ at
time $t_k$, so, to within the accuracy set by $\e$,
describe a first crossing between $t_{k-1}$ and $t_k$. The last term
in Eq.(\ref{6.23}) is the non-crossing class operator, $C_{nc}$. We will
actually assume that $\tau$ is sufficiently large that the wave packet
ends up entirely in $x<0$ at large times, so $C_{nc} | \psi \rangle $
is essentially zero. This means that we effectively have
\beq
\sum_{k=1}^n C_k = 1,
\eeq
as required. We will generally be interested in class operators describing
crossing in intervals $[t_{\a}, t_{\a+1}] $ of size $ \Delta = m \epsilon$,
where $ m $ is a positive integer, and these class operators
are simply obtained by summing,
\beq
C_{\a} = \sum_{k \in \alpha} C_k.
\eeq
Last crossing class operators are similarly constructed but will not be required,
as we shall see below.

Consider the strings of identical projection operators $P$ at different
times appearing in Eq.(\ref{6.17}) (and similarly in Eq.(\ref{6.24})).
Given that the final projection $P(t_n)$ is onto $x>0$ and also that the initial
state is localized in $x>0$, it seems reasonable to suppose that
the projections at times $t_1$ to $t_{n-1}$ do not disturb the
evolving state too much, under the condition $ \e > 1/E$ discussed above.
It therefore seems
reasonable to make the approximation
\beq
 P (t_n) \cdots P(t_2) P(t_1) | \psi \rangle \approx P (t_n) | \psi \rangle
\label{6.21}
\eeq
This is easy to understand in a path integral representation. The right-hand side
is in essence the amplitude from an initial state concentrated in $x>0$ to a final point
in $x>0$ at time $t_n$ (up to overall unitary factors). The sum over paths will be dominated by the straight line
path, which lies entirely in $x>0$ at all intermediate times. It will therefore
be little affected by the insertion of additional projections onto $x>0$
at intermediate time.

Using this approximation, the crossing class operator Eq.(\ref{6.24}) operating
on the given initial state may
be approximated as
\beq
C_{k} \approx  \bar P ( t_{k+1} ) P (t_{k})\label{6.nat}
\eeq
Rearranging and using the approximation Eq.(\ref{6.21}) a second time we obtain
\bea
C_{k} &=& P ( t_k ) - P ( t_{k+1})  P ( t_k )
\nonumber \\
& \approx & P ( t_k ) - P ( t_{k+1} )\label{6.29}
\eea
Which is precisely what we set out to show. We have therefore shown that a simple semiclassical approximation, Eq.(\ref{6.21}), can yield the correct class operators without recourse to complex potentials. However, it is not clear from the analysis above when this semiclassical approximation is valid. For this we do need the results of the previous chapter, which tell us that we must have $t_{2}-t_{1}>>1/E$. 
We also note that the semiclassical approximation used above means that there is no
distinction between first and last crossing, so a last crossing class operator
would yield the same result, see the remarks at the end of Section \ref{sec:2.4}

\section{Arrival Times, Crossing Times and First Crossing Times}\label{sec:6.3}

Given the results of the previous section, it is straightforward to write down the class operators corresponding to crossing or not crossing from either side of the origin. Before we do this however we wish to pause a moment to address a subtle issue in the arrival time problem, which is the difference between crossing times and first crossing times. In classical mechanics, for a particle in a potential, a particle may cross the surface $x=0$ many times. One generally defines the {\it arrival time} as the {\it first} crossing of $x=0$. In fact constructing even the classical arrival time distribution for a particle in a confining potential is somewhat involved. Generally one imposes absorbing boundary conditions on the system, and the arrival time distribution can then be taken as the flux at $x=0$ for the modified system with these boundary conditions. For the case of a free particle, of course, the particle only crosses $x=0$ once, and so there is no distinction between crossing and first crossing. 

In the quantum case, a similar problem arises for a particle in a potential and presumably we could also tackle this problem using the complex potential analysis used in Chapter 5, at least for suitably smooth potentials. However even for a free particle in quantum theory, the set of possible paths summed over in the path integral definition of the propagator is dominated by those that cross $x=0$ an infinite number of times (see Chapter 3.) This means that we must distinguish between notions of crossing and first crossing even for a free particle in quantum theory. 

In practice, however, this issue is unlikely to cause real problems. Since we know from the Zeno effect that we can only define arrival times up to some accuracy $\e>>1/E$, it seems reasonable to assume that we will not resolve multiple crossings of the origin by a free particle\footnote{This assumes that the timescale on which these effects are important is given by $1/E$. Since for a wavepacket this is the genuine quantum timescale this seems reasonable, but this will be studied further elsewhere.}. 

The consequence of this is that although we presented the derivation of the class operators above as if we used a single semiclassical approximation, in fact there are two different approximations being used. The first is the approximation,
\beq
C_k =  \bar P ( t_{k} ) P (t_{k-1}) \cdots P(t_2) P(t_1)\approx \bar P ( t_{k} ) P (t_{k-1})\label{6.31}
\eeq
As claimed, this is based on the idea that the propagator is dominated by the straight line path. Alternatively this may be thought of as expressing the equivalence of first crossing and crossing times. In fact this object is an exact class operator, but for a slightly different question, which is ``What is the probability that the particle crossed the origin from the right during the interval $[t_{k-1},t_{k}]$?'' See Fig.(\ref{fig:cross}) for an illustration of this.

\begin{figure}[htbp]
\centering
\subfloat[First Crossing Times]{\includegraphics[width=0.5\textwidth]{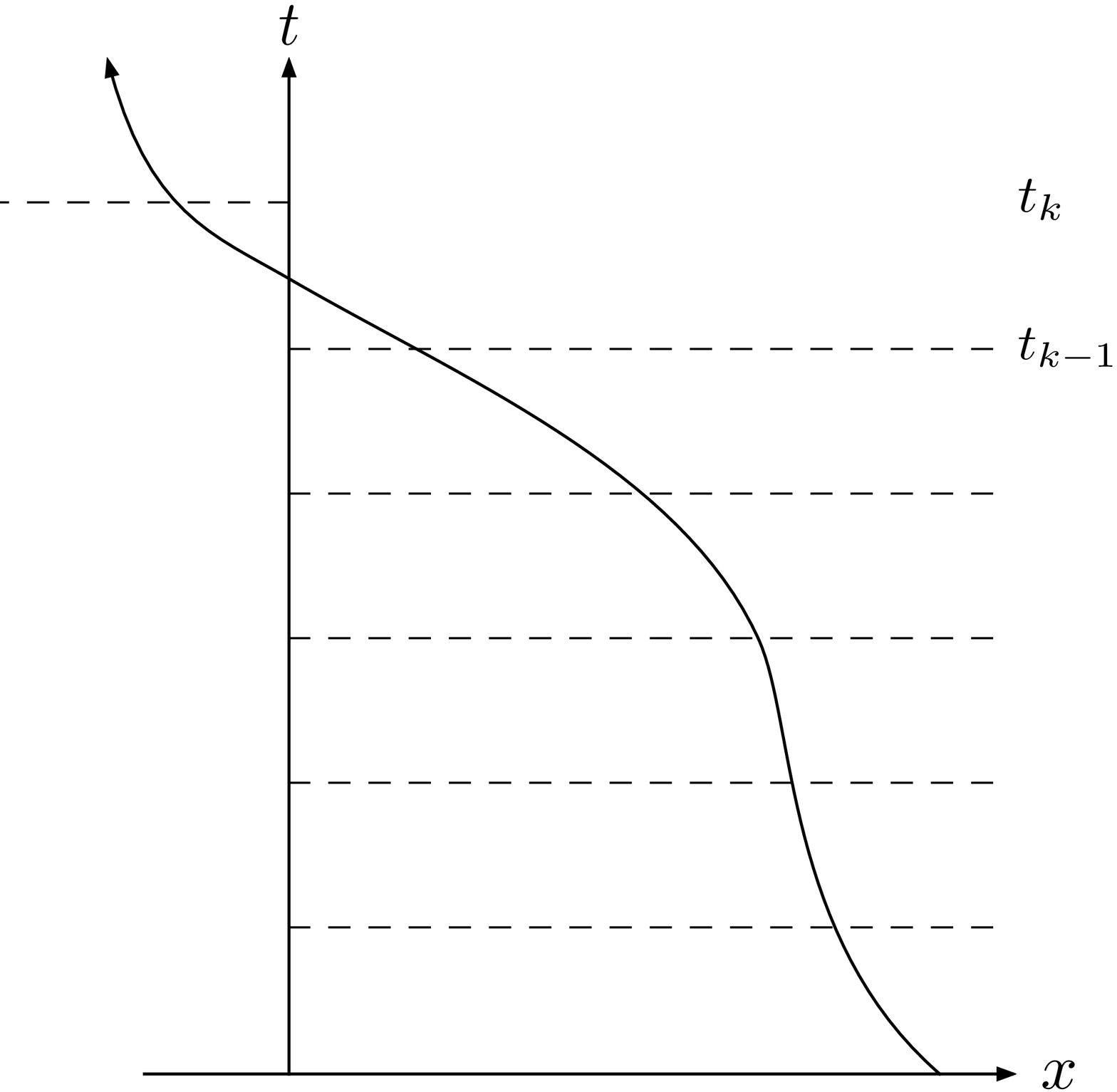} }
\subfloat[Crossing Times]{\includegraphics[width=0.5\textwidth]{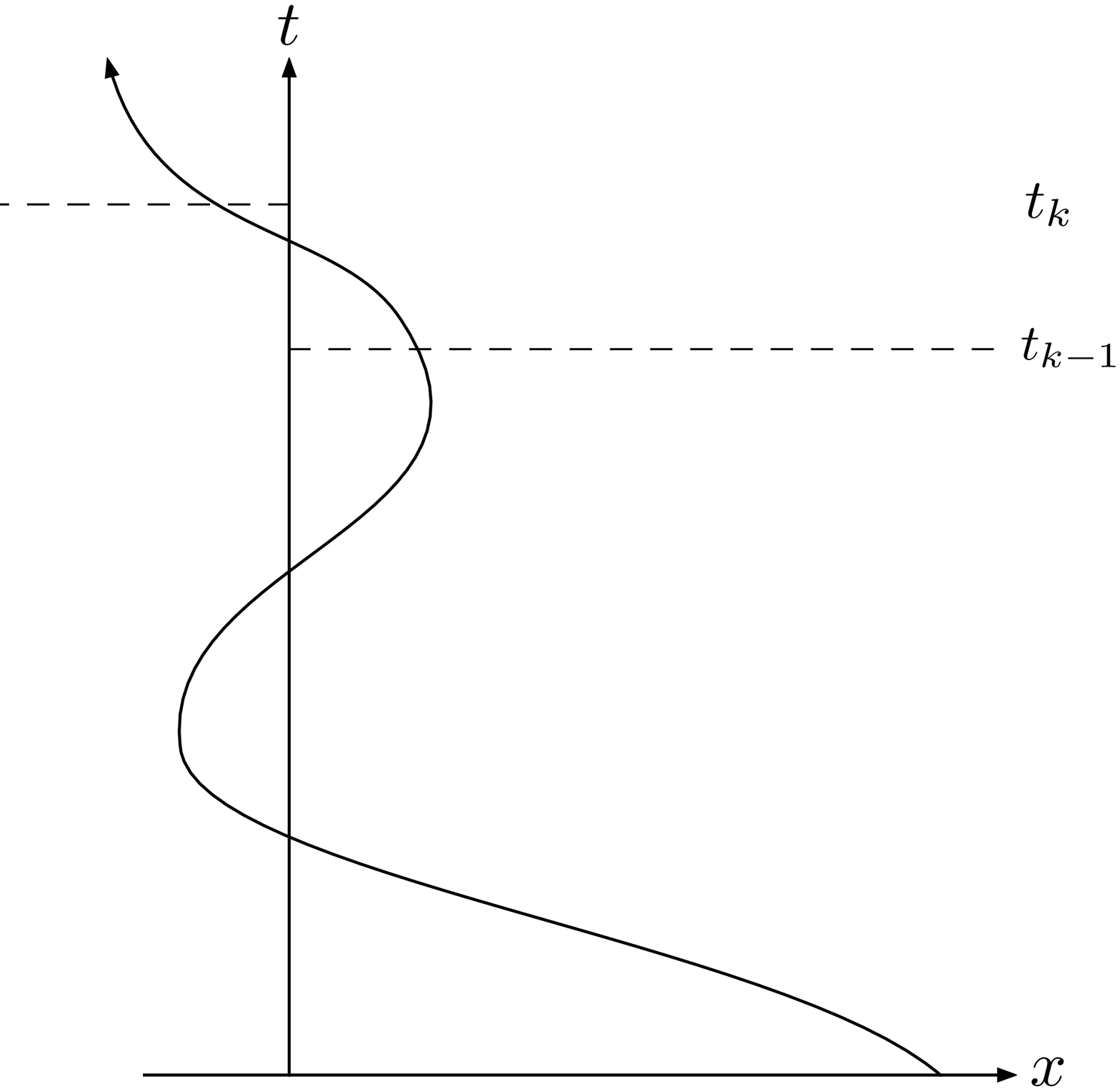}}
\caption[Crossing Times and First Crossing Times]{{\em {\em(a)} First crossing, or arrival time, class operators are sums over paths which {\em first} cross $x=0$ in the interval $[t_{k-1},t_{k}]$. The extra projections at earlier times are included to ensure paths have not previously crossed $x=0$. {\em(b)} In contrast crossing time class operators are sums over paths that cross $x=0$, not necessarily for the first time, in the interval $[t_{k-1},t_{k}]$. }}
\label{fig:cross}
\end{figure}

The second approximation we made was the one leading to Eq.(\ref{6.29}),
\bea
C_{k}&=&\bar P(t_{k+1})P(t_{k})\nonumber\\
&\approx&P(t_{k})-P(t_{k+1})\label{6.cross}.
\eea
This is not so much a {\it semiclassical} approximation (it is not true in the general classical case), rather it depended on us having an entirely a left moving wavepacket. 
In fact the arrival time class operator Eq.(\ref{6.cross}) can be written exactly as,
\beq
P(t_{k})-P(t_{k+1})=\bar P(t_{k+1})P(t_{k})-P(t_{k+1})\bar P(t_{k})\label{6.cross2},
\eeq
so that the approximation Eq.(\ref{6.cross}) corresponds to ignoring crossings from the left. 

These remarks will be important when we come to analyse the class operator for crossing from either side of the origin.

\section{Crossing Class Operators in Decoherent Histories}\label{sec:6.4}

We turn now to the questions of defining class operators for crossing the origin when we have momenta of both signs. The class operators for crossing the origin from the right, $C_{cr}$ and from the left, $C_{cl}$ are clearly given by,
\bea
C_{cr}(t_{k+1},t_{k})&=&\bar P(t_{k+1}) P(t_{k})\label{6.ccr},\\
C_{cl}(t_{k+1},t_{k})&=&P(t_{k+1})\bar P(t_{k})\label{6.ccl}.
\eea
These expressions are simply Eq.(\ref{6.nat}) and its obvious analogue for the case of a right moving wavepacket.
Our provisional class operators for crossing and not crossing from either side during the interval $[t_{k},t_{k+1}]$ are thus,
\bea
\tilde C_{c}(t_{k+1},t_{k})&=&C_{cr}(t_{k+1},t_{k})+C_{cl}(t_{k+1},t_{k})\nonumber\\
&=&\bar P(t_{k+1})P(t_{k})+P(t_{k+1})\bar P(t_{k})\label{6.ccp},\\
\tilde C_{nc}(t_{k+1},t_{k})&=&1-C_{c}(t_{k+1},t_{k})\label{6.cncp},
\eea
Note that, in light of the comments made above, we cannot simplify Eqs.(\ref{6.ccr}) and (\ref{6.ccl}) in the manner of Eq.(\ref{6.cross}). 

In principle Eqs.(\ref{6.ccp}) and (\ref{6.cncp}) provide all the information we need to define crossing probabilities in quantum theory. In practice however these objects are somewhat awkward to work with. What we want is to use the same sort of semiclassical approximations we used to simplify the arrival time class operators to write these crossing class operators in a more useful form.
 
We begin by defining new class operators which are simply the real part of Eqs.(\ref{6.ccp}) and (\ref{6.cncp}),
\bea
C_{c}(t_{1},t_{2})&=&\frac{1}{2}(\tilde C_{c}(t_{1},t_{2})+\tilde C^{\dagger}_{c}(t_{1},t_{2}))\label{6.3.1}\\
C_{nc}(t_{1},t_{2})&=&1-C_{c}(t_{1},t_{2})\label{6.3.2}.
\eea
The use of the Hermitian class operators, Eqs.(\ref{6.3.1}, \ref{6.3.2}) is not strictly necessary, since decoherence will ensure that, when defined, the probabilities computed from the class operators Eqs.(\ref{6.ccp}), (\ref{6.cncp}) are real. However using the Hermitian class operators makes it easier to compare the probabilities with the standard results.

Now consider the class operator for crossing from the right during the interval $[t_{1},t_{2}]$, Eq.(\ref{6.ccr}). Crossing from the right is, semiclassically, the same as restricting to negative momentum and what we want to show is that Eq.(\ref{6.ccr}) is equivalent to some simpler class operator which includes a restriction on the momentum. We begin by noting that Eq.(\ref{6.ccr}) can be written as,
\beq
C_{cr}(t_{1},t_{2})=\bar P(t_{2}) P(t_{1})\theta(\hat p) +\bar P(t_{2})P(t_{1})\theta(-\hat p).
\eeq
We claim that, semiclassically,
\bi
\item $\bar P(t_{2}) P(t_{1})\theta(\hat p)\approx0$,
\item $\bar P(t_{2}) P(t_{1})\theta(-\hat p)\approx(P(t_{1})-P(t_{2}))\theta(-\hat p)$.
\ei
The second approximation follows from Section \ref{sec:6.2}. The first one is clearly true classically, and will be approximately true in the quantum case provided $t_{2}-t_{1}>>1/E$ where $E$ is the energy of the incoming state. These two approximations mean we can write,
\beq
C_{cr}\approx(P(t_{1})-P(t_{2}))\theta(-\hat p).
\eeq
Using the same approximations in $C_{cl}$, we can write down a semiclassical approximation to $\tilde C_{c}$,
\beq
\tilde C_{c}(t_{1},t_{2})=(P(t_{1})-P(t_{2}))(\theta(-\hat p)-\theta(\hat p)).
\eeq
Finally then we take the real part of this expression to obtain, after some rearranging,
\bea
C_{c}(t_{1},t_{2})&\approx&\theta(-\hat p)[P(t_{1})-P(t_{2})]\theta(-\hat p)+\theta(\hat p)[\bar P(t_{1})-\bar P(t_{2})]\theta(\hat p)\label{6.29}\\
C_{nc}(t_{1},t_{2})&=&1-C_{c}(t_{1},t_{2}).\label{6.30}
\eea
Eqs.(\ref{6.29}) and (\ref{6.30}) are the desired class operators for crossing and not crossing the origin from either side during the time interval $[t_{k},t_{k+1}]$. 


As an aside, it is interesting to note that that Eq.(\ref{6.ccp}) may also be obtained from Eq.(\ref{6.cross2}) by the following argument: Start with the class operator corresponding to the integrated current, Eq.(\ref{6.cross2}). Notice that the second term on the right of Eq.(\ref{6.cross2}) is associated with crossing the origin in the opposite direction to the first term. Flip the sign of this second term to correct for this. This leads us to Eq.(\ref{6.ccp}). This is exactly the same argument as leads to Eq.(\ref{6.1.11}) in the classical case.

Let us state this argument again in a different way. The object, $P(t_{1})-P(t_{2})$ is the operator representing the integrated current in standard quantum theory, and is thus the candidate for the arrival or crossing time distribution. Eq.(\ref{6.cross2}) is an identity, but neither of the objects on the right hand side has an obvious interpretation in standard quantum theory (one would have to talk about measurements to give them meaning.) However in decoherent histories we can easily identify these objects as class operators for crossing from the right and from the left. We can therefore perform the classical trick of changing the sign in front of the second term, to arrive at Eq.(\ref{6.ccp}), which is the crossing time class operator.
Eqs.(\ref{6.ccp}) and (\ref{6.cncp}) may therefore be derived either from first principles, or by splitting the object representing the integrated current, Eq.(\ref{6.cross}) into two objects representing crossing from the left and from the right and then making the change of sign in analogy with the classical case, Eq.(\ref{6.1.11}).

\section{Crossing Time Probabilities and Decoherence of Crossing Histories}\label{sec:6.5}

Now we have the class operators for crossing, we can ask about the probabilities for crossing the origin. The easiest way to approach this is to express these probabilities in terms of the momentum representation of the density matrix. For comparison we note that the current can be written in this way as
\beq
J(t)=-\Tr(\dot P(t) \rho)=-\int dp dp' \frac{(p+p')}{2m}\rho(p,p')e^{-i(p^{2}-p'^{2})t/2m}\label{6.4.1}.
\eeq
For a decoherent set of histories, the probabilities are given by,
\bea
p(t_{1},t_{2})&=&\Tr\left(\int_{t_{1}}^{t_{2}}dt\frac{1}{2}\left\{ \theta(\hat p)\dot P(t)\theta(\hat p)-\theta(-\hat p)\dot P(t)\theta(-\hat p)\right\}\rho\right)\nonumber\\
&=&\int_{t_{1}}^{t_{2}}dt \int_{0}^{\infty} dp dp' \frac{(p+p')}{2m} \rho(p,p')e^{-i(p^{2}-p'^{2})t/2m}\nonumber\\
&&-\int_{t_{1}}^{t_{2}}dt \int_{-\infty}^{0} dp dp' \frac{(p+p')}{2m} \rho(p,p')e^{-i(p^{2}-p'^{2})t/2m}\label{6.4.2}.
\eea
This means that the probabilities are in fact functions not of the full density matrix, but rather of the quantity,
\beq
\bar \rho=\theta(\hat p)\rho \theta(\hat p)+\theta(-\hat p)\rho \theta(-\hat p)\label{6.pbar}.
\eeq
This is somewhat unexpected, but very appealing. For a superposition of left and right moving states, our arrival time probabilities are just the sum of the probability for the arrival of the left moving state and that for the right moving one. Interference terms between left and right moving states do not appear. Note the close similarity between the way our class operators partition the current into left and right moving parts and the classical expression, Eq.(\ref{6.1.11}).

We turn now to the question of for which states and intervals we have decoherence. Decoherence between crossing and not crossing during an interval $[t_{1},t_{2}]$ requires,
\beq
D(c,nc)=\Tr(C_{c}(t_{1},t_{2})\rho(1- C_{c}(t_{1},t_{2})))\approx 0
\eeq
which can also be written as
\bea
D(c,nc)&=&\int_{t_{1}}^{t_{2}}dt\Tr\left(\{\theta(-\hat p)\dot P(t)\theta(-\hat p)-\theta(\hat p)\dot P(t)\theta(\hat p)\}\rho\right)\nonumber\\
&&-\int_{t_{1}}^{t_{2}}dtds\Tr\left(\{\theta(-\hat p)\dot P(t)\theta(-\hat p)-\theta(\hat p)\dot P(t)\theta(\hat p)\}\rho\right.\nonumber\\
&&\times \left.\{\theta(-\hat p)\dot P(s)\theta(-\hat p)-\theta(\hat p)\dot P(s)\theta(\hat p)\}\right),\\
D(c,nc)&=&\int_{t_{1}}^{t_{2}}dt\Tr\left(\theta(-\hat p)\dot P(t)\theta(-\hat p)\rho\right)-\int_{t_{1}}^{t_{2}}dtds\Tr\left(\theta(-\hat p)\dot P(t)\theta(-\hat p)\rho\theta(-\hat p)\dot P(s)\theta(-\hat p)\right)\nonumber\\
&&-\int_{t_{1}}^{t_{2}}dt\Tr\left(\theta(\hat p)\dot P(t)\theta(\hat p)\rho\right)+\int_{t_{1}}^{t_{2}}dtds\Tr\left(\theta(\hat p)\dot P(t)\theta(\hat p)\rho\theta(\hat p)\dot P(s)\theta(\hat p)\right)\label{6.5.4}.
\eea
The first line is the decoherence functional for crossing from the right, and the second line is the decoherence functional for crossing from the left. 
This means that a sufficient condition for decoherence of the crossing probabilities is that our initial state is a superposition of a left moving and a right moving wavepacket, each of which satisfy the conditions for decoherence of their corresponding arrival time probabilities. But this is exactly what we set out to prove. Detailed calculation of the two terms in Eq.(\ref{6.5.4}) has been carried out in Chapter 5, and we simply quote the result, that we have decoherence for gaussian wavepackets, and orthogonal superpositions of these, provided they are tightly peaked in momentum, and subject to the requirement
\beq
(t_{2}-t_{1})E>>1,
\eeq
where $E$ is the mean energy of the wavepacket. For more general wavepackets we require an environment to ensure decoherence of the arrival time probabilities \cite{Ye2} and we suspect that the same will be true here. This issue will be addressed elsewhere.

\section{Summary and Discussion}\label{sec:6.6}

Our aims in this chapter were twofold. Firstly we wished to understand whether the origin of the arrival time class operators might be understood by some intuitive argument. We saw that this is indeed the case. Secondly we wished to see how the class operators for the arrival time problem might be extended to deal with wavepackets incident from both sides of the origin. Again we saw that this is possible and that the probabilities defined in this way have the nice feature of being insensitive to the interference terms involving left and right moving wavepackets. 
What is interesting about the analysis of crossing times is that it is not obvious how to do this in the language of complex potentials, yet it seems relatively simple in terms of the class operators defined via strings of projection operators.
In particular it is gratifying that the use of decoherent histories provides a justification for some simple semiclassical manipulations, such as the derivation of Eq.(\ref{6.ccp}), and this suggests it may be usefully employed to address the definition of other time observables. This will be pursued elsewhere.




\chapter{Quantum Arrival Time for Open Systems}
\epigraph{{\it Tempora mutantur, et nos mutamur in illis.}\\ Times change, and we change with them.}{William Harrison}

\section{Introduction}

In the previous two chapters we have discussed how the arrival time distribution may be derived in the framework of decoherent histories. We found that the arrival time distribution may, under the conditions of decoherence, be written as the integral of the current at the origin, thus providing justification for the semiclassical expression we wrote down in Chapter 1, Eq.(\ref{1.16a}). Recall this was,
\bea
p (t_1, t_2)&=&\Tr(P(t_{1})\rho)-\Tr(P(t_{2})\rho)=-\int_{t_1}^{t_2} dt \Tr(P \dot \rho_{t})\nonumber\\
&=&\int_{t_{1}}^{t_{2}}dt J(t),
\label{7.1}\eea
with
\bea
J(t)&=& \int dp dq \frac{(-p)\delta(q)}{m}W_{t}(p,q)\label{7.1w},
\eea
where $P=\theta(\hat x)$, we denote the state at time $t$ by $\rho_{t}$ and the Wigner function $W_{t}(p,q)$ is defined by Eq.(\ref{1.wigdef}).
Recall also the three problems with Eq.(\ref{7.1}), 
\begin{enumerate}
\item It is not generally positive, even when it would be expected to be so classically.
\item It was not derived but postulated.
\item It is not of the general form of the trace of the density matrix times a projection operator or POVM, Eq.(\ref{1.formofprob}). 
\end{enumerate}
Chapters 5 and 6 sought to address these issues by deriving the arrival time distribution from decoherent histories and showing how it reduces to Eq.(\ref{7.1}) in the case where we have decoherence. We succeeded in addressing the first and second points, but no mention was made of the third. In addition we generated the new problem that we could only demonstrate decoherence of histories for particular initial states. 

In this chapter we look at the definition of arrival times for open quantum systems, that is for a particle coupled to some environment. One aim of this is to extend the decoherent histories analysis to a general initial state and we discuss this in more detail below. However the first aim of this chapter is to see whether the semiclassical expression Eq.(\ref{7.1}) is any less problematic for a particle coupled to an environment.
An interesting clue is provided by the expression for the current in terms of the Wigner function, Eq.(\ref{1.8.cur}). Negativity of the Wigner function is a neccessary condition for the negativity of the current (in the sense of backflow, as discussed in \cite{HaYe1}), but it is known that evolution in the presence of an environment typically renders the Wigner function positive after a short time \cite{DoHa}. This suggests that something like Eq.(\ref{7.1}) may be an acceptable, if still heuristic, arrival time distribution for a system coupled to an environment, at least after some time. The first aim of this chapter is to derive the correct analogue of Eq.(\ref{7.1w}) for a system  coupled to an environment, and to prove that it can indeed be regarded as an arrival time distribution after a short time. In particular we show that the arrival time probabilities computed in this way may be written as the trace of the density matrix times a POVM. This then addresses the first and third points above.

However although this may allow us to interpret the current as the arrival time distribution, we are still left with the task of deriving Eq.(\ref{7.1}) from some more fundamental quantity. This was achieved in Chapters 5 and 6 in the context of the decoherent histories approach to quantum theory \cite{GeH1,GeH2,Gri,Omn,Hal2,DoH}, for the case of a free particle and the derivation was shown to hold for states and intervals exhibiting sufficient decoherence. However it was also shown that there exist states for which decoherence of histories cannot be obtained for reasonable levels of coarse graining, and for which therefore we cannot assign a time of arrival probability distribution. States exhibiting backflow are good examples of this. 

Lack of decoherence for a general state is a situation frequently encountered in the literature on decoherent histories. The solution lies in the observation that realistic systems are always coupled to an environment and that as such they are most fundamentally described by open system dynamics. Because by definition an environment consists of degrees of freedom about which we have no knowledge, and over which we have no control, it is natural to coarse-grain over these degrees of freedom. Such coarse-graining generally results in the recovery of approximate classical behaviour, and thus decoherence. We therefore anticipate that decoherence of histories corresponding to arrival times may be achieved for a generic state, provided that particle is coupled to a suitable environment. The second aim of this chapter is to examine this scheme. We show that the probabilities calculated in this way are approximately decoherent and approximately equal to those computed via the analogue of Eq.(\ref{7.1w}), valid in the case of an environment.

Once we introduce an environment, however, the correspondence is  with a classical stochastic theory, rather than with a deterministic one. The classical trajectories may now cross the origin many times due to fluctuations. The arrival time distribution for such a classical theory was computed in Ref.\cite{HaZa}, and is given by
\beq
p(t_{1},t_{2})=\int_{t_{1}}^{t_{2}}dt\int_{-\infty}^{0} dp \int_{-\infty}^{\infty} dq \frac{(-p)}{m}\delta(q) w_{t}^{r}(p,q)\label{stoat},
\eeq
where $w_{t}^{r}(p,q)$ is the initial phase space distribution, evolved with a type of restricted propagator valid in the case of an environment, with boundary conditions, \cite{HaZa}
\beq
w_{t}^{r}(p,0)=0,\quad {\rm for} \quad p>0 \label{wcond}.
\eeq

In Ref.\cite{HaZa} Halliwell and Zafiris presented a decoherent histories analysis of the corresponding quantum case. Although the conclusion they reach is sensible their analysis actually contains a small error. This was the result of a lack of appreciation of the role of the quantum Zeno effect in these calculations, as outlined in Chapter 1 Section 1.6. We will show in this chapter how the analysis can be corrected to give a correct expression for the arrival time probability in decoherent histories.  We shall be interested in a particular limit of this general case where the stochastic trajectories are sharply peaked about the deterministic trajectories we would have in the absence of any environment. In this case the restricted propagator may be replaced with an unrestricted one, and we will show that quantisation in this limit yields an expression of the form Eq.(\ref{7.1}), but with the Wigner function evolved in the presence of an environment. 

It will eventually be necessary to specialise to a particular model of system-environment coupling in order to obtain quantitative results, but we will begin by considering general models which can be written in terms of a master equation of Lindblad form.  This kind of system has been extensively studied in the decoherence literature and the properties of such evolution, such as suppression of interference effects, loss of entanglement and an approximate recovery of classical behaviour have been discussed in \cite{DoHa, HaZo, HaZo1, Zu}. 

This chapter is arranged as follows, in Section \ref{7sec2} we explore some properties of the arrival time distribution for general open systems of Lindblad form, our aim being to derive the corrections to the current resulting from the environmental dynamics. In Section \ref{7sec3} we discuss quantum Brownian motion and specialise the results of Section \ref{7sec2} to this case. In Section \ref{7sec5} we derive some properties of the arrival time distribution for quantum Brownian motion, in particular we prove that the current becomes positive after a finite time. In Sections \ref{7sec6} we briefly recap the expressions for the arrival time distribution that arise in decoherent histories, in particular we show that the decoherence condition is closely linked to the semiclassical approximation. In Section \ref{7sec8} we then examine whether histories corresponding to arriving in different intervals of time are decoherent. We summarise our results in Section \ref{7sec9}. This chapter is based on Ref.\cite{Ye2}.

\section{Arrival Time for Open Quantum Systems}\label{7sec2}

We consider an open quantum system consisting of a free particle coupled to an environment with master equation of the following Lindblad form \cite{lind}, 
\beq
\frac{\partial \rho}{\partial t}=-\frac{i}{\hbar}[H,\rho]-\frac{1}{2}\sum_{j}\left(\{L_{j}^{\dagger}L_{j},\rho\}-2L_{j}\rho L_{j}^{\dagger}\right)\label{master},
\eeq
where $L_{j}$ are the Lindblad operators and $H=H_{0}+H_{1}$ is the free hamiltonian plus a possible extra term that depends on the $L_{j}$. (This term gives rise to frequency renormalisation in oscillating systems.) (In this chapter we explicitly write factors of $\hbar$ to help us see the difference between classical and quantum terms in equations.) Specific forms for the $L_{j}$ may be computed for particular models, but for the moment we leave this general. The Lindblad equation represents the most general master equation for a Markovian system which preserves the properties of the density matrix, in particular positivity.

We now extend the derivation of Eq.(\ref{7.1}) to this system. The probability of crossing during the interval $[t_{1},t_{2}]$ is
\bea
p(t_{1},t_{2})&=&\Tr( P \rho_{t_{1}})-\Tr( P \rho_{t_{2}})=\int_{t_{1}}^{t_{2}}dt \Tr( P\dot \rho_{t})\nonumber \\
&=&\frac{-1}{2m} \int_{t_{1}}^{t_{2}}dt \Tr \left([\hat p \delta(\hat x)+\delta(\hat x)\hat p]\rho_{t} \right)-\frac{i}{\hbar}\int_{t_{1}}^{t_{2}}dt \Tr \left( [H_{1}, P]\rho_{t}\right)\nonumber \\
&& +\frac{1}{2}\sum_{j}\int_{t_{1}}^{t_{2}}dt\Tr\left([L_{j}^{\dagger}, P] L_{j}\rho_{t}+L_{j}^{\dagger}[ P, L_{j}]\rho_{t} \right)\label{curo},
\eea
where $ P=\theta(\hat x)$. The first term is the standard current expression, although with the state evolved according to Eq.(\ref{master}), and we therefore recover Eq.(\ref{7.1}) when all the $L_{j}$ are 0. The second and third terms depend on the Lindblad operators, $L_{j}$ and thus on the form of the system-environment coupling.

To proceed further we specialise to the case where $L$ is a linear combination of $\hat x$ and $\hat p$, $L=a\hat x+i b \hat p$, where $a$ and $b$ are real constants. 
The master equation Eq.(\ref{master}) is then
\beq
\frac{\partial \rho_{t}}{\partial t}=-\frac{i}{\hbar}[H_{0},\rho_{t}]-iab[\hat x,\{\rho_{t},\hat p\}]-\frac{a^{2}}{2}[\hat x,[\hat x,\rho_{t}]]-\frac{b^{2}}{2}[\hat p,[\hat p,\rho_{t}]]\label{me}.
\eeq
Note that this equation is also identical in form to the exact master equation for a particle in a gas environment, given in \cite{Dme}.

To derive the arrival time distribution we could simply substitute $L=a\hat x+ib\hat p$ into Eq.(\ref{curo}), but the algebra is somewhat clumsy, and there are a number of terms which vanish for reasons not immediately apparent from this expression. An equivalent approach is to start from Eq.(\ref{me}), multiplying this expression by $ P$ and taking the trace. Because the second and third terms on the right of this expression have the form $[\hat x,\hat A]$ their contribution is proportional to
\beq
\Tr([\hat x,\hat A] P)=\Tr(\hat A[ P,\hat x])=0,\nonumber,
\eeq
so that only the first and last terms on the right hand side of Eq.(\ref{me}) contribute. The final term on the right of Eq.(\ref{me}) gives a contribution,
\beq
-\frac{b^{2}}{2}\Tr([\hat p,[\hat p,\rho_{t}] P)=-\frac{b^{2}}{2}\Tr([\hat p,[ P,\hat p]]\rho_{t})=-i\hbar b^{2}\Tr([\hat p \delta(\hat x)-\delta(\hat x)\hat p]\rho_{t}),
\eeq
so we arrive at the expression,
\bea
p(t_{1},t_{2})&=& \frac{-1}{2m}\int_{t_{1}}^{t_{2}}dt \Tr \left([\hat p \delta(\hat x)+\delta(\hat x)\hat p]\rho_{t} \right)\nonumber \\
&&-i\hbar b^{2}\int_{t_{1}}^{t_{2}}dt \Tr \left([\hat p \delta(\hat x)-\delta(\hat x)\hat p]\rho_{t} \right)\label{curL}.
\eea

The second term in this expression has a somewhat unusual form and its significance is not immediately clear. We will see below that this term is related to diffusion in position. There is a connection here to a recent paper by Genkin, Ferro and Lindroth \cite{Genkin}, in which the authors sought to examine the effects of an environment on the arrival time distribution. In that paper the authors used the standard expression for the current valid in the unitary case, Eq.(\ref{7.1w}), ignoring the extra terms that arise because of the environment. Although this may be a good approximation when we can neglect the effects of dissipation, it is clear that there may be significant corrections to the current for strong dissipation. They also pointed out that these corrections are equivalent to the presence of extra terms in the continuity equation, although they did not compute these explicitly. For the sake of completeness, and also because it helps to understand the nature of the extra terms in Eq.(\ref{curL}) we will derive them here.

To derive the continuity equation we multiply Eq.(\ref{me}) by $\delta(\hat x-x)$ and perform the trace. If we neglect the final three terms on the right we arrive at the standard result,
\beq
\frac{\partial \rho_{t}}{\partial t}(x,x)+\frac{\partial J}{\partial x}(x,t)=0,
\eeq
with $J(x,t)$ defined by an obvious extension of Eq.(\ref{7.1w}). Turning to the extra terms that result from the inclusion of the environment, the second and third terms vanish in exactly the same way as for the current above, and the correction term is therefore given by,
\beq
-\frac{b^{2}}{2}\Tr([\hat p,[\hat p,\rho_{t}]]\delta(\hat x-x))=2\hbar^{2}b^{2}\frac{\partial^{2} \rho_{t}}{\partial x^{2}}(x,x),
\eeq
so the continuity equation now reads,
\bea
\frac{\partial \rho_{t}}{\partial t}(x,x)+\frac{\partial}{\partial x}\left(J(x,t)+J_{D}(x,t)\right)&=&0,
\eea
where we have identified the diffusive current,
\beq
J_{D}(x,t)=-2\hbar^{2}b^{2}\frac{\partial \rho_{t}}{\partial x}(x,x).
\eeq 
This is a specific example of a modification to a conservation equation resulting from open system dynamics. For a more general discussion of these issues see \cite{Cos}.

The analysis presented above can also be phrased in terms of phase space distributions. The Wigner function corresponding to $\rho_{t}$ is defined as \cite{Wig}
\beq
W_{t}(p,q)=\frac{1}{2\pi\hbar}\int d\z e^{-\frac{i}{\hbar} p \z}\rho_{t}(q+\z/2,q-\z/2).
\eeq 
The Wigner transform of the master equation Eq.(\ref{me}) is
\beq
\frac{\partial W_{t}}{\partial t}=-\frac{p}{m}\dif{W_{t}}{q}+2\hbar^{2}ab\dif{(p W_{t})}{p}+\frac{\hbar^{2}a^{2}}{2}\diff{W_{t}}{p}+\frac{\hbar^{2}b^{2}}{2}\diff{W_{t}}{q}\label{mew}.
\eeq
The first term is the standard unitary term, whilst the second term represents dissipation, and the third and fourth terms represent  diffusion. 
The arrival time distribution can be written in terms of the Wigner function by taking the Wigner transform of Eq.(\ref{curL})
\beq
p(t_{1},t_{2})=\int_{t_{1}}^{t_{2}}dt\int dpdq\left(\frac{(-p)}{m}\delta(q)W_{t}(p,q)+\frac{\hbar^{2}b^{2}}{2}\delta(q) \dif{W_{t}}{q}(p,q)\right)\label{probwe}.
\eeq
Eqs.(\ref{curL}) and (\ref{probwe}) are the sought for generalisation of Eq.(\ref{7.1}) for the case of a particle coupled to an environment.

\section{Quantum Brownian Motion}\label{7sec3}
\label{7sec4}
Quantum Brownian motion \cite{CaLe, HaZo, HaZo1} is a commonly used form for the environment of an open system, partly because it is exactly solvable and partly because in many cases it is a good approximation to a realistic environment. For the quantum Brownian motion model \cite{HaZo} we have Eq.(\ref{me}) with 
\beq
a=\sqrt{2D/\hbar^{2}},\quad b=\frac{\gamma}{\sqrt{2D}}.
\eeq
 Here $D=2m\gamma kT$, where $T$ is the temperature of the environment, and $\gamma$ is a phenomenological damping constant, see \cite{CaLe}. 
Eq.(\ref{me}) may be written as
\bea
\frac{\d \rho_{t}}{\d t}(x,y)&=&\frac{i\hbar}{2m}\left(\frac{\d^{2}}{\d x^{2}}-\frac{\d^{2}}{\d y^{2}}\right)\rho_{t}(x,y)-\frac{D}{\hbar^{2}}(x-y)^{2}\rho_{t}(x,y)\nonumber\\
&&-\gamma(x-y)\left(\dif{}{x}-\dif{}{y}\right)\rho_{t}(x,y)-\frac{\hbar^{2}\gamma^{2}}{D}\left(\dif{}{x}-\dif{}{y}\right)^{2}\rho_{t}(x,y).
\eea

Although we could in principle work with this general case, it is useful to specialise to the case of negligible dissipation. There are two reasons for this, the first is that the analysis is considerably simplified, and this helps us to see the important effects more clearly. The second is that we have a particular aim in mind here, and that is to understand how a sensible classical result emerges from the quantum case. The classical case we have in mind is that of a heavy particle following an essentially classical, deterministic trajectory, but subject to small quantum fluctuations. We therefore restrict our analysis to timescales much shorter than the relaxation time $\gamma^{-1}$, so we can drop the final two terms in the master equation above. 

This master equation may then be solved in terms of the propagator \cite{HaZo1},
\bea
\rho_{t}(x,y)&=&\int dx_{0}dy_{0}J(x,y,t|x_{0},y_{0},0)\rho_{0}(x_{0},y_{0})\\
J(x,y,t|x_{0},y_{0},0)&=&\left(\frac{m}{2\pi\hbar t}\right)\exp\left(\frac{im}{2\hbar t}\left[(x-x_{0})^{2}-(y-y_{0})^{2}\right]\right.\nonumber\\
&&\left.-\frac{Dt}{3}\left[(x-y)^{2}+(x-y)(x_{0}-y_{0})+(x_{0}-y_{0})^{2}\right]\right)\label{qbprop}.
\eea
Taking the same limit in the equation for the Wigner function, Eq.(\ref{mew}), gives
\beq
\frac{\d W_{t}}{\d t}=-\frac{p}{m}\frac{\d W_{t}}{\d q}+D\frac{\d^{2}W_{t}}{\d p^{2}}.
\eeq
Evolution of the Wigner function may also expresed in terms of a propagator \cite{HaZo1},
\bea
W_{t}(p,q)&=&\int dp_{0} dq_{0} K(q,p,t|q_{0},p_{0},0)W_{0}(p_{0},q_{0})\label{wprop}\\
K(q,p,t|q_{0},p_{0},0)&=&N \exp\left( -\a(p-p_{0})^{2}-\b \left(q-q_{0}-\frac{p_{0}t}{m}\right)^{2}\right.\nonumber \\ 
&&\left.+\e(p-p_{0})\left(q-q_{0}-\frac{p_{0}t}{m}\right)\right)\label{prop},
\eea
where $N,\: \a, \: \b,$ and $\e$ are given by
\beq
\a=\frac{1}{D t},\: \b=\frac{3m^{2}}{D t^{3}}, \: \e=\frac{3m}{D t^{2}}, \: N=\left(\frac{3m^{2}}{4 \pi^{2
} D^{2}t^{4}}\right)^{1/2}\label{7prop2}.
\eeq

For later convenience we note that we can make the simple change of variables here, $q_{0}\to q_{0}-p_{0}t/m$, so that
\bea
W_{t}(p,q)&=&\int dp_{0} dq_{0} \tilde K(q,p,t|q_{0},p_{0},0)\tilde W_{0}(p_{0},q_{0})\label{wprop2}\\
\tilde K(q,p,t|q_{0},p_{0},0)&=&N \exp\left( -\a(p-p_{0})^{2}-\b \left(q-q_{0}\right)^{2}+\e(p-p_{0})(q-q_{0})\right)\nonumber\\
\tilde W(p,q)&=&W(p,q-pt/m)\label{wrdef}.
\eea
This form makes it clear that the evolution consists of two effects. The first is a shifting along the classical trajectories, whilst the second is a spreading in phase space. 

It is useful to consider this process in more detail. In the presence of an environment the width of the momentum distribution becomes time dependent, and we have,
\beq
(\Delta p)_{t}^{2}=(\Delta p)_{0}^{2}+Dt\nonumber, 
\eeq 
where $(\Delta p)_{0}$ is the momentum width of the initial state. We recognise an immediate difficulty here. Even for initial distributions consisting entirely of left moving momenta, $W_{t}$ will develop support on $p>0$ under evolution. This means we cannot strictly regard the current as an arrival time distribution, since a typical trajectory will now cross the origin many times. Differently put, the arrival time distribution is strictly a measurement of the first passage time and this is no longer simply related to the current. (Note that although this result is similar to the problems created by backflow, the reasons behind it are very different. The spreading of momentum induced by evolution in an environment is a purely classical effect.)

However, all is not lost. Although the current is no longer strictly the arrival time distribution, it may be a very good approximation to it. This is because the deviation of the current from the ``true'' arrival time distribution will be related to the probability that the state has the ``wrong'' sign momenta. This means that, provided we are in the ``near-deterministic'' limit,
\beq
(\Delta p)_{t}^{2}<<p_{0}^{2},
\eeq 
where $p_{0}$ is the momentum around which the initial state is tightly peaked, the current will still be a very good approximation to the true arrival time distribution. The timescale on which this analysis breaks down is given by the ``stochastic'' time,
\beq
\t_{s}=p_{0}^{2}/D\nonumber.
\eeq
After this time we must therefore revert to using the exact expression for the arrival time given by Eq.(\ref{stoat}). We see from the definition of $D$ that,
\beq
\t_{s}\gamma =\frac{p_{0}^{2}}{2m kT}>>1
\eeq
for the states we are interested in. This means the stochastic time $\t_{s}$ is much longer than the relaxation time $\gamma^{-1}$, so that working in the near-deterministic limit does not impose any further constraints compared with neglecting dissipation.

Returning to our arrival probability, in the limit of negligible dissipation Eq.(\ref{curL}), becomes,
\bea
p(t_{2},t_{1})&=&\int_{t_{1}}^{t_{2}}dt J(t)\label{cat}\\
J(t)&=&\int dq dp \frac{(-p)}{m}\delta(q) W_{t}(p,q)\label{7.3.3}.
\eea
This expression is now identical to the unitary case, Eq.(\ref{7.1w}), but with the Wigner function evolved under quantum Brownian motion. 

We now turn to the question of whether inclusion of an environment helps ensure the positivity of Eq.(\ref{cat}).

\section{Properties of the Arrival Time Distribution in Quantum Brownian Motion}\label{7sec5}

In the introduction to this chapter we noted that evolution in an environment typically renders $W_{t}$ positive after a short time. We now wish to examine the effect of this on our candidate arrival time distribution Eq.(\ref{cat}). To this end we introduce the notation \cite{DoHa}
\beq
\Z=\begin{pmatrix}p\\q \end{pmatrix}=\begin{pmatrix}z_{0}\\z_{1} \end{pmatrix},
\eeq
and also the class of Gaussian phase space functions
\beq
g(\Z;A)=\frac{1}{2\pi|A|^{1/2}}\exp\left(-\frac{1}{2}\Z^{T}A^{-1}\Z \right),
\eeq
where $A$ is a $2\times 2$ positive definite matrix, with determinant $|A|$. $g(\Z;A)$ will be a Wigner function if and only if 
\beq
|A|\geq \frac{\hbar^{2}}{4}.
\eeq
A useful result is that
\beq
\int d^{2}\Z g(\Z_{1}-\Z;A)g(\Z-\Z_{2};B)=g(\Z_{1}-\Z_{2}; A+B)\label{7.4.1}.
\eeq
In this notation we can write the propagation of the Wigner function, Eq.(\ref{wprop}), as 
\beq
W_{t}(\Z)=\int d\Z' g(\Z-\Z';A)\tilde W_{0}(\Z')\label{wpropg},
\eeq
where
 \beq
A=Dt   \begin{pmatrix} 
      2 & t/m \\
      t/m & 2t^{2}/3m^{2} \\
   \end{pmatrix}.
\eeq
Since 
\beq
|A|=\frac{D^{2}t^{4}}{3 m^{2}}\label{deta}
\eeq
after a time
\beq
t=\left(\frac{3}{16}\right)^{1/4}\left(\frac{2m\hbar}{D}\right)^{1/2}=\left(\frac{3}{16}\right)^{1/4}\t_{l}\label{ts}
\eeq
$g(\Z-\Z';A)$ will be a Wigner function, and thus $W_{t}(\Z)$ will be positive because it is equal to the convolution of two Wigner functions. Here $\t_{l}=\sqrt{2m\hbar/D}$ is the localisation time.

This is a useful result. Expressing the current Eq.(\ref{7.3.3}) in this new notation,
\beq
J(t)=\int d\Z \frac{(-z_{0})}{m}\delta(z_{1})W_{t}(\Z)\label{curj},
\eeq
we see that after the time given in Eq.(\ref{ts}), because $W_{t}>0$ the current Eq.(\ref{curj}) will be positive if the state consists purely of negative momenta. This means that after this time Eq.(\ref{cat}) will be a positive arrival time distribution. This holds provided times involved are much smaller than $\t_{s}$, as per the discussion below Eq.(\ref{wrdef}).

Now we turn to examining the properties of the current. We wish to find an expression for the current in the form $\Tr(P \rho)$, where $P$ is a projector or POVM. This would allow us to express the heuristic  arrival time distribution, Eq.(\ref{7.1}), in the same form as other probabilities in quantum theory. Starting from Eq.(\ref{curj}) it is useful to write,
\bea
J(t)&=&\int d\Z\int d\Z' \frac{(-z_{0})}{m}\delta(z_{1}) g(\Z-\Z';A)\tilde W_{0}(\Z'),
\eea
using Eq.(\ref{wpropg}). 
We can deconvolve the propagator into two gaussians using Eq.(\ref{7.4.1}), in particular we will let $A=A_{0}+B$, where $A_{0}$ is a minimum uncertainty gaussian, and $B$ is the remainder.
\beq
A_{0}= \hbar  \begin{pmatrix} 
      s^{2} & 0 \\
      0& 1/4s^{2} \\
   \end{pmatrix}
\eeq
\beq
B=   Dt   \begin{pmatrix} 
      2 & t/m \\
      t/m & 2t^{2}/3m^{2} \\
   \end{pmatrix} - A_{0}.
\eeq
Here $s$ is some real number.
Using the convolution property Eq.(\ref{7.4.1}) we can write the current Eq.(\ref{curj}) as
\bea
J(t)&=& \int d\Z'' \left[\int d \Z \frac{(-z^{0})}{m}\delta(z^{1})g(\Z-\Z'';B)\right]\left[\int d\Z'\tilde W_{0}(\Z')g(\Z''-\Z';A_{0})\right]\nonumber\\
&=& \int d\Z'' \left[\int d \Z \frac{(-z^{0})}{m}\delta(z^{1}+z_{0}t/m)g(\Z-\Z'';B)\right]Q(\Z'')\label{4a},
\eea
where $Q(\Z)$ is the Q-function \cite{Qfn}, and we have undone the change of variables implied in Eq.(\ref{wrdef}). The Q-function can be written as
\beq
Q(\Z)=\frac{1}{\pi}\bra{\Z}\rho \ket{\Z}\nonumber.
\eeq
We can therefore express the current as,
\beq
J(t)=\Tr(F \rho)\label{opcur},
\eeq
where
\bea
F&=&\frac{1}{\pi} \int d\Z'' \ket{\Z''}\bra{\Z''}\left(\int d\Z \frac{(-z_{0})}{m}\delta(z_{1}+z_{0}t/m)g(\Z-\Z'';B)\right)\nonumber\\
&=&\int d\Z P_{\Z}\frac{(-z_{0})}{m}\delta(z_{1}+z_{0}t/m),
\eea
and we have defined the POVM
\beq
P_{\Z}=\frac{1}{\pi}\int d\Z'' \ket{\Z''}\bra{\Z''}g(\Z-\Z'';B),
\eeq
which is clearly a phase space operator localised around $\Z$. $F$ is therefore a smeared version of the  object used to compute the current classically, $-p\delta(x_{t})/m$.
This holds for times
\beq
\left(\frac{3}{16}\right)^{1/4}\t_{l}\leq t<<\t_{s}.
\eeq

Assume for a moment that $B=0$, and so $P_{\Z}=\ket{\Z}\bra{\Z}$. The current would then be given by
\beq
J(t)=\int d\Z \frac{(-z_{0})}{m}\delta(z_{1}-z_{0}t/m)Q(\Z)\label{crq}.
\eeq
Since the Q-function is positive by construction the current computed in this way will be positive to the extent that the Q-function has vanishing support on $p>0$ ($Q(\Z)$ cannot strictly vanish for $p>0$ but it can be exponentially small, and this suffices here. See the comments below Eq.(\ref{wrdef}).) For $B>0$ the Q-function is simply smeared further, and this will preserve the property of positivity. In fact one could imagine {\it postulating} Eq.(\ref{crq}) as the definition of the arrival time even in the unitary case, since it clearly satisfies all the conditions required.

The probability of arriving in the interval $[t_{1},t_{2}]$ is given by the integral of Eq.(\ref{opcur}), and because $P_{\Z}$ is time dependent (via $B$) this will not in general have a simple form. However there is a separation of time scales here, the time scale of evolution of $P_{\Z}$ is seen from Eq.(\ref{deta}) to be $\t_{l}$. So, if $t_{2}-t_{1}<<\t_{l}$ we have approximately
\beq
p(t_{1},t_{2})=\int_{t_{1}}^{t_{2}} dt \Tr(F \rho)\approx \int d\Z\Tr\left[\rho P_{\Z} \int_{t_{1}}^{t_{2}}dt \frac{(-z_{0})}{m}\delta(z_{1}+z_{0}t/m)\right]=\Tr(E\rho)\label{atpovm},
\eeq
where 
\beq
E=\int d\Z P_{\Z}[\theta(z_{1}+z_{0}t_{1}/m)-\theta(z_{1}+z_{0}t_{2}/m)]\label{defE}
\eeq
is a POVM representing arrival at $x=0$ between $t_{1}$ and $t_{2}$.
Compare this expression with Eq.(\ref{7.1}), which may be written,
\beq
p(t_{1},t_{2})=\Tr([\theta(\hat x_{t_{1}})-\theta(\hat x_{t_{2}})]\rho). 
\eeq
The two expressions are very similar, but crucially Eq.(\ref{defE}) is positive for $z_{0}<0$ (ie $p<0$).
In this case therefore, the effect of the environment is simply to smear the unitary result over a region of phase space, given by $P_{\Z}$ computed at $t_{1}$. 

Eqs.(\ref{atpovm}) and (\ref{defE}) form the first significant result of this chapter. Eq.(\ref{atpovm}) expresses the heuristic arrival time distribution, Eq.(\ref{7.1}), as the trace of an operator times a POVM, and thus has the same form as standard expressions for probability in quantum theory. The POVM, Eq.(\ref{defE}), arises because the environment effectively measures the system.

We have therefore discovered a range of times for which the current, Eq.(\ref{curj}), gives a positive arrival time distribution. After a time of order $\t_{l}$ interference effects vanish and we can regard Eq.(\ref{curj}) as the arrival time distribution, which can also be written in the form Eq.(\ref{opcur}). Eventually, however, on a time scale of order $\t_{s}$ the environment causes diffusion in momentum to such an extent that the trajectories of the particle are no longer sharply peaked around the classical trajectory computed in the absense of an environment. Since the particle is still behaving classically after this time there will exist an arrival time distribution of the form Eq.(\ref{stoat}), but our simpler expression Eq.(\ref{curj}) will no longer be a good approximation to this. It is easy to show that $\t_{s}/\t_{l}=E \t_{l}/\hbar$ where $E$ is the energy of the initial state, and thus these expressions are valid for an large interval if $E \t_{l}>>\hbar$.

\section{The Decoherent Histories Approach to the Arrival Time Problem} \label{7sec6}

We turn now to the question of defining arrival times for open systems via decoherent histories. In Chapters 5 and 6  the class operators corresponding to a first crossing of the origin in the time interval $[t_{k-1}, t_{k}]$ were claimed to be
\beq
C_k =  \bar P ( t_{k} ) P (t_{k-1}) \cdots P(t_2) P(t_1),
\label{7.24}
\eeq 
for $ k \ge 2 $, with $C_1 = \bar P (t_1)$ and where $\bar P(t)=\theta(-\hat x_{t})$. These clearly describe histories
which are in $ x>0$ at times $ t_1, t_2, \cdots t_{k-1}$ and in $x<0$ at
time $t_k$, so, approximately,
describe a first crossing between $t_{k-1}$ and $t_k$. 

In Chapter 6 these class operators were then simplified with the help of the following semiclassical approximation
\beq
 P (t_n) \cdots P(t_2) P(t_1) | \psi \rangle \approx P (t_n) | \psi \rangle.
\label{7.21}
\eeq
From this we obtain the class operators for first crossing between $t_{k-1}$ and $t_k$ as
\bea
C_{k}& \approx & \bar P ( t_{k} ) P (t_{k-1})\label{ap1} \\
&=& P ( t_{k-1} ) - P ( t_{k})  P ( t_{k-1} )
\nonumber \\
& \approx & P ( t_{k-1} ) - P ( t_{k} )\label{ap2},
\eea
where we use the semiclassical approximation again to arrive at the last line. 


The key step in making contact with the simple heuristic result of Eq.(\ref{7.1}) was the approximation of the class operators  Eq.(\ref{7.24}) by Eq.(\ref{ap2}). We now want to further explore this approximation, and prove that it is in fact closely related to the decoherence condition. The aim here is partly to make the the argument in Section \ref{sec:6.5} more precise and partly to check whether the inclusion of an environment adds anything new. 

For a free particle, the error in the semiclassical approximation comes from spreading in momentum caused by the chopping of the wavepacket. We have seen in Chapter 4 however, that for projections spaced by a time greater than $\hbar/E$ the probability of the measurements causing the particle to be reflected is very low. 

We want to find some way of quantifying the error induced by this approximation, for this purpose it is sufficient to consider the approximation as used to go from Eq.(\ref{ap1}) to Eq.(\ref{ap2}).
Writing this in terms of the density matrix,
\bea
\Tr( P(t_{k+1})P(t_{k})\rho P(t_{k}))&=&\Tr(P(t_{k+1})\rho)-\Tr(P(t_{k+1})\bar P(t_{k})\rho)-\Tr(\rho \bar P(t_{k}) P(t_{k+1}))\nonumber \\
& &+\Tr(P(t_{k+1})\bar P(t_{k})\rho \bar P(t_{k}))\nonumber  \\
&\approx& \Tr(P(t_{k+1})\rho)\nonumber,
\eea
where the last line defines the semi-classical approximation. Noting that, with $t_{m}\geq t_{k}$
\beq
\left| \Tr(P(t_{m}) \bar P(t_{k})\rho)\right|^{2}\leq \Tr(P(t_{m}) \bar P(t_{k})\rho \bar P(t_{k})).
\eeq A sufficient condition for the validity of the semi-clasical approximation is therefore that the object
\beq
\Delta_{k,m} := \Tr(P(t_{m})\bar P(t_{k})\rho \bar P(t_{k}))\label{delta}
\eeq
is much smaller than 1 for all $t_{m}>t_{k}$.

Now assume for a moment that the semiclassical approximation holds, so that the class operators are given by Eq.(\ref{ap2}). The  probability of first crossing between $t_{k}$ and $t_{k+1}$ is given by
\bea
\Tr(C_{k}\rho C^{\dagger}_{k})&=& \Tr(C_{k}\rho)+2\Tr(\rho  P(t_{k+1}))-\Tr(\rho P(t_{k}) P(t_{k+1}))-\Tr(P(t_{k+1})P(t_{k})\rho)\nonumber.
\eea
However since we are assuming that the semi-classical approximation holds, the last three terms cancel and we obtain
\beq
\Tr(C_{k}\rho C^{\dagger}_{k})=\Tr(C_{k}\rho)=\Tr(\rho C^{\dagger}_{k})\nonumber.
\eeq
Furthermore, consider a general off-diagonal term in the decoherence functional, where without loss of generality we take $t_{k+1}<t_{m}$
\bea
\Tr(C_{k}\rho C^{\dagger}_{m})&=&\Tr(  P(t_{m}) P(t_{k})\rho)-\Tr( P(t_{m}) P(t_{k+1})\rho)\nonumber\\
&&-\Tr( P(t_{m+1}) P(t_{k})\rho)+\Tr( P(t_{m+1}) P(t_{k+1})\rho).
\eea
We see that this vanishes if the semi-classical approximation holds.

What we have shown therefore, is that the semi-classical condition and the decoherence conditions are closely related. If the object given in Eq.(\ref{delta}) is much smaller than 1 then both conditions are satisfied. This means firstly that the class operators are approximately given by Eq.(\ref{ap2}) and secondly that the histories described by these class operators are approximately decoherent.

In general, the probabilities for time of arrival as computed by decoherent histories differ from the heuristic ones obtained by the standard quantum mechanical analysis. We don't expect agreement since, as discussed in the Introduction to this chapter, the heuristic formula, Eq.(\ref{7.1}), is not in general of the canonical form required for genuine quantum mechanical probabilities. It is also the case that decoherent histories only ascribes probabilities to certain sets of histories. However if the object $\Delta_{k,m}$ is small, then we have shown that the decoherent histories analysis reproduces the probabilities computed in the standard, heuristic, way.

As well as clarifying the analysis of \cite{HaYe2} and Chapter 6, note that all of the statements above apply to the case of a particle coupled to an environment. In this case the trace is to be taken over the system and environment, and $P(t_{k})=P(t_{k})_{s}\otimes 1_{\e}$ is a projection onto the degrees of freedom of the system, $s$, tensored with the identity operator on the environmental degrees of freedom, $\e$. This is the case of interest in this chapter.

Since the object $\Delta_{k,m}$ plays a central role in our analysis it is interesting to ask if it has any physical interpretation. The answer is that it does. Recall that our class operator for crossing between $t_{k-1}$ and $t_{k}$ is given by Eq.(\ref{ap2}). In Chapter 6 Eq.(\ref{6.cross2}) we noted that
\beq
P(t_{k-1})-P(t_{k})=\bar P(t_{k})P(t_{k-1})-P(t_{k})\bar P(t_{k-1})\label{kterms}.
\eeq
The approximation leading from Eq.(\ref{ap1}) to Eq.(\ref{ap2}) is equivalent to dropping the second term on the right hand side of Eq.(\ref{kterms}). The error in this approximation can be estimated by computing the probability associated with the class operator  $P(t_{k})\bar P(t_{k-1})$. We see
\beq
\Tr([P(t_{k})\bar P(t_{k-1})]\rho[P(t_{k})\bar P(t_{k-1})]^{\dagger})=\Tr(P(t_{k})\bar P(t_{k-1})\rho \bar P(t_{k-1}))=\Delta_{k-1,k}\nonumber.
\eeq
However Eq.(\ref{kterms}) has a simple physical interpretation: it is the decomposition of the total current into right and left moving parts. The object $\Delta_{k,m}$ is therefore just the probability associated with the right moving current. Classically this is small by construction, since we will choose to work with wavepackets tightly peaked in negative momentum. Quantum mechanically however this term need not be small, indeed the existence of the backflow effect \cite{BrMe} shows that this term can sometimes be larger than the left moving current, even for wavepackets composed entirely of negative momenta. 

The semiclassical condition we are imposing, and by extension the decoherence condition, is therefore stronger than the condition for the current to be positive. Decoherence requires the absence of interference between crossings at different times, whilst the standard heuristic analysis includes some interference effects, provided they are not so large as to render the arrival time probabilities negative. This is a consequence of the fact that although for a suitably semiclassical state histories peaked about the classical trajectories are decoherent, the set of sets of decoherent histories contains many more histories than simply the classical ones. The significance of this will be explored elsewhere.

\section{Decoherence of Arrival Times in Quantum Brownian Motion}\label{7sec8}

\subsection{General Case}

In the previous section we have shown that an arrival time distribution may be derived from decoherent histories and that it agrees with the current provided there is decoherence. We therefore turn now to the question of determining for which states and intervals decoherence is achieved. Recall that our aim is to show that the inclusion of an environment gives rise to decoherence of arrival time probabilities for generic initial states of the system.
We will start our decoherent histories analysis with a discussion of the general case considered in Ref.\cite{HaZa}, but most of our detailed results will concern the near-deterministic limit discussed in the introduction and in the discussion of the Wigner function above.
 
The decoherence functional may be written in path integral form as
\beq
\D(\a,\a')=\int_{\a} \D x \int_{\a'} \D y \exp\left(\frac{i}{\hbar} S[x]- \frac{i}{\hbar}S[y]+iW[x,y]\right) \rho_{0}(x,y)\label{7.3.1},
\eeq
where $\a,\a'$ represent the restriction to paths that are, for example, in $x>0$ at $t_1$ and in $x<0$ at $t_{2}$.  $W[x,y]$ is the Feynman-Veron influence functional phase which summarises the effect of the environment, and is given in the case of negligible dissipation by
\beq
W[x,y]=\frac{i D}{\hbar^{2}}\int dt(x-y)^{2}.
\eeq 
It is this phase which is responsible for the suppression of interference between paths $x(t)$ and $y(t)$ that differ greatly, and produces decoherence. 

The decoherent histories analysis we are about to perform was first attempted by Halliwell and Zafiris in Ref.\cite{HaZa}. Their conclusions are reasonable, but the analysis leading to them in fact contains a small error. This error arose due to a lack of appreciation of the role of the Zeno effect, as discussed in the Introduction. We aim to show here how the analysis may be modified in line with the treatment of the arrival time problem presented here and in Ref.\cite{HaYe2}.

We start from the expression for the density matrix for states that do not cross the origin in a time interval $[0,t]$ (Ref.\cite{HaZa}, Eq.(4.34))
\beq
\rho_{rr}(x_{f},y_{f})=\int_{r}\D x\int_{r}\D y \exp\left[\frac{im}{2\hbar}\int dt (\dot x^{2}-\dot y^{2})-\frac{D}{\hbar^{2}}\int dt(x-y)^{2}\right]\rho_{0}(x_{0},y_{0}).
\eeq
Here the restriction is interpreted as $x,y>0$ at times $\e, 2\e...$. The probability that the state does not cross the origin in this interval is then given by the trace of this expression. There are two interesting limits to be taken here. The first is letting $\e\to0$, and this recovers the notion of the restricted propagator. The second limit is that of strong decoherence effects, so that we can assume that the path integral is tightly peaked in $(x-y)$, this recovers the classical limit. The claim in Ref.\cite{HaZa} is that we can take these limits simultaneously. To see why this is problematic we write the restricted path integral as a product of propagators,
\beq
\rho_{rr}(x_{f},y_{f})=\prod_{k=0}^{n}\int_{0}^{\infty}  dx_{k} dy_{k}J(x_{k+1},y_{k+1},t_{k+1}|x_{k},y_{k},t_{k})\rho_{0}(x_{0},y_{0}),
\eeq
where $t_{k}=k\e$, $t_{0}=0$, $t_{n+1}=t$.
Changing variables to $X=x+y$, $\z=\frac{1}{2}(x-y)$ gives
\bea
\rho_{rr}(x_{f},y_{f})&=&\prod_{k=0}^{n} \int_{0}^{\infty} dX_{k} \int_{-X_{k}}^{X_{k}} d\z_{k} \left(\frac{m}{\pi\hbar \e}\right)\exp\left[\frac{im}{\hbar \e}[(\z_{k+1}-\z_{k})(X_{k+1}-X_{k})]\right. \nonumber\\
&&\left.-\frac{D\e}{3\hbar^{2}}[\z_{k+1}^{2}+\z_{k+1}\z_{k}+\z_{k}^{2}]\right]\rho_{0}(X_{0},\z_{0}).
\eea
Now in order for the product of propagators to be equal to a restricted propagator, we need to take $\e\to0$. However to recover the classical result we need to replace the limits of integration for $\z_{k}$ with $\pm \infty$, which requires that $D\e>>\hbar^{2}/l^{2}$ where $l$ is some length scale. Clearly these two limits are incompatible for finite $D$.

We see the difficulty here immediately, trying to take $\e\to0$ will lead to the Zeno effect. We know from Chapter 4 that for a free particle projections separated by times greater than $\hbar/E$ will not give rise to the Zeno effect, but the issue  is more subtle here, since we have non-unitary evolution. One would expect that the time scale on which projections give rise to reflection to be {\it reduced} in the present case, since a generic system coupled to an environment tends to display classical behavior and the Zeno effect should vanish in the classical limit. The claim in Ref.\cite{HaZa} is essentially (although they did not express it in this way), that the inclusion of the environment is enough to eliminate the Zeno effect entirely. This seems unlikely to be true. This issue will be taken up elsewhere \cite{HaYeZeno}. The more general point is that it is not possible to take $\e\to0$ and still get a good account of arrival times, we should always expect to have to settle for some minimum resolution. For the purposes of recovering classical behaviour, however, this is of no significance, since the timescale on which a classical ``wavepacket'' would cross the origin is $\sim \hbar/\Delta H$ and this is much longer than $\hbar/E$. We therefore do not loose anything classically by leaving $\e$ small but not zero. 

In any case it seems clear that there exists a short timescale on which we can still approximate the limits in the $\z$ integrals above in the manner indicted, but that is short enough to give non-trivial crossing probabilities. Following the steps in Ref.\cite{HaZa} then gives 
\beq
p(t_{1},t_{2})=1-\Tr(\rho_{rr})=\int_{t_{1}}^{t_{2}}dt\int_{-\infty}^{0} dp \int_{-\infty}^{\infty} dq \frac{(-p)}{m}\delta(q) W_{t}^{r}(p,q)\label{stoat2},
\eeq
where the Wigner function, $W^{r}_{t}$ obeys,
\beq
W_{t}^{r}(p,q)=\int dp_{0} dq_{0}K_{r}(p,q,t|p_{0},q_{0},0) W_{0}(p_{0},q_{0}),
\eeq
where $K_{r}$ is the ``restricted'' propagator, defined by
\beq
K_{r}(p,q,t|p_{0},q_{0},0)=\prod_{k=1}^{n}\int dp_{k}\int _{0}^{\infty}dq_{k}K(p_{k},q_{k},t_{k}|p_{k-1},q_{k-1},t_{k-1}),
\eeq
with $t_{k}=k\e$. However because we are now dealing with an essentially classical system, this propagator is well approximated by its continuum limit. 
Understood in this sense, we see that the conclusions of Ref.\cite{HaZa} are in fact correct, even if the argument leading to them is not. The interesting subtleties that arise in this argument relate to the classical limit of the Zeno effect, this will be explored elsewhere \cite{HaYeZeno}.

As important as this general case is, it is of interest to examine the simpler case of near-deterministic evolution, where the classical limit of the arrival time probability is simply expected to be the time integral of the current density. The analysis in this case simplifies considerably and we will explicitly exhibit decoherence of histories for suitable intervals.

\subsection{The Near-Deterministic Limit}
We have previously seen that, in the near-deterministic limit, if there is decoherence then the arrival time probabilities derived from decoherent histories agree with those computed from the current. We therefore turn now to discussing the conditions under which we have decoherence of crossing probabilities.  The general picture we have in mind is illustrated in Fig.(\ref{7fig1}), we have an initial wavepacket defined at $t=0$, evolved in the presence of an environment until time $t_{1}$, and then we wish to compute the probability of crossing between $t_{1}$ and $t_{2}$.
\begin{figure}
    \centering
        \includegraphics[width=4.7in]{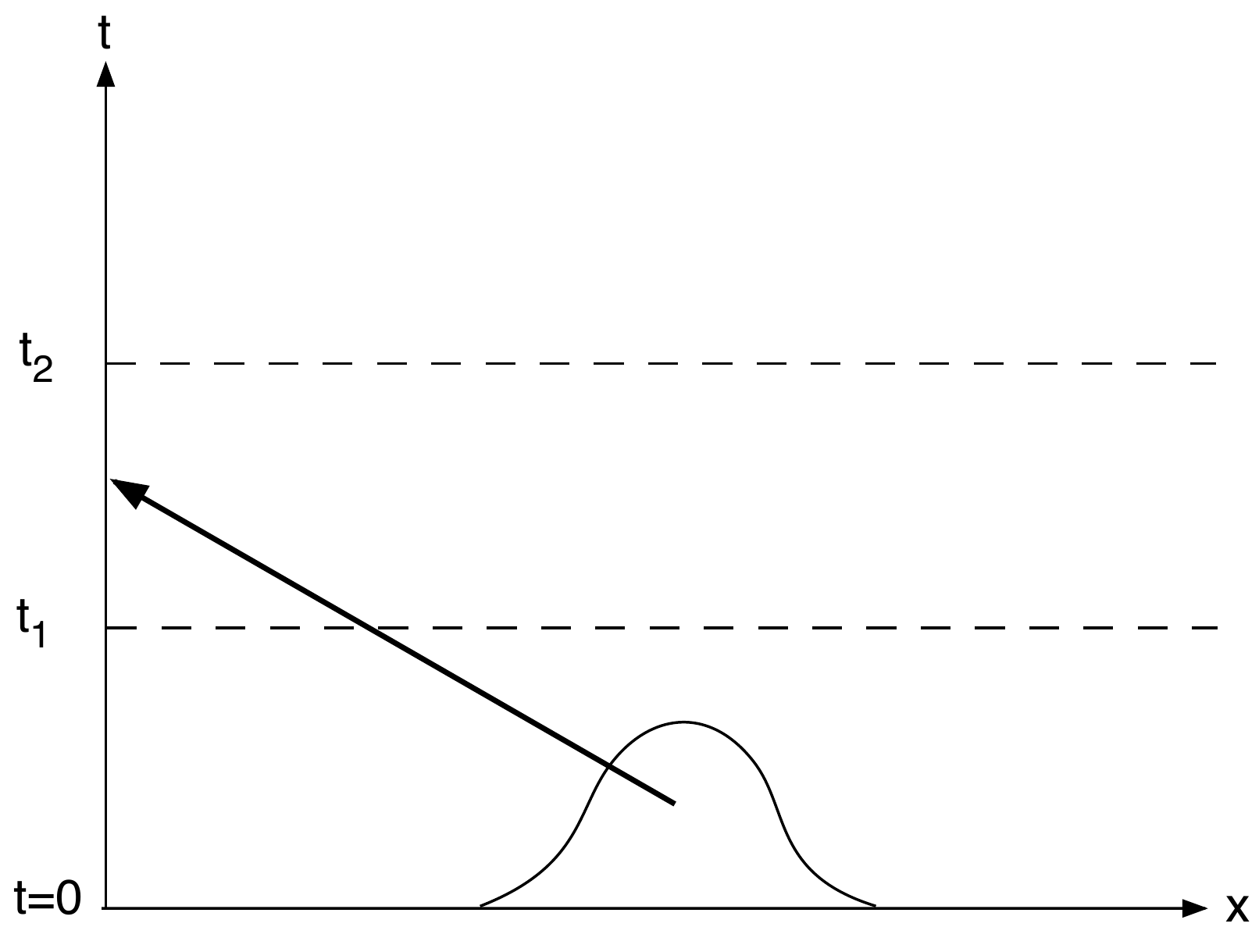}
    \caption[Arrival Times in Terms of Initial States and Propagators.]{The probability that a generic wavepacket arrives between $t_{1}$ and $t_{2}$ can be expressed in terms of the state at $t_{1}$ or, via the propagator, in terms of an initial time $t=0$.}\label{7fig1}
\end{figure}

Our task is to compute the off-diagonal elements of Eq.(\ref{7.3.1}), and this will be rather involved in general. However we are saved from having to do this by the observation in Section \ref{7sec6} about the relation between the semiclassical condition and the decoherence condition. It suffices therefore to compute the quantity $\Delta_{k,m}$ defined in Eq.(\ref{delta}), in the presence of an environment. We anticipate that this will be small simply from the form of Eq.(\ref{7.3.1}). This is because the effect of the environment is to cause the density matrix to become tightly peaked around the classical path and classically, for $p<0$ the probability given by $\Delta_{k,m}$ is zero. We will see how this works in a specific example below. 

We take $k=1, m=2$ without loss of generality, and we drop the subscript from now on
\beq
\Delta=\Tr(P(t_{2})\bar P(t_{1})\rho \bar P(t_{1}))=\int_{\a}\D x \int_{\a} \D y \exp\left(\frac{i}{\hbar} S[x]-\frac{i}{\hbar} S[y]- \frac{D}{\hbar^{2}}\int dt(x-y)^{2}\right) \rho_{t_{1}}(x_{1},y_{1}),
\eeq
where the histories $\a$ are those that start at $x_{1},y_{1}<0$ and finish at $x_{2}>0$. In terms of the density matrix propagator, Eq.(\ref{qbprop})
\bea
\Delta&=&\int_{0}^{\infty}dx_{2}\int_{-\infty}^{0}dx_{1}\int_{-\infty}^{0}dy_{1}J(x_{2},x_{2},t_{2}|,x_{1},y_{1},t_{1})\rho_{t_{1}}(x_{1},y_{1})\nonumber \\
&=& \int_{0}^{\infty}dx_{2}\int_{-\infty}^{0}dx_{1}\int_{-\infty}^{0}dy_{1} \left(\frac{m}{\pi\hbar t}\right)\nonumber \\ &&\exp\left(\frac{im}{2\hbar(t_{2}-t_{1})}((x_{2}-x_{1})^{2}-(x_{2}-y_{1})^{2})-\frac{D(t_{2}-t_{1})}{3\hbar^{2}}(x_{1}-y_{1})^{2}\right)\rho_{t_{1}}(x_{1},y_{1})\nonumber.
\eea
Transforming to new variables
\beq
X=\frac{x_{1}+y_{1}}{2},\quad \z=x_{1}-y_{1},
\eeq
and writing the density matrix in terms of the Wigner function via
\beq
\rho(x,y)=\int_{-\infty}^{\infty}dp e^{\frac{i}{\hbar}p(x-y)}W(p,\frac{x+y}{2}),
\eeq
we obtain
\bea
\Delta&=&-\int_{0}^{\infty}dx_{2}\int_{-\infty}^{0}dX\int_{-X}^{X}d\z \int_{-\infty}^{\infty}dp \left(\frac{m}{2\pi\hbar (t_{2}-t_{1})}\right) \nonumber \\ 
&&\exp\left(\frac{im}{\hbar(t_{2}-t_{1})}\z(X-x_{2})+\frac{i}{\hbar} p\z-\frac{D(t_{2}-t_{1})}{3\hbar^{2}}\z^{2}\right)W_{t_{1}}(X,p)\label{int}.
\eea
We see from this expression that there is a time scale, $\t_{l}=\sqrt{2m\hbar/D}$, set by the environment on which localisation effect are important. This timescale is the same as that on which the current becomes positive, as we saw earier.

There are three cases to explore here, the first is where there is no environment, $D=0$. This is the case covered in Ref.\cite{HaYe1}. 
The second case is the intermediate one, $t_{1}/\t_{l}>>1$ but $(t_{2}-t_{1})/\t_{l}<<1$. This is the most general case in which we can expect to have environmentally induced decoherence.
The final case is where $t_{1}/\t_{l}, (t_{2}-t_{1})/\t_{l}>>1$. This is the case of very strong environmental coupling.  

For the free case, $D=0$ the integrals over $\z$ and $x_{2}$ in Eq.(\ref{int}) may be carried out to give,
\beq
\Delta=\int_{-\infty}^{0}dX \int_{-\infty}^{\infty}dp W_{t_{1}}(X,p)f\left[\frac{X}{\hbar}\left(\frac{mX}{(t_{2}-t_{1})}+p\right)\right]\label{deltaw},
\eeq
where 
\beq
f(u)=\frac{1}{\pi}\int_{u}^{\infty}dy \frac{\sin{y}}{y}.
\eeq 
This expression for $\Delta$ is now identical to Eq.(\ref{5.7.40}) of Chapter 5 and the same conclusions apply. We briefly repeat the analysis here however, since it turns out to be relevant for the second case. 

We note firstly that $f(+\infty)=0, f(-\infty)=1, f(0)=1/2$. Now we assume that our state at $t_{1}$ is of the form
\beq
W_{t_{1}}(X,p)=\frac{1}{\pi\hbar}\exp\left(-\frac{(X-X_{0}-p_{0}t_{1}/m)^{2}}{2\s^{2}}-\frac{2\s^{2}}{\hbar^{2}}(p-p_{0})^{2}\right),
\eeq
and that $\s^{2}$ is large, so the state is tightly peaked in momentum. We can therefore integrate out $p$, setting $p=p_{0}$ to obtain,
\beq
\Delta=\int_{-\infty}^{0}dX \frac{1}{\sqrt{2\pi\s^{2}}} \exp\left(-\frac{(X-X_{0}-p_{0}t_{1}/m)^{2}}{2\s^{2}}\right)f\left[\frac{X}{\hbar}\left(\frac{mX}{(t_{2}-t_{1})}+p_{0}\right)\right].
\eeq
In Chapter 5 it was noted that this type of integral is dominated by values of $X$ for which
 \beq
 0\leq X\leq -\frac{\pi\hbar}{4|p_{0}|},
 \eeq
 providied that $p_{0}^{2}(t_{2}-t_{1})/2m>>1$, and in this range the exponential term is approximately constant. We can approximate this integral then by intergrating from $0$ to $-\pi/4|p_{0}|$, taking $X=0$ in the exponential term and approximating,
 \beq
 f(u)\approx \frac{1}{2}-\frac{u}{\pi}+O(u^{3}).
 \eeq
This gives
\beq
\Delta \approx \sqrt{\frac{\pi}{2}}\frac{\hbar}{8\s |p_{0}|}\exp\left(-\frac{(X_{0}+p_{0}t_{1}/m)^{2}}{2\s^{2}}\right)<<1,
\eeq
there will therefore be decoherence for gaussian wavepackets tightly peaked in momentum, provided their momentum is such that $p_{0}^{2}(t_{2}-t_{1})/2m>>\hbar$, ie $E_{0}(t_{2}-t_{1})>>\hbar$. In Chapter 5 it was argued that this conclusion also holds for orthogonal superpositions of gaussians.

Turning to the intermediate case, since $(t_{2}-t_{1})/\t_{l}<<1$ we can set $D=0$ in Eq.(\ref{int}) and we obtain Eq.(\ref{deltaw}) again. Now however, $W_{t_{1}}(X,P)$ is the initial state evolved with an environment, and since $t_{1}/\t_{l}>>1$ this should be significant. To proceed we write the Wigner function at time $t_{1}$ in terms of the initial state at $t=0$ using Eq.(\ref{wprop})
\beq
\Delta= \int dX_{0} dp_{0} W_{0}(X_{0},p_{0}) F[X_{0},p_{0}]
\eeq
\beq
F[X_{0},p_{0}]=\int_{-\infty}^{0}dX  \int_{-\infty}^{\infty}dp K(X,p,t_{1}|X_{0},p_{0},0) f\left[\frac{X}{\hbar}\left(\frac{mX}{(t_{2}-t_{1})}+p\right)\right].
\eeq
Because the Wigner function propagator is a gaussian the analysis is similar to the first case. Recall that we are assuming $t<<\t_{s}$ and so $p_{0}^{2}/Dt_{1}>>1$, this means we can integrate out $p$, setting $p=p_{0}$. This gives us
\beq
F[X_{0},p_{0}]=\int_{-\infty}^{0}dX N\sqrt{\frac{\pi}{\a}} \exp\left(  -\frac{\b}{4}(X-X_{0}-p_{0}t_{1}/m)^{2})\right)f\left[\frac{X}{\hbar}\left(\frac{mX}{(t_{2}-t_{1})}+p_{0}\right)\right],
\eeq
where, $N,\a,\b$ are defined in Eq.(\ref{7prop2}). Now,
\beq
p_{0}^{2}/Dt_{1}>>1
\eeq
implies that
\beq
\frac{1}{p_{0}^{2}}<<\frac{1}{\hbar^{2}\b}\left(\frac{\t_{l}}{t_{1}}\right)^{4},
\eeq
so that for $t_{1}>\t_{l}$, $1/p_{0}^{2}<<1/\b\hbar^{2}$. This means, again like the case of $D=0$, that the exponential term is roughly constant compared to $f$.  Since we have $p_{0}^{2}(t_{2}-t_{1})/2m>>\hbar$, we follow the same procedure as the $D=0$ case, arriving finally at
\beq
F[X_{0},P_{0}]\approx \frac{\sqrt{\pi}\hbar}{8|p_{0}|}\sqrt{\frac{\b}{4}}\exp\left(-\frac{\b}{4}(X_{0}+p_{0}t_{1}/m)^{2}\right).
\eeq
The value of $\Delta$ now depends on the relationship between the width of the initial state $\s$ and $\b$. However we can obtain an upper bound by ignoring the effects of the exponential term in $F$, this gives
\beq
\Delta \leq\frac{\sqrt{\pi}\hbar}{8|p_{0}|}\sqrt{\frac{\b}{4}}=\frac{1}{16}\sqrt{\frac{2m\hbar}{p_{0}^{2}t_{1}}}\left(\frac{\t_{l}}{t_{1}}\right)<<1.
\eeq

Finally we have the case of strong system-environment coupling.
Since $(t_{2}-t_{1})/\t_{l}>>1$ the integral over $\z$ in Eq.(\ref{int}) will be peaked around $\z=0$, and we can therefore extend the limits of integration to $(-\infty,\infty)$. This integral and the one over $x_{2}$ may then be carried out and we obtain
\beq
\Delta=\int_{-\infty}^{0}dX\int_{-\infty}^{\infty}dp \frac{1}{2}{\rm Erfc}\left[-\sqrt{\frac{3m^{2}}{D (t_{2}-t_{1})^{3}}}\left(X+\frac{p(t_{2}-t_{1})}{m}\right)\right]W_{t_{1}}(X,p),
\eeq
where ${\rm Erfc}$ is the complementary error function \cite{Amb}. 
For large positive values of the argument we have that,
\beq
{\rm Erfc}[u]\approx \frac{e^{-u^{2}}}{u\sqrt{\pi}}\left(1+O\left(\frac{1}{u}\right)\right).
\eeq
Therefore, since the Wigner function is peaked around $p<0$, $\Delta$ will be very small, provided $(t_{2}-t_{1})<<\t_{s}$. This gives us an {\it upper} bound on the time interval, $[t_{1},t_{2}]$, rather than a lower one. The lower time scale is provided by the condition $t_{2}-t_{1}>>\t_{l}=\sqrt{2m\hbar/D}$. This lower time scale is compatible with the condition that the current be positive.

Note however that this lower limit is state independent. There will be states for which arrival time probabilities decohere on much shorter time scales than this, for example the simple cases which decohere in the absence of any environment will continue to do so in the presence of an environment, at least until a time $\sim \t_{s}$.

The key point is that whether or not one can assign arrival time probabilities in decoherent histories depends on the form of the state at the time it crosses the origin.  Environmentally induced decoherence produces mixtures of gaussian states from generic initial ones and thus, after a short time, arrival time probabilities can be defined whatever the initial state. Crucially however it is not necessary for the system to be monitored whilst it is crossing the origin. The smallest time interval,  $\delta t$, over which we can define a decoherent arrival time probability is therefore set by the energy of the system and not the details of the environment and we must have $E \delta t>>\hbar$. This is in agreement with Chapter 5 and Ref.\cite{HaYe1}, and also with the results of earlier works, concerning the accuracy with which a quantum system may be used as a clock \cite{clock}.

In conclusion then, for a general state, decoherence of histories requires that we evolve for a time much greater than $\t_{l}=\sqrt{2m\hbar/D}$ before the first crossing time. This is because this is the time scale on which quantum correlations disappear and our initial state begins to resemble a mixture of gaussian states. After this time, we may define arrival time probabilities to an accuracy $\delta t$, provided only that $E\delta t>>\hbar$. States which start as gaussians may be assigned arrival time probabilities without this initial period of evolution. This is in line with the general result that some coarse-graining is always required to achieve a decoherent set of histories in quantum theory.

\section{Summary and Discussion}\label{7sec9}

In this chapter we have been concerned with deriving an arrival time distribution for open quantum systems and comparing this with the classical result. We began by discussing the generalisation of the current, which is the classical arrival time distribution, to open quantum systems and in particular to the case of quantum Brownian motion. We found that in general the inclusion of an environment leads to extra terms appearing in the expressions for the current, compared with the those valid in the unitary case. However we have shown that in the limit of negligible dissipation these correction terms may be dropped. We then explored the resulting arrival time distribution and showed that it is non-negative after a time of order $\t_{l}$ and that after this time it can be written as the trace of the density matrix times a POVM.

We then turned to the question of deriving this arrival time distribution from the decoherent histories approach to quantum theory.  We extended the decoherent histories analysis of the arrival time problem to the case of a particle coupled to an environment. As expected, the inclusion of an environment produces decoherence of arrival time probabilities for a generic initial state. There are, however, some limitations to the permitted class of histories. For a generic state arrival times can only be specified after an initial time $t>>\t_{l}$. Even after this time arrival times cannot be specified with arbitrary precision, coarse graining over intervals $\delta t>>\hbar/E$ is required to ensure decoherence. 
We showed that the decoherence condition is very closely related to a semiclassical approximation for the evolution of the state, and that both conditions are satisfied if the time between projections is sufficiently large. This is a specific case of a more general connection between decoherence and classical behavior.

Our approach has proceeded at two levels. At the heuristic level the simple generalisation of Eq.(\ref{7.1w}) to the case of a particle coupled to an environment, Eq.(\ref{curj}), is a positive arrival time distribution after a time of order $\t_{l}$. This can also be written as the expectation value of a POVM, Eq.(\ref{opcur}), and thus has the same form as other probabilities in quantum theory. On a more fundamental level, these expressions can be {\it derived} from the decoherent histories approach to quantum theory, where they are seen to be valid for times much later than $\t_{l}$. 

Although the arrival time probabilities computed from decoherent histories agree with the heuristic ones when we have decoherence, their derivation in this way represents a significant advance in our understanding. The great difficulty with regarding the current as the arrival time distribution is the arbitrary way in which one accepts these ``probabilities'' when they are positive, but declines to do so when they are not. Because decoherence is an essential part of the histories formalism, this arbitrariness is replaced with a consistent set of rules governing when probabilities may or may not be assigned. Whilst this may be of no consequence in the setting of a laboratory, it may prove hugely important in the analysis of closed systems, in particular the study of quantum cosmology \cite{jjhqc}.   

Another very interesting result we have presented is that the current becomes strictly positive after a finite time whilst assignment of probabilities in decoherent histories is only possible asymptotically. In some ways this difference between the heuristic analysis and decoherent histories is to be expected. Indeed, the current represents a linearly positive history \cite{Gold}, and it is essentially the condition of linear positivity that we have proven holds after a time of order $\t_{l}$.  It is known that linear positivity is a weaker condition than decoherence \cite{Harlp, Haqp}, and thus it is not surprising that demanding decoherence leads to a stricter limit on the assignment of probabilities than the heuristic analysis. It would be interesting to examine what is gained in this context by imposing decoherence rather than linear positivity. 

There are similarities here with the relationship between the current and Kijowski's arrival time distribution \cite{Kij}, which may be shown to agree in the classical limit, but not more generally. Indeed because it is manifestly positive, one might expect the arrival time distribution computed from decoherent histories to be more closely related to Kijowski's distribution on shorter time scales. These issues will be discussed elsewhere.

\chapter{Quantum Arrival and Dwell Times via an Ideal Clock}
\epigraph{But what minutes!  Count them by sensation, and not by calendars, and each moment is a day.}{Benjamin Disraeli}

\section{Introduction}
\subsection{Opening Remarks}
So far in this thesis we have been concerned with defining time observables without reference to the specific mechanisms by which they might be measured. This is of course reasonable, we do not need to specify the means by which the position of a particle is to be measured, for example, to be able to state that the probability distribution for finding it at a given point $x$ at time $t$ is $|\psi(x,t)|^{2}$. However since obtaining a satisfactory definition of these observables has proven rather tricky, it is worth asking to what extent the {\em ideal} quantities we have been discussing so far are related to the {\em measured} quantities obtained in some particular measurement set up. As well as this, we mentioned in Chapter 1 that one way of {\em defining} time observables was via a particular measurement model. The basic theory is outlined in Section 1.5, we look for an arrival or dwell time distribution of the form Eq.(\ref{1.8.best}), and then try to extract the ideal distribution which is independent of the measurement model. In this chapter we use a clock model to define arrival and dwell times and compare the results with standard semiclassical expressions. 

\subsection{Dwell Time}

Although we have talked at length in this thesis about the arrival time problem, we have not so far mentioned the dwell time problem in any detail. Since we will use our model clock in this chapter to  look at both arrival and dwell times, it is useful to summerise some basic facts about dwell time here. For more details see Ref.\cite{dwell}.

The dwell time distribution is the probability $ \Pi (t) dt $ that
the particle spends a time between $[t,t+dt]$ in the interval
$[-L,L]$. One approach to defining this is to use the dwell time
operator,
\beq \hat
T_{D}=\int_{-\infty}^{\infty}dt  \chi(\hat x_{t})\label{8.dwellop},
\eeq
where  $\chi(x)$ is the characteristic function of the region
$[-L,L]$ \cite{dwell}. Since $T_{D}$ is self-adjoint the distribution $\Pi (t)$ can then be written as
\beq
\Pi(t)=\bra{\psi_{0}}\delta(t-\hat T_{D})\ket{\psi_{0}}.\label{8.dwell} 
\eeq
In the limit $|p|L\gg1$, where $p$ is the momentum of
the incoming state, the dwell time operator has the approximate
form $ \hat T_{D}\approx {2mL } / {|\hat p|} $ so that the
expected
semiclassical form for the dwell time distribution is
\beq
\Pi(t)=\bra{\psi_{0}}\delta \left( t- \frac {2mL }  {|\hat p|}
\right)\ket{\psi_{0}}.
\label{8.dwellsemi}
\eeq
This is the form against which we shall compare our clock model.


\subsection{Clock Model}

In this chapter we will derive arrival and dwell time
distributions by coupling the particle to a model clock. We denote the
particle variables by $(x,p)$ and those of the clock by
$(y,p_{y})$.  We denote the initial states of the particle and clock
by $\ket{\psi}$, $\ket{\phi}$, respectively, and the total system state by $\ket{\Psi}$.
We couple this clock to the particle via the
interaction $H_{I}=\l  \chi(\hat x) H_C$. The total Hamiltonian
of the system plus clock is therefore given by,
\beq
H=H_{0}+\l  \chi(\hat x) H_C \label{8.clk}
\eeq
where $H_{0}$ is the Hamiltonian of the particle. Here $\chi$ is
the characteristic function of the region where we want our clock
to run, so that $ \chi (x) = \theta (x)$ for the arrival time
problem and $\chi (x) = \theta (x+L ) \theta (L-x)$ for the
dwell time problem.
The operator,
\beq
H_{c}=H_{c}(\hat y, \hat p_{y})
\eeq
describes the details of the dynamics of the clock and we assume
that it is such that the clock position $y$ is the measured time.
The physical situation is depicted in Figure (\ref{fig:8.1}).
\begin{figure}[htbp] 
   \centering
   \includegraphics[width=5in]{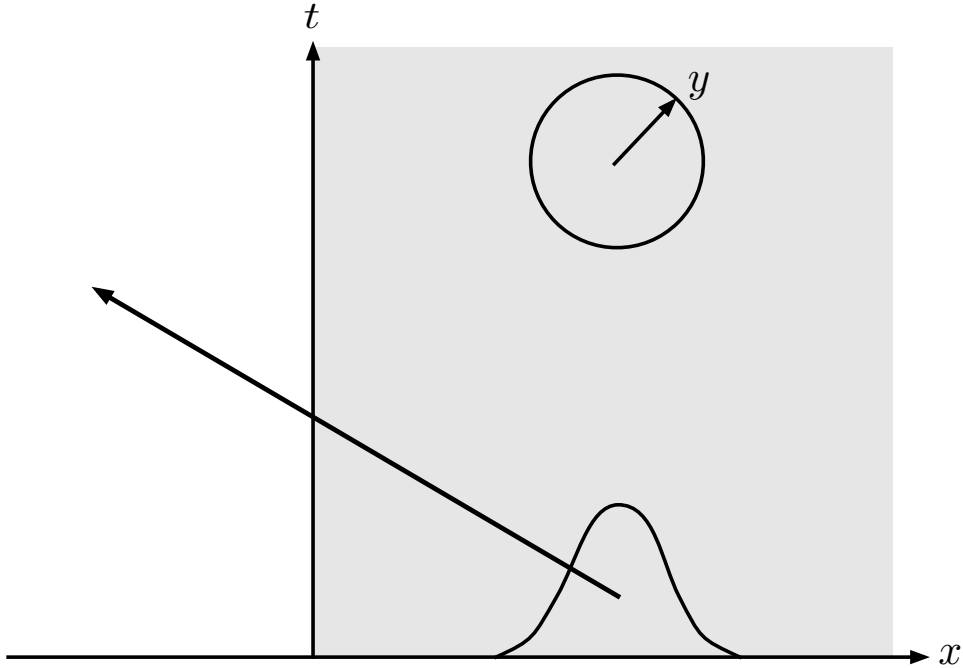}
   \caption[The Arrival Time Problem Defined Using a Model Clock.]{{\em The arrival time problem defined using a model clock. The clock runs while the particle is in $x>0$.}}
   \label{fig:8.1}
\end{figure}
For the moment we will assume only that the clock Hamiltonian is
self adjoint, so that it may be written in the following form,
\beq
H_{c}=\int \ d\h \ \h\ket{\h}\bra{\h}
\eeq
where the $\ket{\h}$ form an orthonormal basis for the Hilbert
space of the clock. Later on we will restrict $H_{c}$ further by
considering the accuracy of the clock. We will also quote some
results for the special choice of $H_{c}=\hat p_{y}$, whose action
is to simply shift the pointer position $y$ of the clock in
proportion to time. This is the
simplest and most frequently used choice for the clock
Hamiltonian.
The physical relevance of this and other clock models is discussed
in Refs. \cite{clockchapter, clock}.

Our aim, for both arrival and dwell times,
is to first solve for the evolution of the combined system of
particle and clock. We write this as
\bea
\Psi(x,y,\t)&=&\bra{x,y}e^{-iH\t}\ket{\Psi_{0}}
\nonumber \\
&=&\bra{x,y}\exp\left(-iH_{0}\t-i\l \chi(\hat x)H_{c}\t\right)\ket{\psi_{0}}\ket{\phi_{0}}\nonumber\\
&=&\int d\h\brak{y}{\h}\brak{\h}{\phi_{0}}\bra{x}\exp(-iH_{0}\t-i\l \chi(\hat x)\h \t)\ket{\psi_{0}}\label{8.psifinal}
\eea
and we then solve for the propagator in the integrand using path
integral methods. We will take the total time $\t$ to be
sufficiently large that the wave packet has left the region
defined by $\chi (x)$.
We will then compute the final distribution of the pointer
variable $y$, which is
\beq
\Pi(y)=\int_{-\infty}^{\infty} dx|\Psi(x,y,\t)|^{2}\label{8.pred}.
\eeq
Our main aim is to show that the predictions of the clock model
Eq.(\ref{8.pred}) reduce, in certain limits, to the standard forms
described above. Our main tools in solving Eq.(\ref{8.psifinal}) will be those we have used throughout this thesis, namely the path decomposition expansion and the semiclassical approximation it suggests. These results are covered in Chapter 2.

\subsection{Connections to Earlier Work}

Clock models of the type Eq.(\ref{8.clk}) for arrival and dwell
times have been studied by  numerous previous authors, including
Peres \cite{clock}, Aharanov et al
\cite{clock2}, Hartle \cite{Hartleclock} and Mayato et al
\cite{clockchapter}. These studies are largely focused on the
characteristics of clocks.  Refs.\cite{clock,clock2} are the works
perhaps most closely related to the present work. They  concentrate 
on the case of a clock Hamiltonian linear in momentum, with some elaborations
on this basic model in the case of Ref.\cite{clock2}. Here, we focus on a different issue
not addressed by these works, namely
the dependence of the distribution Eq.(\ref{8.pred}) on
the initial state of the particle, for reasonably general clock
Hamiltonians. In particular, we determine the extent to which the standard
semiclassical forms derived above are obtained for general initial
states of the particle.
We also use path integral methods to perform the calculations,
in contrast to the scattering methods used in most of the previous
works.

\subsection{This Chapter}

The rest of this chapter is arranged as follows. In Section \ref{sec8.3}
we compute Eqs.(\ref{8.psifinal}), (\ref{8.pred}) for the arrival
time problem, and similarly in Section \ref{sec8.4} for the dwell time problem.
We conclude in Section \ref{con8}. This chapter is based on Ref.\cite{Yeclocks}.


\section{Arrival Time Distribution from an Idealised Clock}\label{sec8.3}

We now turn to the calculation of the arrival time distribution, Eq.(\ref{8.pred}),
recorded by our model clock. Using the path decomposition expansion in the form Eq.(\ref{PDX81})
the state of the system Eq.(\ref{8.psifinal}) can be written as,
\bea
\Psi(x,y,\t)&=&\bra{x,y}e^{-i(H_{0}+\l \theta(\hat x) H_{c})\t}\ket{\Psi_{0}}\nonumber\\
&=&\frac{1}{m}\int d\h\brak{y}{\h}\brak{\h}{\phi_{0}}\nonumber\\
&&\times\int_{0}^{\t}dt \bra{x}\exp(-i(H_{0}+\l \h\theta(\hat x))(\t-t))\delta(\hat x)\hat p\exp(-i(H_{0}+\l \h)t)\ket{\psi_{0}}.\label{8.atstart}
\eea
We can  simplify this expression in two different regimes,
the weak coupling regime of $E\gg\l \h$, and the strong coupling regime of $E\ll\l \h$.

\subsection{Weak Coupling Regime}\label{sec8.3.1}

In the limit $E\gg\l \h$ we can make use of the semiclassical approximation to the PDX formula. This yields,
\bea
\Psi(x,y,\t)&=&\frac{1}{m}\int d\h\brak{y}{\h}\brak{\h}{\phi_{0}}\nonumber\\
&&\times\int_{0}^{\t}dt \bra{x}\exp(-iH_{0}(\t-t))\delta(\hat x)\hat p\exp(-i(H_{0}+\l \h)t)\ket{\psi_{0}}.\label{8.begin}
\eea
This means that the arrival time distribution is
\bea
\Pi(y)&=&\frac{1}{m^{2}}\int d\h d\h'\brak{\phi_{0}}{\h'}\brak{\h'}{y}\brak{y}{\h}\brak{\h}{\phi_{0}}\nonumber\\
&&\times \int_{0}^{\t}dt dt'\bra{\psi_{0}}\exp(i(H_{0}+\l \h')t')\hat p \delta(\hat x)\exp(-iH_{0}(t'-t))\delta(\hat x)\nonumber\\
&&\times\hat p \exp(-i(H_{0}+\l \h)t)\ket{\psi_{0}}\label{8.3.3}
\eea
To proceed, we first note that for any operator $\hat A$, we have
 \beq
 \delta(\hat x)\hat A\delta(\hat x)=\delta(\hat x)\bra{0}\hat A\ket{0}.
 \eeq
Using this in Eq.(\ref{8.3.3}) gives,
\bea
\Pi(y)&=&\frac{1}{m^{2}}\int d\h d\h'\brak{\phi_{0}}{\h'}\brak{\h'}{y}\brak{y}{\h}\brak{\h}{\phi_{0}}\nonumber\\
&&\times \int_{0}^{\t}dt dt'\bra{\psi_{0}}\exp(i(H_{0}+\l \h')t')\hat p \delta(\hat x)\hat p \exp(-i(H_{0}+\l \h)t)\ket{\psi_{0}}\nonumber\\
&&\times\bra{0}\exp(-iH_{0}(t'-t))\ket{0}\label{8.3.5}.
\eea
We see here the appearance of the combination $\hat p \delta(\hat x)\hat p$, and the main challenge is to show how this turns into the current operator, $\delta(\hat x)\hat p + \hat p \delta(\hat x)$.

Next we rewrite the integrals using
\beq
\int_{0}^{\t}dt dt'=\int_{0}^{\t}dt\int_{t}^{\t}dt'+\int_{0}^{\t}dt'\int_{t'}^{\t}dt.
\eeq
In the first term we set $u=t$, $v=t'-t$, and in the second we set $u=t'$, $v=t-t'$ to obtain,
\bea
\Pi(y)&=&\frac{1}{m^{2}}\int d\h d\h'\brak{\phi_{0}}{\h'}\brak{\h'}{y}\brak{y}{\h}\brak{\h}{\phi_{0}}\int_{0}^{\t}du\int_{0}^{\t-u}dv\nonumber\\
&&\times\left\{\bra{\psi_{0}}\exp(i(H_{0}+\l \h')u)\hat p \delta(\hat x)\hat p\exp(-i(H_{0}+\l \h)(u+v))\ket{\psi_{0}}\bra{0}\exp(iH_{0}v)\ket{0}\right.\nonumber\\
&&\left.+\bra{\psi_{0}}\exp(i(H_{0}+\l \h')(u+v))\hat p\delta(\hat x)\hat p \exp(-i(H_{0}+\l \h)u)\ket{\psi_{0}} \bra{0}\exp(-iH_{0}v)\ket{0}\right\}\nonumber\\
\eea
Since we take the time $\t$ to be large, we can extend the upper limits of the integrals to infinity. The integral over $v$ can then be carried out, to give,
\bea
\Pi(y)&=&\int d\h d\h'\brak{\phi_{0}}{\h'}\brak{\h'}{y}\brak{y}{\h}\brak{h}{\phi_{0}}\nonumber\\
&&\times\frac{(-1)}{2m}\int_{0}^{\infty}du \bra{\psi_{u}}e^{i\l \h'u}\big(\hat p \delta(\hat x)+\delta(\hat x) \hat p\big) e^{-i\l \h u}\ket{\psi_{u}}\nonumber\\
\Pi(y)&=&\frac{(-1)}{2m}\int_{0}^{\infty}du |\Phi(y,u)|^{2}\bra{\psi_{u}}\big(\hat p \delta(\hat x)+\delta(\hat x) \hat p\big) \ket{\psi_{u}}\nonumber\\
&=&\int_{0}^{\infty} dt |\Phi(y,t)|^{2} J(t)
\label{8.3.8}
\eea
where
 \beq
\Phi(y,t)=\int d\h\brak{y}{\h}\brak{\h}{\phi_{0}}e^{-i\l \h t}=\bra{y}e^{-i\lambda H_{c}t}\ket{\phi_{0}}
\eeq
is the wavefunction of the clock, and $J(t)$ is the current, Eq.(\ref{1.8.cur}).

This form shows that, in the weak coupling limit, our arrival time probability
distribution yields the current, but smeared with a function depending on the clock state.
We thus get agreement with the expected result, Eq.(\ref{1.8.best}).
Note that the physical quantity measured, the current, is not
affected by the form of the clock Hamiltonian.

Although the form Eq.(\ref{8.3.8}) holds for a wide class of clock Hamiltonians, not all choices make for equally good clocks. To further restrict the coupling $H_{c}$ we require that different arrival times may be distinguished up to some accuracy $\delta t$. For this to be the case we require that the clock wavefunctions corresponding to different arrival times are approximately orthogonal, so that
\beq
\int dy \Phi^{*}(y,t') \Phi(y,t)\approx\begin{cases}1 &\mbox{ if } t\approx t'\\
0 &\mbox{ otherwise}\end{cases}
\eeq
We easily see that,
\bea
\int dy \Phi^{*}(y,t') \Phi(y,t)&=&\int d\h d\h'dy \brak{\phi_{0}}{\h'}\brak{\h'}{y}\brak{y}{\h}\brak{\h}{\phi_{0}}e^{i\l (t' \h'-t \h)}\nonumber\\
&=&\int d\h|\phi_{0}(\h)|^{2}e^{-i\l \h\delta t}\label{8.acc}
\eea
where $\delta t=t-t'$. Clearly this expression is equal to 1 if $\delta t=0$. Suppose now that $|\phi_{0}(\h)|^{2}$ is peaked around some value $\h_{0}$ with width $\sigma_{\h}$. This integral will approximately vanish if
\beq
\l  \sigma_{\h} \delta t>>1,
\eeq
and so the resolution of the clock is given by $1/\l\sigma_{\h}$.
The relationship between $t$ and the pointer variable $y$
will depend on the specific model.
It is easily seen that a clock with good characteristics may be
obtained using, for example, a free particle with a Gaussian
initial state. But clocks with more general Hamiltonians
can also be useful if they evolve an initial Gaussian along
an approximately classical path (as many Hamiltonians do).
See Refs.\cite{clock,clock2,Hartleclock,clockchapter}
for further discussion of clock characteristics.

For the special case $H_{c}=\hat p_{y}$, $\ket{\h}=\ket{p_{y}}$,  the expression for the arrival time distribution simplifies, since
\beq
\Phi(y,t)=\int \frac{dp_{y}}{\sqrt{2\pi}}e^{ip_{y}(y-\l t)}\tilde\phi_{0}(p_{y})=\phi_{0}(y-\l t)
\eeq
The time is related to $y$ by $ t = y/\lambda$ and the
expected form Eq.(\ref{1.8.best}) then becomes a simple
convolution.

\subsection{Strong Coupling Regime}
\subsubsection{Special Case: $H_{c}=\hat p_{y}$}

We now turn to the limit of strong coupling between the particle and clock. The analysis of the case of general clock Hamiltonian is rather subtle, so before we tackle this we first examine the special case where the clock Hamiltonian is linear in the momentum. That is, we have,
\beq
H_{c}=\hat p_{y}=\int dp_{y}\;p_{y}\ket{p_{y}}\bra{p_{y}}.
\eeq
We start from Eq.(\ref{8.atstart}), and insert a complete set of momentum states for the particle, $p$ to obtain,
\bea
\Psi(x,y,\t)&=&\frac{1}{m}\int \frac{dp_{y} dp}{\sqrt{2\pi}}\brak{y}{p_{y}}\tilde\phi_{0}(p_{y})\exp(-i(E+\l p_{y})\t)p\brak{p}{\psi_{0}}\nonumber\\
&&\times\int_{0}^{\t}dt \bra{x}\exp(-i(H_{0}-\l p_{y}\theta(-\hat x))t)\ket{0} \exp(iEt),
\eea
where $\tilde \phi_{0}(p_{y})$ is the initial momentum space wavefunction of the clock, and $E=p^{2}/2m$ is the kinetic energy of the particle.
Note the appearance of the momentum $p$ in the integrand.
The expression involving the integral over $t$ has been computed previously using the final crossing PDX \cite{Hal2, HaYe1}. In the limit $\t\to\infty$ it is given by,
\bea
\int_{0}^{\infty}dt \bra{x}\exp(-i(H_{0}-\l p_{y}\theta(-\hat x))t)\ket{0} \exp(iEt)
&=&\sqrt{\frac{2m}{\l p_{y}}}\exp\left(-ix\sqrt{2m(E+\l p_{y})}\right)\nonumber.
\eea

We can now write our probability distribution for $y$. Carrying out the $x$ integral we obtain,
\bea
\Pi(y)&=&\int dp_{y} dp_{y}'dpdp'\tilde \phi_{0}^{*}(p_{y}')\tilde \phi_{0}(p_{y})\brak{p_{y}'}{y}\brak{y}{p_{y}}\exp(-i\l(p_{y}-p_{y}')\t)\nonumber\\
&&\times \frac{pp'}{m}\brak{\psi_{0}}{p'}\brak{p}{\psi_{0}}\exp(-i(E-E')\t)\nonumber\\
&&\times \frac{2}{\sqrt{\l^{2}p_{y} p_{y}'}}\delta\left(\sqrt{2m(E+\l p_{y})}-\sqrt{2m(E'+\l p_{y}')}\right).
\eea
Using the formula $\delta(f(x))=\delta(x)/f'(0)$, we can carry out the $p_{y}'$ integral to give,
\bea
\Pi(y)
&\approx&\int dp_{y} dpdp'|\tilde \phi_{0}(p_{y})|^{2}\frac{pp'}{m^{2}}\brak{\psi_{0}}{p'}\brak{p}{\psi_{0}}\exp\left(-i\frac{(E-E')}{\l}y\right)\frac{2}{\l}\sqrt{\frac{2m}{\l p_{y}}}\label{8.3.17}\\
&=&\int dp_{y}|\tilde \phi_{0}(p_{y})|^{2}\frac{2}{ m^{2}}\sqrt{\frac{2m}{\l p_{y}}}\bra{\psi_{0}}\exp\left(iH_{0}\frac{y}{\l}\right)\hat p\delta(\hat x)\hat p\exp\left(-iH_{0}\frac{y}{\l}\right)\ket{\psi_{0}}\nonumber\\
&=& A\bra{\psi_{0}}\exp\left(iH_{0}\frac{y}{\l}\right)\hat p\delta(\hat x)\hat p\exp\left(-iH_{0}\frac{y}{\l}\right)\ket{\psi_{0}}
\eea
where $A$ is some constant whose explicit form is not required,
and we have used the fact that $E<<\l p_{y}$.
We see therefore that in this limit the probability of
finding the clock at a position $y$ is given by the kinetic
energy density of the system at the time $t=y/\l$, in agreement
with Eqs.(\ref{1.8.Zeno}) and (\ref{1.8.normalized}).

Note that there is no response function involved in this case, as
one might have expected from the general form Eq.(\ref{1.8.best}).
(A similar feature was noted in the complex potential model of
Ref.\cite{Hal2}). It seems likely that this is because the strong
measurement prevents the particle from leaving $x>0$ until the
last moment, so that
the response function $R(t,s)$ is effectively a delta-function
concentrated around the latest time.

\subsubsection{General Case}

As well as the approximations valid for $E<<\l p_{y}$, the key to the analysis in the special case presented above is that the position space eigenfunction of the clock Hamiltonian with eigenvalue $p_{y}$ takes the simple form,
\beq
\brak{y}{p_{y}}=\frac{1}{\sqrt{2\pi}}\exp(i y p_{y}).
\eeq
This greatly simplifies the resulting calculation. For the case of a more general clock Hamiltonian, the eigenstates will not have this simple form. Instead we make a standard WKB approximation for the eigenstates of the clock,
\beq
\brak{y}{\h}=C(y,\h)\exp(i S(y,\h))
\eeq
where $S(y,\h)$ is the Hamilton Jacobi function of the clock at fixed energy.  This means Eq.(\ref{8.3.17}) becomes,
\bea
\Pi(y)
&\approx&\int d\h dpdp'|\brak{\h}{\phi_{0}}|^{2}\frac{pp'}{m^{2}}\brak{\psi_{0}}{p'}\brak{p}{\psi_{0}}\brak{\h+\frac{(E-E')}{\l}}{y}\brak{y}{\h}\frac{2}{\l}\sqrt{\frac{2m}{\l \h}}\nonumber\\
&\approx&\int d\h|\brak{\h}{\phi_{0}}|^{2}\frac{2}{ m^{2}}\sqrt{\frac{2m}{\l \h}}|C(y,\h)|^{2}\nonumber\\
&&\times\bra{\psi_{0}}\exp\left(iH_{0}\frac{1}{\l}\frac{\partial S(y,\h)}{\partial \e}\right)\hat p\delta(\hat x)\hat p\exp\left(-iH_{0}\frac{1}{\l}\frac{\partial S(y,\h)}{\partial \h} \right)\ket{\psi_{0}}\label{8.zfin}
\label{8.Pistrong}
\eea
where we have used,
\beq
\brak{\h+\frac{(E-E')}{\l}}{y}\brak{y}{\h}\approx |C(y,\h)|^{2}\exp\left(-i\frac{(E-E')}{\l} \frac{\partial S(y,\h)}{\partial \h}\right),
\eeq
which is valid for $E-E'<<\l \h$.

We now suppose that the clock state is a simple Gaussian in $y$,
or equivalently in $p_y$. It follows that it will be peaked in
$\h$ about some value $\h_0$. This means that the integral over $\h_0$ in
Eq.(\ref{8.Pistrong}) may be carried out. The result for $\Pi(t)$ will again be
proportional to the kinetic energy density, of the form
Eq.(\ref{1.8.Zeno}), where the relationship between $t$ and the
pointer variable $y$ is defined by the equation
\beq
t=\frac{1}{\l}\frac{\partial S(y,\h_0)}{\partial \h}
\eeq
as one might expect from Hamilton-Jacobi theory \cite{Goldstein}.
Hence the arrival time distribution has the expected general form,
Eq.(\ref{1.8.Zeno}) (and therefore Eq.(\ref{1.8.normalized}) also holds),
but the precise definition of the time variable depends on the
properties of the clock.

\section{Dwell Time Distribution from an Idealised Clock}\label{sec8.4}

We now turn to the related issue of dwell times. Here the aim is
to measure the time spent by the particle in a given region of
space which, for simplicity, we take to be the region $[-L,L]$.
This is portrayed in Figure (\ref{fig:8.3}). In
this section we will work exclusively in the weak coupling regime
where $E>>\l \h$.

\begin{figure}[htbp] 
   \centering
   \includegraphics[width=5in]{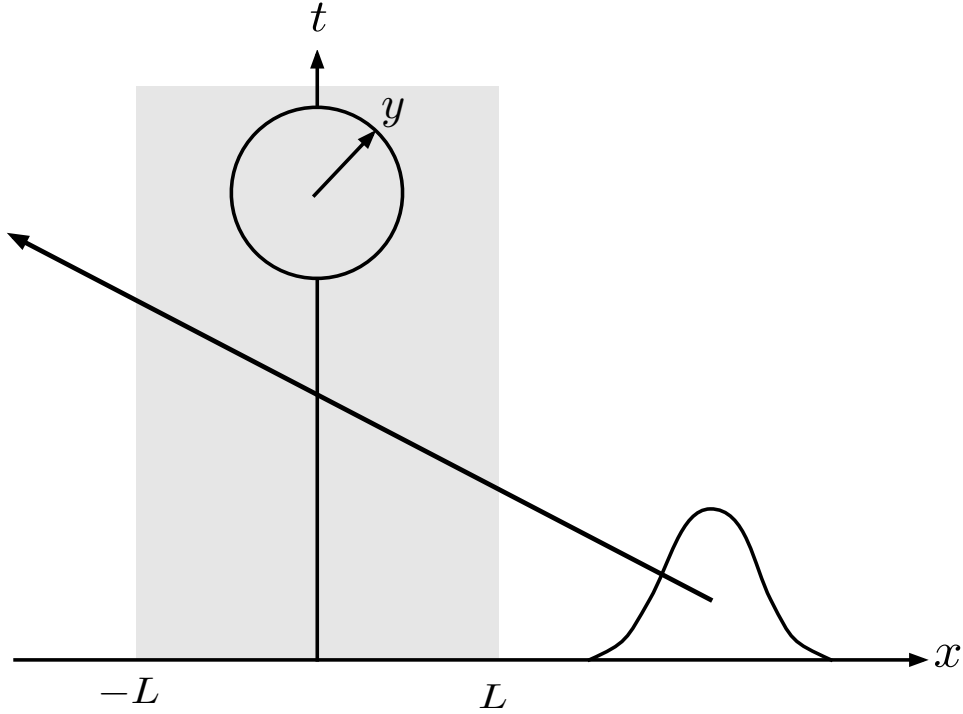}
   \caption[The Dwell Time Problem, Defined Using a Model Clock.]{{\em The dwell time problem, defined using a model clock. The clock runs while the particle is in the region $[-L,L]$.}}
   \label{fig:8.3}
\end{figure}

The starting point is the final state of the particle plus clock,
Eq.(\ref{8.psifinal}), which we write as
\beq
\Psi (x,y, \tau )=\bra{x,y}e^{-i(H_{0}+\l H_{c}\chi(\hat x))\t}\ket{\Psi_{0}}.
\eeq
where $\chi(\hat x)=\theta(\hat x+L)\theta(L-\hat x)$.
We wish to re-express this using the path decomposition in a similar way to Eq.(\ref{8.atstart}).
For this case we need a PDX which is more general than the one
used for the arrival time, since there are now crossings of two
surfaces.
One way to proceed is to use the path integral expression for the first crossing of $x=L$ and
$x=-L$, which is
\bea
\Psi (x,y, \tau)&=&\frac{1}{m^{2}}\int d\h\brak{\h}{\phi_{0}}\brak{y}{\h}\int_{0}^{\infty}ds\int_{-\infty}^{\t-s}dt \bra{x}\exp(-i(H_{0}+\l\chi(\hat x)\h)(\t-t-s))\ket{-L}\nonumber\\
&&\bra{-L}\hat p\exp(-i(H_{0}+\l\chi(\hat x)\h)s)\ket{L}\bra{L}\hat p\exp(-iH_{0}t)\ket{\psi_{0}},\nonumber
\eea
This is shown in Figure (\ref{fig:8.4}).
\begin{figure}[htbp] 
   \centering
   \includegraphics[width=5in]{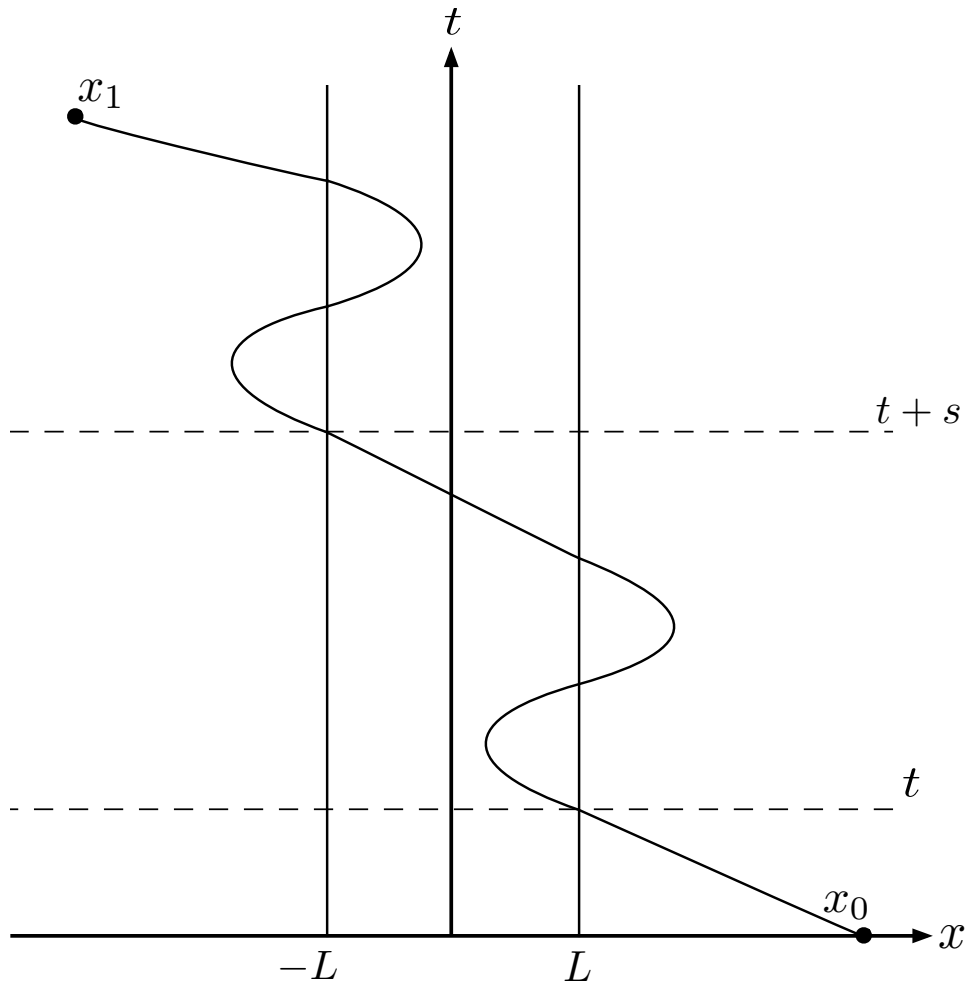}
   \caption[The PDX Used for the Dwell Time Problem.]{{\em The PDX used for the dwell time problem: Paths from $x_{0}>0$ to $x_{1}<0$ have a first crossing of $x=L$ at a time $t$ and a first crossing of $x=-L$ at a time $t+s$.} }
   \label{fig:8.4}
\end{figure}
However there are other choices we could make.
We could consider the first crossing of $x=L$ and the last crossing of $x=-L$ for example.
In the semiclassical limit these choices lead to equivalent expressions for the dwell time.
It would be interesting to explore what differences do arise in other regimes. This will be addressed elsewhere.

It will prove more useful to work with the wavefunction in position space for the clock and momentum space for the particle. Changing to this representation, and making use of the semiclassical approximation we obtain,
\bea
\brak{p,y}{\Psi_{\t}}&\approx&\frac{1}{m^{2}}\int d\h\brak{y}{\h}\brak{\h}{\phi_{0}}\int_{0}^{\infty}ds\int_{-\infty}^{\t-s}dt \bra{p}\exp(-iH_{0}(\t-t-s))\ket{-L}\nonumber\\
&\times&\bra{-L}\hat p\exp(-i(H_{0}+\l \h)s)\ket{L}\bra{L}\hat p\exp(-iH_{0}t)\ket{\psi_{0}}
\eea
Now we make the standard scattering approximation of letting the upper limits of the integrals go to infinity. This means we can carry out the $t$ and $s$ integrals to obtain,
\bea
\brak{p,y}{\Psi_{\t}}&\approx&\int d\h\brak{y}{\h}\brak{\h}{\phi_{0}} \exp(-iE\t)\exp(-i2Lm\l \h/|p|)\brak{p}{\psi_{0}},
\eea
where we have used the standard integral \cite{HaYe1},
\beq
\int_{0}^{\infty} ds \bra{x}\exp(-iH_{0}s)\hat p\ket{0}e^{iEs}=m\exp(i|x|\sqrt{2mE})
\eeq
and
\beq
\int_{-\infty}^{\infty}\frac{dt}{2\pi}\bra{x}\hat p \exp(-i(H_{0}-E)t)\ket{\psi_{0}}=\bra{x}\hat p \delta(H_{0}-E)\ket{\psi_{0}}=m\bra{x}\delta(\hat p-p)\ket{\psi_{0}},
\eeq
Here, we have neglected the term involving $\delta(\hat p+p)$ since this corresponds to reflection, which will be negligible in this semiclassical limit. We have also use the fact that $E\gg\l \h$.
We therefore obtain the distribution for $y$ as,
\bea
\Pi(y)&=&\int dp \left|\brak{p,y}{\Psi_{\t}}\right|^{2}\nonumber\\
&=&\int d\h d\h'\brak{\phi_{0}}{\h'}\brak{\h'}{y}\brak{y}{\h}\brak{\h}{\phi_{0}}\int dp|\psi_{0}(p)|^{2}\exp\left(\frac{i2Lm\l}{|p|}(\h'-\h)\right)\nonumber\\
&=&\int dp|\psi_{0}(p)|^{2}|\Phi(y,2Lm/|p|)|^{2}
\eea
where,
\bea
\Phi(y,2Lm/|p|)&=&\int d\h\brak{y}{\h}\brak{\h}{\phi_{0}}\exp\left(\frac{-i2Lm\l}{|p|}\h\right)\nonumber\\
&=&\bra{y}\exp\left(\frac{-i2Lm\l}{|p|} H_{c}\right)\ket{\phi_{0}}
\eea
is the clock wavefunction.
We may rewrite this as
\beq
\Pi (y) = \int dt \ |\Phi(y,t)|^{2} \langle \psi_0 | \delta \left( t
-\frac { 2 m L } { | \hat p | } \right) | \psi_0 \rangle
\eeq
It is therefore of precisely the desired form,
Eqs.(\ref{8.dwellsemi}), (\ref{1.8.best}), with $ |\Phi(y,t)|^{2} $
playing the role of the response function. The discussion of clock
characteristics is then exactly the same as the arrival time case
discussed in Section 3.

\section{Conclusion}\label{con8}

We have studied the arrival and dwell time problems defined using
a model clock with a reasonably general Hamiltonian. We found that
in the limit of weak particle-clock coupling, the time
of arrival probability distribution is given by the probability
current density Eq.(\ref{1.8.cur}), smeared with some function depending on the
initial clock wave function, Eq.(\ref{1.8.best}). This is expected semiclassically,
agrees with previous studies and is independent
the precise form of the clock Hamiltonian.

In the regime of strong coupling, we found that the arrival time
distribution is proportional to the kinetic energy density of the
particle, in agreement with earlier approaches using a complex
potential. The fact that two very different models give the same
result in this regime suggests that the form Eq.(\ref{1.8.normalized})
is the generic result in this regime, independent of the method
of measurement. It would be of interest to develop a general
argument to prove this. (See Ref.\cite{Ech} for further discussion
of this regime).

For the case of dwell time, we have shown that the dwell time
distribution measured by our model clock may be written in terms
of the dwell time operator in semiclassical form, smeared
with some convolution function, Eqs.(\ref{8.dwellsemi}),
(\ref{1.8.best}).

In all of these cases, the precise form of the
clock Hamiltonian and clock initial state determine
the relationship between time $t$ and the pointer variable
$y$ and they determine the form of the response function
$R$ in the general form Eq.(\ref{1.8.best}). These are particularly
simple for the special case $H_C = \hat p_y $ explored
previously. However, what is
important is that, once the definition of the time variable is fixed,
the clock characteristics do
not effect the form of the underlying distributions -- the $\Pi
(s)$ in Eq.(\ref{1.8.best}). The $\Pi (s) $ are always one of the
general forms
Eqs.(\ref{1.8.cur}), (\ref{1.8.Zeno}) and (\ref{8.dwellsemi}), no matter what
the clock characteristics are.
This means that these general forms will always play
central
role, irrespective of how they are measured.

\chapter{Summary and Some Open Questions}
\epigraph{This is the first thing\\ I have understood:\\ Time is the echo of an axe\\ Within a wood.}{Philip Larkin}

\section{Summary}

The main result presented in this thesis is an analysis of the arrival time problem in the framework of decoherent histories. This analysis has given us an axiomatic way of understanding how probabilities may be assigned to this observable, as well as the conditions under which these probabilities are meaningful.  We have carried out this analysis both for a free particle, where decoherence is achieved for a limit class of states, and for a particle coupled to a environment, where decoherence is achieved for a general class of states after a given time. The main tool which allowed us to proceed was the equivalence between pulsed and continuous measurements set out in Chapter 4. This result allows us to construct the appropriate class operators either by writing them in terms of complex potentials, or by giving us the confidence to use semiclassical approximations provided the energies and times involved satisfy the conditions laid out in Chapters 4 and 5. The major advance achieved compared to previous work is a decoherent histories based framework that respects the Zeno limit. In addition we have also shown how the semiclassical, heuristic formulae may be derived from framework. In Chapter 8 we also demonstrated that the probabilities obtained via decoherent histories agree with those computed from model clocks. This general scheme of decoherent histories, using path integral methods and complex potentials to define class operators, seems to be general enough to encompass any time observable in quantum theory. We encourage the reader to test this for themselves.

\section{Open Questions and Future Work}

We list here, with some short comments, a few obvious questions which arise from the work presented in this thesis, but which time constraints have prevented the author from studying in any depth. The majority of these questions would form the basis of interesting further study.

\subsection*{Decoherent Histories Analysis of the Dwell Time Problem}

The most obvious question one might ask of the work presented here is whether it can be extended to cover other time observables. The most obvious such extension would be to the definition of dwell times. Dwell times were discussed very briefly in Chapter 8 where they were defined using a model clock, but no attempt has been made here to perform a decoherent histories analysis. The key step in carrying this out lies in the construction of the relevant class operators. One might imagine doing this either via the introduction of a suitable complex potential or potentials in the manner of Chapter 5, or via some semiclassical approximation in the manner of Chapter 6. What is particularly interesting about the dwell time problem is that there exists a relatively straightforward operator formalism, see Ref.\cite{dwell}, and it is not clear at the moment how the decoherent histories analysis fits in with this.

\subsection*{Quantum Zeno Effect for Open Systems}

Another topic which we touched on briefly in Chapter 7 was the occurrence of the quantum Zeno effect in an open system. Whilst we studied the evolution of a particle undergoing continuous measurement in detail in Chapter 4, it seems at first sight a difficult task to extend the analysis to a particle coupled to some environment. As we pointed out in Chapter 7, this is an important question since it is the Zeno effect that limits the accuracy with which we can define time observables in quantum theory. On the face of it one might expect that the Zeno effect should somehow vanish in the classical limit, so that inclusion of an environment should reduce the ``Zeno time'' from $1/E$.  However it is not at all obvious how this might come about, since one generally assumes that all time scales associated with the environment are much longer than $1/E$. In part, this is related to the issues that for a genuine step potential the reflection and transmission coefficients, Eqs.(\ref{4.7}) and (\ref{4.7b}), do not depend on $\hbar$ and thus have no naive classical limit (see Chapter 2).

\subsection*{Time Observables and the Backflow Effect}

Another question which we have encountered but not discussed in any detail in this thesis concerns the role of the backflow effect in the definition of time observables. The backflow effect causes the current Eq.(\ref{1.8.cur}) to be negative even when classically it should be positive, and thus prevents decoherence of histories, see Section \ref{sec:5.5.4}. However unlike the quantum Zeno effect, we do not have a good grasp of the time scale on which this effect occurs. Whilst it is true that one can always find a state which displays backflow for an arbitrarily long period of time \cite{BrMe}, one suspects that for a given state this time interval is determined by some property of the state such as the moments of the Hamiltonian. Since it it the backflow effect and the Zeno effect that together determine the accuracy with which time observables may be defined, further investigation of these effects would be useful. Work on this has already begun \cite{YeBack}.


\chapter*{Appendix: Some Properties of the Current}
\addcontentsline{toc}{chapter}{Appendix: Some Properties of the Current}

We have derived the expression
\beq
p (0, T) = \int^{T}_{0} dt \ J(t)
\eeq
as the approximate probability for crossing the origin during the time interval $[0,T]$, where
$J(t)$ is the usual quantum-mechanical current. The current itself is not necessarily positive
due to backflow. Here we explore the possibility that averaging it over time might improve
the situation. On the one hand, the results of Bracken and Melloy \cite{BrMe} show that
there is always {\it some} state for which $p(0,T)$ defined above is negative, for any $T$.
On the other hand, for a {\it given} state, one might hope that $p(0,T)$ will be positive
for sufficiently large $T$. Here we give a brief argument for this, which also makes
contact with the negativity of the Wigner function.

The current can be written in terms of the Wigner function $ W(p,x)$ as
\beq
J(t)= - \int dp \ \frac{p}{m}\  W(p,0,t).
\eeq
The Wigner function evolves freely according to $W(p,x,t)=W(p,x-pt/m,0)$. We assume
it has support only on negative momentum states, with average momentum $p_0 <0$
and momentum width $\sigma_p$.

Consider a time interval $0<t < T$ over which backflow occurs. It is clear that in order
for this to occur the Wigner function must be negative for at least some of this interval.
We can write,
\bea
p(0,T) &=& - \int^{T}_{0}dt\ \int dp \ \frac{p}{m}  \ W(p,-pt/m,0)
\nonumber \\
&=&  \int dp \ \int_0^{|p|T/m}  dx\   \ W(p,x,0)
\eea
So $p(0,T)$ is given by the average of the Wigner function over a region of
phase space. We now recall a standard property of the Wigner function
which is that, broadly speaking, it will tend to be positive when averaged
over a region of phase space of size greater than order $1$ (in the units used
here where $\hbar = 1 $).
This region is of size of order $ | p_0 | T / m $ in the $x$-direction but
infinite in the $p$-direction. However, the Wigner function has momentum
spread $\sigma_p$, so the effective size averaged over is $\sigma_p p_{0} T/m$
which is approximately the same as $\Delta H\ T$. This means that we expect that
$p(0,T)$ will be positive as long as
\beq
T  \ > \ \frac {1} { \Delta H}
\eeq
Hence, as expected, the integrated current will be positive for $T$ sufficiently
large and the key timescale is the Zeno time. This heuristic argument
will be revisited in more detail elsewhere.

\bibliography{apssamp}

\end{document}